\documentclass[12pt,letterpaper,english,aps,tightenlinesletterpaper,groupaddress,secnumarabic,nofootinbib]{revtex4}
\usepackage{mathptmx}

\usepackage[T1]{fontenc}
\usepackage[latin9]{inputenc}
\usepackage{babel}
\usepackage{amsmath}
\usepackage{cancel}
\usepackage{graphicx}
\usepackage{esint}
\usepackage[unicode=true,pdfusetitle,
 bookmarks=true,bookmarksnumbered=false,bookmarksopen=false,
 breaklinks=false,pdfborder={0 0 1},backref=section,colorlinks=false]
 {hyperref}
\hypersetup{
 citecolor=blue}

\makeatletter

\pdfpageheight\paperheight
\pdfpagewidth\paperwidth

\@ifundefined{textcolor}{}
{%
 \definecolor{BLACK}{gray}{0}
 \definecolor{WHITE}{gray}{1}
 \definecolor{RED}{rgb}{1,0,0}
 \definecolor{GREEN}{rgb}{0,1,0}
 \definecolor{BLUE}{rgb}{0,0,1}
 \definecolor{CYAN}{cmyk}{1,0,0,0}
 \definecolor{MAGENTA}{cmyk}{0,1,0,0}
 \definecolor{YELLOW}{cmyk}{0,0,1,0}
}

\usepackage{babel}
\usepackage{tensind}

\makeatother

\begin{document}

\title{Non-Abelian basis tensor gauge theory}

\author{Edward E. Basso}

\email{ebasso@wisc.edu}

\affiliation{Department of Physics, University of Wisconsin-Madison, Madison,
WI 53706, USA}

\author{Daniel J. H. Chung}

\email{danielchung@wisc.edu}

\affiliation{Department of Physics, University of Wisconsin-Madison, Madison,
WI 53706, USA}
\begin{abstract}
Basis tensor gauge theory is a vierbein analog reformulation of ordinary
gauge theories in which the difference of local field degrees of freedom
has the interpretation of an object similar to a Wilson line. Here
we present a non-Abelian basis tensor gauge theory formalism. Unlike
in the Abelian case, the map between the ordinary gauge field and
the basis tensor gauge field is nonlinear. To test the formalism,
we compute the beta function and the two-point function at the one-loop
level in non-Abelian basis tensor gauge theory and show that it reproduces
the well-known results from the usual formulation of non-Abelian gauge
theory.
\end{abstract}
\maketitle

\section{Introduction}

The Standard Model (SM) of particle physics~\cite{Glashow:1961tr,Weinberg:1967tq,Salam:1968rm,Gross:1973id,Politzer:1973fx,Weinberg:1996kr,Ramond:1999vh,Langacker:2010zza,Aad:2012tfa,Chatrchyan:2012xdj}
is usually formulated with gauge fields that transform inhomogeneously
under the gauge group: i.e.~they are connections on principal bundles
(see e.g.\,\cite{Nakahara:2003nw,Wu:1975es}). This mechanism is used
to construct covariant derivatives acting on matter fields, which
allows a simple recipe for constructing kinetic terms for local field
theories living on principal bundles. Gauge theories of this sort
have a long history (see e.g.\,\cite{Weyl:1919fi,Weyl:1929fm,Yang:1954ek,Abers:1973qs,Itzykson:1980rh,Polyakov:1987ez,Sterman:1994ce,'tHooft:1995gh,Weinberg:1996kr})
and are very economical in describing the physics locally at the cost
of introducing redundancies into the system. Despite this long history,
rewriting gauge theories in novel formalisms continue to offer insights
into both computational techniques and ideas for physics beyond the
SM (see e.g.\,\cite{Arkani-Hamed:2013jha,Arkani-Hamed:2017jhn,Badger:2005zh,Elvang:2013cua,Henn:2014yza,Christensen:2018zcq,Witten:1998qj,Aharony:1999ti}).

The work of~\cite{Chung:2016lhv} gives a reformulation of $U(1)$
gauge theories in analogy with the vierbein formalism of general relativity.
In that paper, it was shown that the vierbein analog field $G_{\phantom{\alpha}\beta}^{\alpha}$
transforms homogeneously under the $U(1)$ gauge group and satisfies
certain constraints, in contrast with the ordinary formulation in
which the gauge field transforms inhomogeneously. The nonlinear map
between the ordinary $A_{\mu}$ field and $G_{\phantom{\alpha}\beta}^{\alpha}$
can be changed to a linear relationship using a set of $N$ unconstrained
scalar fields $\theta_{a}(x)$ in $N$ dimensions.\footnote{In~\cite{Chung:2016lhv}, we used upper indices to denote the components
of $\theta^{a}(x)$ field. In this work, the analogous index will
appear as a lower index. } The field theory of $\theta_{a}(x)$ is called \emph{basis tensor
gauge theory}~(BTGT), which can be viewed as a theory of Wilson lines
(e.g.\,\cite{Wilson:1974sk,Giles:1981ej,Migdal:1984gj,Terning:1991yt,Gross:2000ba,Kapustin:2005py,Cherednikov:2008ua,Mandelstam:1962mi}
and references therein) modded by a particular symmetry that is required
to allow only couplings equivalent to ordinary gauge theories. In
\cite{Chung:2017zck}, the Ward identities of the theory were constructed
and the theory was explicitly shown to be one-loop stable.

In this work, we present a non-Abelian version of basis tensor gauge
theory. Just as in the Abelian case, the interpretation of the basis
tensor gauge field is similar to a Wilson line. This means that the
basis tensor field $\theta_{a}^{A}(x)$ is more non-local when expressed
in terms of the ordinary gauge potential $A_{\mu}^{B}$. Unlike in
the Abelian case, the map between $\theta_{a}^{A}(x)$ and $A_{\mu}^{B}$
is nonlinear. A perturbation theory can be defined in powers of $\theta_{a}^{A}$
that allows us to have a finite power expansion map between $\theta_{a}^{A}$
and $A_{\mu}^{B}$. Just as in the Abelian case, we can impose a symmetry
(BTGT symmetry) to eliminate charge violating couplings and enforce
positivity of the Hamiltonian.

As the map between $\theta_{a}^{A}$ and $A_{\mu}^{B}$ is nonlinear,
unlike in the Abelian case, the choice of $\theta_{a}^{A}$ variables
to parameterize the gauge manifold target space is not motivated by
simplicity. On the other hand, this motivation still exists since
the number of functional degrees of freedom between $A_{\mu}^{B}$
theories and $\theta_{m}^{B}$ theories naturally match without imposing
additional constraints on the vierbein-like field that would make
it difficult to quantize. The basis choice is also a natural generalization
of the Abelian construction (i.e. both are gauge group manifold target
space fields), and it has the same relationship with the Wilson line
as in the Abelian case. Furthermore, the BTGT symmetry representation
that stabilizes the theory (e.g.~enforces charge conservation and
bounds the Hamiltonian from below) naturally generalizes the Abelian
theory's representation.

To test the formalism we perturbatively compute the $\beta$-function
and find that it matches the usual result non-Abelian gauge theory
at one loop. We also compute the one-loop divergent contribution to
the $\langle A^{\mu}(x)A^{\nu}(y)\rangle$ correlator, where $A_{\mu}\left[\theta\right]$
is now treated as a local composite operator. We find that before
introducing the counter terms, the divergence that is obtained using
the $\theta_{a}^{A}$ formalism is the same as in the usual $A_{\mu}^{A}(x)$
formalism. This is an indication that the UV structure of ordinary
gauge theories are faithfully reproduced by the non-Abelian BTGT theory.

The order of presentation is as follows. In Section 2, we present
the definition of non-Abelian basis tensor gauge theory. In Section
3, we present the path integral formulation of the BTGT theory. This
includes the perturbative expansion terms similar to what is done
in nonlinear sigma models. To check that the quantum formulation of
BTGT is stable and computable, in section 4, we compute the $\beta$-function
explicitly by renormalizing the two-point functions of the BTGT field
$\theta_{a}^{A},$ the ghost fields $c\bar{c}$, and the $\theta c\bar{c}$
vertex functions. In section 5, we compute the two-point function
$\langle A_{\mu}^{A}(x)A_{\nu}^{B}(y)\rangle$ at one-loop using the
BTGT formalism. We check the transversality of the divergent contribution
consistent with gauge invariance and check that introducing the appropriate
composite operator counter terms allow both $\langle\theta_{a}^{A}(x)\theta_{b}^{B}(y)\rangle$
and $\langle A_{\mu}^{A}(x)A_{\nu}^{B}(y)\rangle$ to be finite. In
section 6, we make a conjecture regarding what the relationship will
be for the infinite number of renormalization constants based on the
computations done in this paper. In section 7, we present our conclusions.
In Appendix A, we collect some of the less-standard notation and conventions
used in this paper. In Appendix B, we derive the relationship between
the non-Abelian basis tensor field and the ordinary gauge field. In
Appendix C, we discuss the representations of gauge and BTGT symmetry
transformations. In Appendix D, we list the Feynman rules for the
theory.

\section{Non-Abelian BTGT basis definition }

In this section, we construct an explicit relationship between the
vierbein analog field $G$ and ordinary non-Abelian gauge field $A$.
We will work with 4 spacetime dimensions throughout this paper to
maintain simplicity and obvious physical relevance even though generalizations
to different spacetime dimensions are straightforward. All repeated
indices will be summed unless specified otherwise. For example, whenever
one side of an equation has indices specified, the other side of the
equation may have repeated indices that are not summed.

Given a field $\phi$ that is a complex scalar transforming under
gauge transformations as 
\begin{equation}
\phi^{k}(x)\rightarrow\left[g(x)\right]^{ks}\phi^{s}(x)
\end{equation}
\begin{equation}
\left[g(x)\right]^{ks}\equiv\left(e^{i\Gamma^{C}(x)T^{C}}\right)^{ks},
\end{equation}
where $(T^{C})^{ab}$ are Hermitian generators of the gauge group
in representation $R$, we define a Lorentz tensor $G_{(f)\,\beta}^{\alpha}$
that exhibits the gauge group transformation property 
\begin{equation}
\left[G_{(f)\,\beta}^{\alpha}(x)\right]^{i}\rightarrow\left[G_{(f)\,\beta}^{\alpha}(x)\right]^{j}\left[g^{-1}\left(x\right)\right]^{ji},\label{eq:Gtransform}
\end{equation}
such that $G_{(f)\,\beta}^{\alpha}\phi$ is gauge invariant, where
$f$ is a basis index that specifies a fixed direction in the gauge
group representation space. The requirement of rank 2 comes from having
enough functional degrees of freedom to match the gauge field functional
degrees of freedom as explained in~\cite{Chung:2016lhv}. More formally,
$G_{(f)\,\beta}^{\alpha}(x)$ is a field that transforms as an $\bar{R}$
from the right under the non-Abelian gauge group representation and
as a rank 2 Lorentz projection tensor. The index $(f)$ in $G_{(f)\,\beta}^{\alpha}$
spans the dimension of the representation. Hence, $G_{(f)\,\beta}^{\alpha}$
contains $2\times16D(R)$ real functional degrees of freedom (in 4-spacetime
dimensions), where $D(R)$ is the dimension of the representation.
The analogy with gravitational vierbeins $(e_{a})_{\mu}$ can be identified
as follows (similar to the Abelian case of~\cite{Chung:2016lhv}):
the indices $\{f,\alpha,\beta\}$ are the analogs of the fictitious
Minkowski space index $a$ of $(e_{a})_{\mu}$, and the representation
of Eq.\,(\ref{eq:Gtransform}) is the analog of the diffeomorphism
acting on the $\mu$ index of ($e_{a})_{\mu}$.

To reproduce ordinary gauge theory with $G_{(f)\,\beta}^{\alpha}$,
we must be able to path integrate over unconstrained functions that
match the number of degrees of freedom in $A_{\mu}$. This means that
we must eliminate the number of field degrees of freedom either by
imposing a constraint through an introduction of an auxiliary field
or explicitly solving such a matching constraint. Since the gauge
field real functional degrees of freedom necessary for constructing
covariant derivatives on fundamental representation fields is $4D(A)$
(where $D(A)$ is the dimension of the adjoint representation), we
need to eliminate $32D(R)-4D(A)$ degrees of freedom. We can accomplish
this by choosing the field degrees of freedom that represent $G_{(f)\,\beta}^{\alpha}$
to live on the target space of the gauge manifold, which will cause
the $D(A)$ dimension matching condition to be satisfied. We can then
construct 4 such sets with the help of a projection tensor (just as
in the Abelian BTGT) to match $4D(A)$ degrees of freedom in $A_{\mu}$:
the gauge manifold target space fields are $\theta_{a}^{C}$ where
$a\in\{0,1,2,3\}$ and $C\in\{1,2,...,D(A)\}$. 

To find a map between $G_{(f)\,\beta}^{\alpha}$ and $\theta_{a}^{C}$,
define an orthonormal set of spacetime-independent vectors $\xi_{(f)}^{l}$
for $f\in\{1,...,{\rm dim}R\}$ that span the group representation
vector space such that the following completeness relationship is
satisfied: 
\begin{equation}
\delta^{kl}=\sum_{f}\xi_{(f)}^{k}\xi_{(f)}^{*l}.\label{eq:normalbasis}
\end{equation}
The $\xi_{\left(f\right)}$ are defined to be invariant under gauge
transformations.

In the spirit of the Abelian case of~\cite{Chung:2016lhv}, the vierbein
analog in the non-Abelian gauge theory can be defined as
\begin{equation}
\boxed{\left(\left[G_{(f)}(x)\right]_{\phantom{\gamma}\delta}^{\gamma}\right)^{j}=\xi_{(f)}^{*l}\left[\left(\exp\left[-i\theta_{a}^{M}(x)H^{a}T^{M}\right]\right)_{\phantom{\gamma}\delta}^{\gamma}\right]^{lj}.}\label{eq:nonabelianvierbein}
\end{equation}
Here the objects $H^{a}$ with $a\in\{0,...,3\}$ are $4\times4$
real matrices that transform under Lorentz transformations as an (1,1)
tensor satisfying $[H^{a},H^{b}]=0$, which satisfy the completeness
relationship 
\begin{equation}
\sum_{a=0}^{3}(H^{a})_{\phantom{\mu}\nu}^{\mu}=\delta_{\phantom{\mu}\nu}^{\mu}
\end{equation}
and the orthonormality condition 
\begin{equation}
{\rm Tr}\left(H^{a}H^{b}\right)=\delta^{ab},
\end{equation}
(just as in the Abelian case of~\cite{Chung:2016lhv}). These matrices
can be chosen to have the following orthonormal projection property
\begin{equation}
(H^{a})_{\phantom{\mu}\nu}^{\mu}(H^{b})_{\phantom{\nu}\beta}^{\nu}=\delta^{ab}(H^{a})_{\phantom{\mu}\beta}^{\mu}\,\,\,\,\,\,\mbox{no sum on }a\label{eq:projection}
\end{equation}
and symmetry property 
\begin{equation}
(H^{a})^{\mu\nu}=(H^{a})^{\nu\mu}.\label{eq:symmetry}
\end{equation}
The fields $\theta_{a}^{M}(x)$ are real scalar fields which transform
under gauge transformations as 
\begin{equation}
U_{a}\rightarrow e^{i\Gamma}U_{a}\label{eq:gaugetransform}
\end{equation}
where 
\begin{equation}
U_{a}\equiv\exp\left[i\theta_{a}^{A}T^{A}\right]\label{eq:uvariable}
\end{equation}
\begin{equation}
\Gamma\equiv\Gamma^{B}T^{B}.
\end{equation}
The reason why $\theta_{a}^{A}$ is easier to work with than $G_{(f)}(x)$
is that it is unconstrained, similar to the $\pi$ variable being
easier to work with compared to $U(\pi)$ in sigma models~\cite{Weinberg:1996kr}.

There are several salient features to note regarding Eq.\,(\ref{eq:nonabelianvierbein}).
Given the representation identity 
\begin{equation}
\psi^{C}\rightarrow(g_{{\rm adj}})^{CS}\psi^{S},
\end{equation}
if 
\begin{equation}
\psi^{C}T^{C}\rightarrow g[\psi^{C}T^{C}]g^{-1},
\end{equation}
where $g_{{\rm adj}}$ is the adjoint representation group element
(independent of the representation of $g$), we might naively expect
that $\theta_{a}^{M}$ has its $M$ index transforming as an adjoint.
However, this is not true because the transformation property of $\theta^{M}$
is 
\begin{equation}
\xi_{(f)}^{*}\left(\exp\left[-i\theta_{a}^{M}(x)H^{a}T^{M}\right]\right)_{\phantom{\gamma}\delta}^{\gamma}\rightarrow\xi_{(f)}^{*}\left(\exp\left[-i\theta_{a}^{M}(x)H^{a}T^{M}\right]\right)_{\phantom{\gamma}\delta}^{\gamma}g^{-1}(x),
\end{equation}
and \emph{not} 
\begin{equation}
\xi_{(f)}^{*}\left(\exp\left[-i\theta_{a}^{M}(x)H^{a}T^{M}\right]\right)_{\phantom{\gamma}\delta}^{\gamma}\rightarrow\xi_{(f)}^{*}g(x)\left(\exp\left[-i\theta_{a}^{M}(x)H^{a}T^{M}\right]\right)_{\phantom{\gamma}\delta}^{\gamma}g^{-1}(x)
\end{equation}
in Eq.\,(\ref{eq:nonabelianvierbein}). Another aspect is that the
index $f$ in Eq.\,(\ref{eq:nonabelianvierbein}) runs from 1 to ${\rm dim}(R)$
components in $G_{(f)}(x)$, but the number of independent scalar
field degrees of freedom of $G_{(f)}(x)$ in terms of $\theta_{m}^{A}$
is the rank of the group times the spacetime dimension $4$ (spanned
by $m\in\{0,...,3\}$). This is similar to the ordinary gauge field
having ${\rm dim}(R)$ components of the $f$ index in $A_{\mu}^{M}(T^{M})^{fk}$
but counting in terms of $A_{\mu}^{M}$, the index $M$ runs through
the rank of the group.

Another interesting relationship is the map between the ordinary non-Abelian
gauge field and $\left[G_{(f)}(x)\right]_{\phantom{\gamma}\delta}^{\gamma}$.
As shown in Appendix \ref{sec:Construction-of-Nonabelian}, the relationship
is 
\begin{equation}
A_{\mu}=i\left[G^{-1\alpha\beta}\right]\left[\partial_{\alpha}G_{\beta\mu}\right]\label{eq:agrelation}
\end{equation}
where $G_{\beta\mu}$ are related to the basis tensor as 
\begin{equation}
\left[G_{\beta\mu}\right]^{qm}=\sum_{f}^{{\rm dim}R}\xi_{(f)}^{q}\left[G_{(f)\beta\mu}\right]^{m}.
\end{equation}
We note that the relationship of $U_{a}$ and $G_{\phantom{\alpha}\beta}^{\alpha}$
is 
\begin{equation}
G_{\phantom{\mu}\lambda}^{\mu}=[H^{b}]_{\phantom{\mu}\lambda}^{\mu}U_{b}^{\dagger}\label{eq:gintermsofu}
\end{equation}
according to Eq.\,(\ref{eq:projection}). Owing to the projection
property of Eq.\,(\ref{eq:projection}) in a conveniently normalized
basis, the ordinary non-Abelian gauge field can also be rewritten
as 
\begin{equation}
A_{\mu}=iU_{a}\tilde{\partial}_{\mu}^{a}U_{a}^{\dagger},\label{eq:amurewrite}
\end{equation}
where 
\begin{equation}
\tilde{\partial}_{\nu}^{a}\equiv(H^{a})_{\phantom{\mu}\nu}^{\mu}\partial_{\mu}.
\end{equation}
This can be seen simply by using Eq.\,(\ref{eq:projection}) and Eq.\,(\ref{eq:gintermsofu});
\begin{eqnarray}
A_{\mu} & = & i\sum_{a}U_{a}(H^{a})^{\alpha\beta}\sum_{b}\partial_{\alpha}(H^{b})_{\beta\mu}U_{b}^{\dagger}\\
 & = & i\sum_{a}\sum_{b}\delta_{ab}(H^{a})_{\phantom{\alpha}\mu}^{\alpha}U_{a}\partial_{\alpha}U_{b}^{\dagger}\\
 & = & i\sum_{a}U_{a}\tilde{\partial}_{\mu}^{a}U_{a}^{\dagger}\,.
\end{eqnarray}
As discussed in Appendix \ref{sec:Construction-of-Nonabelian}, the
relationship between the $\theta_{a}^{A}$ field and the ordinary
non-Abelian gauge fields can be written explicitly as 
\begin{equation}
A_{\mu}^{Q}=\sum_{c}\left(\left(\left[\theta_{c}^{J}f^{J}\right]^{-1}\right)^{QR}\left(e^{\theta_{c}^{K}f^{K}}-1\right)^{RB}\tilde{\partial}_{\mu}^{c}\theta_{c}^{B}\right),\label{eq:nonabelianamu}
\end{equation}
where $f^{J}$ is a structure constant matrix having the components
$(f^{J})^{AB}=f^{JAB}$. The non-Abelian Eq.\,(\ref{eq:nonabelianamu})
reduces to the Abelian case of~\cite{Chung:2016lhv} in the limit
that the structure constant matrix $f\rightarrow0$. Note that the
map between $\theta_{c}^{B}$ and $A$ differ by a minus sign compared
to the original Abelian BTGT paper~\cite{Chung:2016lhv} because the
sign convention for $\theta$ has been flipped (see Eq.\,(23) of that
paper and Eq.\,(\ref{eq:nonabelianvierbein}) above).\footnote{Note that Ref.\,\cite{Chung:2016lhv} uses the notation of having
the basis tensor index $c$ of $\theta^{c}$ instead of $\theta_{c}^{B}$
as in Eq.\,(\ref{eq:nonabelianvierbein}).} As we see in this expression, one key difference between the Abelian
BTGT and the non-Abelian BTGT is that the map between the ordinary
gauge field $A$ and the $\theta$ field is linear in the Abelian
case and nonlinear in the non-Abelian case. On the other hand, since
$\theta_{c}^{B}$ represents a solution to a first order differential
equation, it still does have the interpretation of a type of object
similar to a Wilson line.

As noted in~\cite{Chung:2016lhv}, because gauge invariance is insufficient
to impose global charge conservation (unlike in the usual gauge theory
formulation), we must impose a new symmetry introduced in~\cite{Chung:2016lhv}
called a BTGT symmetry. The BTGT transformation in the non-Abelian
case is 
\begin{equation}
U_{a}\rightarrow U_{a}e^{iZ_{a}}\label{eq:BTGTtransform}
\end{equation}
\begin{equation}
Z_{a}\equiv Z_{a}^{B}T^{B},
\end{equation}
where $Z_{a}^{B}$ satisfies 
\begin{equation}
(H^{a})_{\phantom{\lambda}\mu}^{\lambda}\partial_{\lambda}Z_{a}^{B}=0.
\end{equation}
Because this transformation will not transform the gauge field variable
when written in terms of the ordinary $A_{\mu}^{M}$ basis, this transformation
is independent of the usual gauge transformations. Infinitesimally,
Eqs.\,(\ref{eq:Gtransform}) and (\ref{eq:BTGTtransform}) can be
rewritten as 
\begin{equation}
\delta\theta_{a}^{A}=\left(\frac{f\cdot\theta_{a}}{\exp\left[f\cdot\theta_{a}\right]-1}\right)^{AB}\Gamma^{B}+\left(\frac{f\cdot\theta_{a}}{1-\exp\left[-f\cdot\theta_{a}\right]}\right)^{AB}Z_{a}^{B}
\end{equation}
to linear order in $\Gamma^{B}$ and $Z_{a}^{B}$, where $\left(f\cdot\theta_{a}\right)^{MN}\equiv f^{CMN}\theta_{a}^{C}$.
The derivation of this linearized transformation is presented in Appendix
\ref{sec:Gauge-and-BTGT}. Finally, note that we can also write the
combined gauge and BTGT transformations acting on $G_{\phantom{\alpha}\beta}^{\alpha}(x)$
as 
\begin{equation}
[H^{f}]_{\phantom{\mu}\mu}^{\psi}G_{\phantom{\mu}\lambda}^{\mu}\rightarrow e^{-iZ_{f}^{B}(x)T^{B}}[H^{f}]_{\phantom{\mu}\mu}^{\psi}G_{\phantom{\mu}\lambda}^{\mu}e^{-i\Gamma^{C}(x)T^{C}}\label{eq:combinedtransform1}
\end{equation}
and 
\begin{equation}
G_{\phantom{\mu}\lambda}^{\mu}[H^{f}]_{\phantom{\lambda}\nu}^{\lambda}\rightarrow e^{-iZ_{f}^{B}(x)T^{B}}G_{\phantom{\mu}\lambda}^{\mu}[H^{f}]_{\phantom{\lambda}\nu}^{\lambda}e^{-i\Gamma^{C}(x)T^{C}}\label{eq:combinedtransform2}
\end{equation}
This means that it is convenient to write gauge invariant and BTGT
invariant fields in terms of $\left(H^{a}\right)_{\phantom{\mu}\alpha}^{\beta}G_{\phantom{\alpha}\beta}^{\alpha}(x)$
because of these simple transformation properties.

\section{Path integral formulation}

We define the quantized theory of $G$ in this section using a path
integral over the $\theta_{a}^{A}$ variable in this section. To this
end, we begin by writing down the BTGT and gauge invariant action
in terms of $U_{a}$ variable (defined in Eq.\,(\ref{eq:uvariable})).
Next, we define a coupling constant expansion that allows us to match
perturbative gauge theory computations. Afterwards, we construct the
path integral over $\theta_{a}^{A}$.

\subsection{Non-perturbative action}

In this section, we construct the action for the basis tensor field
$\theta_{a}^{A}$. Because of Eq.\,(\ref{eq:nonabelianamu}), any
non-Abelian gauge theory with finite powers of $A_{\mu}$ will map
to a field theory with an infinite power series in $\theta_{c}^{K}$.
In this section, we construct the action of the usual Yang-Mills theory
in terms of $\theta_{a}^{A}$.

Recall that $A_{\mu}$ is a BTGT transformation invariant (which we
will refer to as a BTGT invariant for short). Hence, we can construct
BTGT invariant objects involving just $\theta_{a}^{A}$ fields if
we work with our knowledge of the usual gauge kinetic terms. Using
Eq.\,(\ref{eq:amurewrite}), we can write the action in the usual
way as 
\begin{equation}
\mathcal{L}=\frac{-1}{4g^{2}T(R)}{\rm Tr}\left(F^{\mu\nu}F_{\mu\nu}\right),\label{eq:puregluon}
\end{equation}
where the field strength is 
\begin{equation}
F_{\mu\nu}=i[D_{\mu},D_{\nu}]
\end{equation}
and the covariant derivative in terms of $U_{a}$ is 
\begin{equation}
D_{\mu}=\partial_{\mu}+\sum_{a=0}^{3}U_{a}\tilde{\partial}_{\mu}^{a}U_{a}^{\dagger}.
\end{equation}
More explicitly, we can expand the the field strength tensor as 
\begin{equation}
F_{\mu\nu}=i\left(\partial_{\mu}\sum_{a=0}^{3}U_{a}\tilde{\partial}_{\nu}^{a}U_{a}^{\dagger}-\partial_{\nu}\sum_{a=0}^{3}U_{a}\tilde{\partial}_{\mu}^{a}U_{a}^{\dagger}\right)+i\sum_{a,b=0}^{3}[U_{a}\tilde{\partial}_{\mu}^{a}U_{a}^{\dagger},U_{b}\tilde{\partial}_{\nu}^{b}U_{b}^{\dagger}].
\end{equation}
When written in terms of components, we can identify 
\begin{eqnarray}
\mathcal{L} & = & \frac{-1}{2g^{2}}\left(\partial_{\mu}A_{\nu}^{A}\partial^{\mu}A^{A\nu}-\partial_{\mu}A_{\nu}^{A}\partial^{\nu}A^{A\mu}\right)-\frac{1}{g^{2}}f^{ABC}\partial_{\mu}A_{\nu}^{A}A^{B\mu}A^{C\nu}\nonumber \\
 &  & -\frac{1}{4g^{2}}f^{ABC}f^{AB_{2}C_{2}}A_{\mu}^{B}A_{\nu}^{C}A_{\mu}^{B_{2}}A_{\nu}^{C_{2}}
\end{eqnarray}
with 
\begin{equation}
A_{\mu}^{A}=\frac{i}{T(R)}\sum_{a=0}^{3}{\rm Tr}\left(T^{A}U_{a}\tilde{\partial}_{\mu}^{a}U_{a}^{\dagger}\right).\label{eq:amucomponentintermsoftr}
\end{equation}
Just as in the Abelian BTGT theory, we see that the theory has a 4-derivative
kinetic term structure, which begs the question of whether the Hamiltonian
is bounded from below~\cite{Woodard:2006nt,Hawking:2001yt,Antoniadis:2006pc,Chen:2012au,Salvio:2015gsi}.
Just as in the Abelian case~\cite{Chung:2016lhv}, the Hamiltonian
is indeed bounded from below because the BTGT symmetry gives rise
to only field dependence on $A_{\mu}^{A}[U_{a}]$.

The matter coupling can be written down by noting that under BTGT
transformations, we have 
\begin{equation}
\partial_{\psi}\left[\left(H^{f}\right)_{\phantom{\psi}\alpha}^{\psi}G_{\phantom{\alpha}\beta}^{\alpha}\phi\right]\rightarrow e^{-iZ_{f}^{B}(x)T^{B}}\partial_{\psi}\left[\left(H^{f}\right)_{\phantom{\psi}\alpha}^{\psi}G_{\phantom{\alpha}\beta}^{\alpha}\phi\right].
\end{equation}
This means we can construct a gauge, Lorentz, and BTGT invariant combination
\begin{equation}
\sum_{f}\left(\partial_{\psi_{2}}\left[\left(H^{f}\right)_{\phantom{\psi}\alpha_{2}}^{\psi_{2}}G_{\phantom{\alpha}\beta_{2}}^{\alpha_{2}}\phi\right]\right)^{\dagger}g^{\beta_{2}\beta}\partial_{\psi}\left[\left(H^{f}\right)_{\phantom{\psi}\alpha}^{\psi}G_{\phantom{\alpha}\beta}^{\alpha}\phi\right].
\end{equation}
It is easy to check using Eqs.\,(\ref{eq:gintermsofu}), (\ref{eq:projection}),
and (\ref{eq:symmetry}) that this is equivalent to the usual gauge
coupling to matter $D^{\mu}\phi^{\dagger}D_{\mu}\phi$: 
\begin{equation}
D^{\mu}\phi^{\dagger}D_{\mu}\phi=\left[\partial^{\mu}\phi+\sum_{a=0}^{3}(H^{a})^{\lambda_{2}\mu}U_{a}\partial_{\lambda_{2}}U_{a}^{\dagger}\phi\right]^{\dagger}\left[\partial_{\mu}\phi+\sum_{b=0}^{3}(H^{b})_{\phantom{\lambda}\mu}^{\lambda}U_{b}\partial_{\lambda}U_{b}^{\dagger}\phi\right].
\end{equation}
We can of course write down a similar coupling for the fermions charged
under the non-Abelian gauge group: 
\begin{equation}
\mathcal{L}_{fK}=\overline{\Psi}\left[i\cancel{\partial}+i\gamma^{\mu}\sum_{b=0}^{3}(H^{b})_{\phantom{\lambda}\mu}^{\lambda}U_{b}\partial_{\lambda}U_{b}^{\dagger}\right]\Psi.
\end{equation}

We note that because of BTGT invariance, couplings of the form 
\begin{equation}
\sum_{f}\left[G_{(f)\,\beta}^{\alpha}\phi\right]\left[G_{(f)\,\alpha}^{\beta}\phi\right]
\end{equation}
cannot be written down because they violate BTGT symmetry. There exists
gauge and BTGT invariant terms of the form 
\begin{equation}
\sum_{a}{\rm Tr}\left(U_{a}U_{a}^{\dagger}\right)
\end{equation}
that we might worry about. However, owing to their group representation
structure, these are constants and will not contribute nontrivially
in flat spacetime.

\subsection{Perturbative expansion}

Written in terms of the $\theta_{a}^{A}$ fields of Eq.\,(\ref{eq:nonabelianvierbein}),
the Lagrangian is a power series in $\theta_{a}^{A}$. For perturbative
computations, we only require a consistent truncation in the coupling
constant. The usual perturbation theory proceeds through the identification
\begin{equation}
A_{\mu}^{A}\rightarrow gA_{\mu}^{A}.\label{eq:usualpert}
\end{equation}
Motivated by this and a need to truncate the power series of Eq.\,(\ref{eq:nonabelianvierbein}),
we make the change of variables 
\begin{equation}
\theta_{a}^{A}\rightarrow g\theta_{a}^{A}\label{eq:newpert}
\end{equation}
and expand perturbatively about $g\rightarrow0$. However, given that
Eqs.\,(\ref{eq:usualpert}) and (\ref{eq:newpert}) match only to
linear order in $g$, the perturbative expansion of the $A_{\mu}$
theory with $g\rightarrow0$ will match the perturbative expansion
of $\theta_{a}^{A}$ theory with $g\rightarrow0$ only if we deal
with composite operators.

For example, if we want to match the $A_{\mu}^{A}\rightarrow gA_{\mu}^{A}$
perturbation theory to $\theta_{a}^{A}\rightarrow g\theta_{a}^{A}$
perturbation theory to quadratic order in $g$, we must make the identification
\begin{eqnarray}
gA_{\mu}^{A} & = & g\sum_{a}\left[\frac{e^{gf\cdot\theta_{a}}-1}{gf\cdot\theta_{a}}\right]^{AB}\tilde{\partial}_{\mu}^{a}\theta_{a}^{B}\label{eq:thetaAexplicit}\\
 & \approx & g\tilde{\partial}_{\mu}^{a}\theta_{a}^{A}+\frac{g^{2}}{2}f^{ABC}\left(\tilde{\partial}_{\mu}^{a}\theta_{a}^{B}\right)\theta_{a}^{C}+O(g^{3})
\end{eqnarray}
at least to quadratic order in $g$. We explicitly then see a quadratic
field identification with $A_{\mu}$. In this case, a two-point function
in $A_{\mu}$ becomes
\begin{align}
\langle A_{\mu}^{A}(x)A_{\nu}^{B}(y)\rangle= & \sum_{a,b}\left\langle (\tilde{\partial}_{\mu}^{a}\theta_{a}^{A}(x)+\frac{g}{2}f^{AC_{1}D_{1}}\theta_{a}^{D_{1}}(x)\tilde{\partial}_{\mu}^{a}\theta_{a}^{C_{1}}(x)+\dots)\right.\nonumber \\
 & \left.\times(\tilde{\partial}_{\nu}^{b}\theta_{b}^{B}(y)+\frac{g}{2}f^{BC_{2}D_{2}}\theta_{b}^{D_{2}}(y)\tilde{\partial}_{\nu}^{b}\theta_{b}^{C_{2}}(y)+\dots)\right\rangle .
\end{align}
Although this nonlinearity seems undesirable from the perspective
of matching to ordinary non-Abelian field theory perturbative expansion
in terms of $A_{\mu}^{A}$, there may be an advantage since it allows
us to map nontrivial composite non-local operator correlators in the
language of $A_{\mu}^{A}$ field in terms of correlators of the elementary
$\theta_{a}^{A}$ correlator. We will defer the exploration of this
feature to a future work.

The power series can be explicitly written as
\begin{align}
A_{\mu}^{A} & =\sum_{a}\left[\frac{e^{gf\cdot\theta_{a}}-1}{gf\cdot\theta_{a}}\right]^{AB}\tilde{\partial}_{\mu}^{a}\theta_{a}^{B}\label{eq:A=00005Btheta=00005D}\\
 & =\tilde{\partial}_{\mu}^{a}\theta_{a}^{A}+\frac{g}{2}f^{ABC}\left(\tilde{\partial}_{\mu}^{a}\theta_{a}^{B}\right)\theta_{a}^{C}+\frac{g^{2}}{6}f^{ABE}f^{CDE}\theta_{a}^{B}\theta_{a}^{C}\left(\tilde{\partial}_{\mu}^{a}\theta_{a}^{D}\right)+O\left(g^{3}\right).
\end{align}

With the proper addition of the gauge fixing term, Eq.\,(\ref{eq:puregluon})
takes the form 
\begin{equation}
\mathcal{L}_{\mathrm{gauge}}=-\frac{1}{4}F^{A,\mu\nu}F_{\mu\nu}^{A}-\frac{1}{2\xi}\partial^{\mu}A_{\mu}^{A}\partial^{\nu}A_{\nu}^{A}.
\end{equation}
With Eq.\,(\ref{eq:thetaAexplicit}) the gauge boson sector becomes
\begin{equation}
\mathcal{L}_{\mathrm{gauge}}=\mathcal{L}_{\theta^{2}}+\mathcal{L}_{\theta^{3}}+\mathcal{L}_{\theta^{4}}+\dots=\sum_{n=2}^{\infty}\mathcal{L}_{\theta^{n}}
\end{equation}
where
\begin{equation}
\mathcal{L}_{\theta^{2}}=-\frac{1}{2}\left(\partial^{\mu}\tilde{\partial}_{a}^{\nu}\theta_{a}^{A}\right)\delta^{AB}\left(\partial_{\mu}\tilde{\partial}_{\nu}^{b}\theta_{b}^{B}-\left(1-\tfrac{1}{\xi}\right)\partial_{\nu}\tilde{\partial}_{\mu}^{b}\theta_{b}^{B}\right),\label{eq:L_theta^2}
\end{equation}
\begin{align}
\mathcal{L}_{\theta^{3}}= & -gf^{ABC}\left(\partial^{\mu}\tilde{\partial}_{a}^{\nu}\theta_{a}^{A}\right)\left(\tilde{\partial}_{\mu}^{b}\theta_{b}^{B}\right)\left(\tilde{\partial}_{\nu}^{c}\theta_{c}^{C}\right)\nonumber \\
 & -\frac{g}{2}f^{ABC}\left(\partial^{\mu}\tilde{\partial}_{a}^{\nu}\theta_{a}^{A}-\left(1-\tfrac{1}{\xi}\right)\partial^{\nu}\tilde{\partial}_{a}^{\mu}\theta_{a}^{A}\right)\left(\partial_{\mu}\left(\left(\tilde{\partial}_{\nu}^{b}\theta_{b}^{B}\right)\theta_{b}^{C}\right)\right)
\end{align}
and 
\begin{align}
\mathcal{L}_{\theta^{4}}= & -\frac{g^{2}}{4}f^{EAB}f^{ECD}\left(\tilde{\partial}_{\mu}^{a}\theta_{a}^{A}\right)\left(\tilde{\partial}_{\nu}^{b}\theta_{b}^{B}\right)\left(\tilde{\partial}_{c}^{\mu}\theta_{c}^{C}\right)\left(\tilde{\partial}_{d}^{\nu}\theta_{d}^{D}\right)\nonumber \\
 & -\frac{g^{2}}{2}f^{EAB}f^{ECD}\left(\tilde{\partial}_{\mu}^{a}\theta_{a}^{A}\right)\left(\tilde{\partial}_{\nu}^{b}\theta_{b}^{B}\right)\partial^{\mu}\left(\left(\tilde{\partial}_{c}^{\nu}\theta_{c}^{C}\right)\theta_{c}^{D}\right)\nonumber \\
 & -\frac{g^{2}}{2}f^{EAB}f^{ECD}\left(\partial^{\mu}\tilde{\partial}_{a}^{\nu}\theta_{a}^{A}-\partial^{\nu}\tilde{\partial}_{a}^{\mu}\theta_{a}^{A}\right)\left(\tilde{\partial}_{\mu}^{b}\theta_{b}^{B}\right)\left(\tilde{\partial}_{\nu}^{c}\theta_{c}^{C}\right)\theta_{c}^{D}\nonumber \\
 & -\frac{g^{2}}{8}f^{EAB}f^{ECD}\left(\partial_{\mu}\left(\left(\tilde{\partial}_{\nu}^{a}\theta_{a}^{A}\right)\theta_{a}^{B}\right)-\left(1-\tfrac{1}{\xi}\right)\partial_{\nu}\left(\left(\tilde{\partial}_{\mu}^{a}\theta_{a}^{A}\right)\theta_{a}^{B}\right)\right)\partial^{\mu}\left(\left(\tilde{\partial}_{c}^{\nu}\theta_{c}^{C}\right)\theta_{c}^{D}\right)\nonumber \\
 & -\frac{g^{2}}{6}f^{EAB}f^{ECD}\left(\partial^{\mu}\tilde{\partial}_{a}^{\nu}\theta_{a}^{A}-\left(1-\tfrac{1}{\xi}\right)\partial^{\nu}\tilde{\partial}_{a}^{\mu}\theta_{a}^{A}\right)\partial_{\mu}\left(\theta_{b}^{B}\theta_{b}^{C}\left(\tilde{\partial}_{\nu}^{b}\theta_{b}^{D}\right)\right).
\end{align}

If gauge fixing is accomplished using the Faddeev-Popov procedure,
we can write down the ghost Lagrangian coming from the delta-function
involving the $A_{\mu}^{A}$ in the usual way:
\begin{eqnarray}
\mathcal{L}_{{\rm gh1}} & = & -\partial^{\mu}\bar{c}^{A}D_{\mu}^{AB}c^{B}\\
 & = & -\partial^{\mu}\bar{c}^{A}\delta^{AB}\partial_{\mu}c^{B}+gf^{ABC}\partial^{\mu}\bar{c}^{A}c^{B}A_{\mu}^{C}\label{eq:Lghost1}
\end{eqnarray}
where $A_{\mu}^{C}$ is given in terms of $\theta_{a}^{A}$ explicitly
in Eq.\,(\ref{eq:thetaAexplicit}). To second order in $g$, the explicit
expansion is 
\begin{equation}
\mathcal{L}_{{\rm gh1}}=-\partial^{\mu}\bar{c}^{A}\partial_{\mu}c^{A}+gf^{ABC}\tilde{\partial}_{\mu}^{a}\theta_{a}^{A}\partial^{\mu}\bar{c}^{B}c^{C}+\frac{g^{2}}{2}f^{ABE}f^{CDE}(\tilde{\partial}_{\mu}^{a}\theta_{a}^{A})\theta_{a}^{B}\partial^{\mu}\bar{c}^{C}c^{D}+O\left(g^{3}\right).
\end{equation}

The ghost field couples to the gauge sector with quartic and higher
power couplings unlike in the usual vector potential formalism. If
we formulate the path integral measure in terms of $A_{\mu}$ and
view the path integral in terms of $\theta_{a}^{A}$ as a change of
variables, then there will be additional ghost contributions from
\begin{equation}
\mathcal{D}A=\mathcal{D}\theta_{{\rm nz}}\det\left[\frac{\delta A_{\mu}^{A}(x)}{\delta\theta_{{\rm nz},b}^{B}(y)}\right],
\end{equation}
where $\theta_{{\rm nz},b}^{B}$ stands for functions that are not annihilated by
\begin{equation}
(H^b)^{\alpha}_{\phantom{\alpha}\beta}\frac{\partial}{\partial x^{\alpha}}.
\end{equation}
The functional determinant can be written as usual as a Grassmannian
integral yielding an additional ghost Lagrangian: 
\begin{equation}
\mathcal{L}_{{\rm gh}2}=\bar{d}_{a}^{A}\mathcal{O}_{ab}^{AB}d_{b}^{B}=\bar{d}_{a}^{A}\tilde{\partial}_{a}^{\mu}\mathcal{O}_{\mu b}^{AB}d_{b}^{B}=-\left(\tilde{\partial}_{a}^{\mu}\bar{d}_{a}^{A}\right)\mathcal{O}_{\mu b}^{AB}d_{b}^{B}\label{eq:Lghost2}
\end{equation}
where we define the operator 
\begin{eqnarray}
\mathcal{O}_{\mu b}^{AB} & = & \left[\int_{0}^{1}dt\,e^{tg\theta_{b}\cdot f}\right]^{AB}\left(H^{b}\right)_{\phantom{\lambda}\mu}^{\lambda}\vec{\partial}_{\lambda}\nonumber \\
 &  & +\left[\int_{0}^{1}dt\int_{0}^{1}dse^{\left(1-s\right)tg\theta_{b}\cdot f}tgf^{B}e^{stg\theta_{b}\cdot f}\right]^{AD}\left(H^{b}\right)_{\phantom{\lambda}\mu}^{\lambda}\left(\partial_{\lambda}\theta_{b}^{D}\right)\\
 & = & \left[\delta^{AB}+\frac{g}{2}f^{ABC}\theta_{b}^{C}+\frac{g^{2}}{6}f^{AEC}\theta_{b}^{C}f^{EBD}\theta_{b}^{D}\right]\tilde{\partial}_{\mu}^{b}\nonumber \\
 &  & +\left[\int_{0}^{1}dt\int_{0}^{1}dse^{\left(1-s\right)tg\theta_{b}\cdot f}tgf^{B}e^{stg\theta_{b}\cdot f}\right]^{AD}\left(\tilde{\partial}_{\mu}\theta_{b}^{D}\right)+O\left(g^{3}\right).
\end{eqnarray}
We next work out the explicit Feynman rule factors.

\subsubsection{Gauge propagator}

The inverse of the propagator in momentum space can be written as
\begin{align}
-iV_{ab}^{AB}\left(k\right) & =\frac{\partial^{2}\left(i\mathcal{L}_{\theta^{2}}\right)}{\partial\theta_{a}^{A}\left(k\right)\partial\theta_{b}^{B}\left(-k\right)}\\
 & =-i\left(k^{\mu}\tilde{k}_{a}^{\nu}\right)\delta^{AB}\left(k_{\mu}\tilde{k}_{\nu}^{b}-\left(1-\tfrac{1}{\xi}\right)-k_{\nu}\tilde{k}_{\mu}^{b}\right)\\
 & =-i\delta^{AB}\left(\delta_{ab}k^{2}k\star_{a}k-\left(1-\tfrac{1}{\xi}\right)\left(k\star_{a}k\right)\left(k\star_{b}k\right)\right),
\end{align}
where we define the star product as 
\begin{equation}
k_{1}\star_{a}k_{2}=(H^{a})_{\mu\nu}k_{1}^{\mu}k_{2}^{\nu}.
\end{equation}

The gauge propagator $\Delta_{ab}^{AB}\left(k\right)$ is given implicitly
by
\begin{equation}
\sum_{c}V_{ac}^{AC}\left(k\right)\Delta_{cb}^{CB}\left(k\right)=\delta^{AB}\delta_{ab},
\end{equation}
the solution to which is
\begin{equation}
-i\Delta_{ab}^{AB}\left(k\right)=\frac{-i\delta^{AB}}{k^{2}k\star_{a}k-i\epsilon}\left(\delta_{ab}-\left(1-\xi\right)\frac{k\star_{a}k}{k^{2}}\right),
\end{equation}
where the $i\epsilon$ is the solution Feynman propagator pole prescription.
If we assume a diagonal basis for $H^{a}$ and a Wick rotation to
Euclidean space, then this can be written as
\begin{equation}
-i\Delta_{ab}^{AB}\left(k\right)=\frac{-i\delta^{AB}}{k^{2}k_{a}k_{b}}\left(\delta_{ab}-\left(1-\xi\right)\frac{k_{a}k_{b}}{k^{2}}\right).
\end{equation}
In position space the propagator can be written as
\begin{equation}
\Delta_{ab}^{AB}\left(x-y\right)=\int\frac{d^{4}k}{\left(2\pi\right)^{4}}e^{ik\cdot\left(x-y\right)}\Delta_{ab}^{AB}\left(k\right).
\end{equation}

\subsubsection{Cubic gauge self-coupling}

For Feynman rules with momenta satisfying $k_{1}+k_{2}+k_{3}=0$,
the vertex function $iV_{abc}^{ABC}\left(k_{1},k_{2},k_{3}\right)$
can be written as 
\begin{eqnarray}
iV_{abc}^{ABC}\left(k_{1},k_{2},k_{3}\right) & = & \frac{\partial^{3}\left(i\mathcal{L}_{\theta^{3}}\right)}{\partial\theta_{a}^{A}\left(k_{1}\right)\partial\theta_{b}^{B}\left(k_{2}\right)\partial\theta_{c}^{C}\left(k_{3}\right)}\\
 & = & igf^{ABC}\left\{ \delta_{bc}\left(k_{2}\star_{b}k_{3}\right)k_{1}\star_{a}\left(k_{2}-k_{3}\right)+\delta_{ac}\left(k_{1}\star_{c}k_{3}\right)k_{2}\star_{b}\left(k_{3}-k_{1}\right)\right.\nonumber \\
 &  & +\delta_{ab}\left(k_{1}\star_{b}k_{2}\right)k_{3}\star_{c}\left(k_{1}-k_{2}\right)+\frac{1}{2}\delta_{abc}\left[k_{1}^{2}k_{1}\star_{a}\left(k_{2}-k_{3}\right)+k_{2}^{2}k_{2}\star_{a}\left(k_{3}-k_{1}\right)\right.\nonumber \\
 &  & \left.+k_{3}^{2}k_{3}\star_{a}\left(k_{1}-k_{2}\right)\right]-\frac{1}{2}\left(1-\tfrac{1}{\xi}\right)\left[\delta_{bc}\left(k_{1}\star_{a}k_{1}\right)k_{1}\star_{b}\left(k_{2}-k_{3}\right)\right.\nonumber \\
 &  & \left.\left.+\delta_{ac}\left(k_{2}\star_{b}k_{2}\right)k_{2}\star_{c}\left(k_{3}-k_{1}\right)+\delta_{ab}\left(k_{3}\star_{c}k_{3}\right)k_{3}\star_{a}\left(k_{1}-k_{2}\right)\right]\right\} .
\end{eqnarray}
If we assume a diagonal basis for $H^{a}$, then we get 
\begin{equation}
iV_{abc}^{ABC}\left(k_{1},k_{2},k_{3}\right)=igf^{ABC}\left(\sum_{i=1}^{2}V_{abc}^{(i)}\left(k_{1},k_{2},k_{3}\right)+\left(1-\tfrac{1}{\xi}\right)V_{abc}^{(3)}\left(k_{1},k_{2},k_{3}\right)\right)
\end{equation}
with 
\begin{align}
V_{abc}^{(1)}\left(k_{1},k_{2},k_{3}\right) & =+k_{1a}k_{2b}k_{3c}\left(\delta_{bc}\left(k_{2a}-k_{3a}\right)+\delta_{ac}\left(k_{3b}-k_{1b}\right)+\delta_{ab}\left(k_{1c}-k_{2c}\right)\right)\\
V_{abc}^{(2)}\left(k_{1},k_{2},k_{3}\right) & =+\frac{1}{2}\delta_{abc}\left(k_{1}^{2}k_{1a}\left(k_{2a}-k_{3a}\right)+k_{2}^{2}k_{2a}\left(k_{3a}-k_{1a}\right)+k_{3}^{2}k_{3a}\left(k_{1a}-k_{2a}\right)\right)\\
V_{abc}^{(3)}\left(k_{1},k_{2},k_{3}\right) & =-\frac{1}{2}\left(\delta_{bc}k_{1a}^{2}k_{1b}\left(k_{2b}-k_{3b}\right)+\delta_{ac}k_{2b}^{2}k_{2a}\left(k_{3a}-k_{1a}\right)+\delta_{ab}k_{3c}^{2}k_{3a}\left(k_{1a}-k_{2a}\right)\right).
\end{align}
Setting $\xi=1$ with the Feynman gauge simplifies calculations because
$V_{abc}^{(3)}$ can be ignored. Tree level $\xi$-dependent vertex
terms are an interesting distinction from the usual vector potential
gauge theory. The numbering here is organized according to powers
of $A_{\mu}$ that contribute to these $\theta_{a}$ vertices in the
following way:
\begin{align}
f^{ABC}\partial^{\mu}A^{A\nu}A_{\mu}^{B}A_{\nu}^{C} & \rightarrow V^{(1)}\\
\partial_{\mu}A_{\nu}^{A}\partial^{\mu}A^{A\nu} & \rightarrow V^{(2)}\\
\left(1-\tfrac{1}{\xi}\right)\partial^{\mu}A_{\mu}^{A}\partial^{\nu}A_{\nu}^{A} & \rightarrow\left(1-\tfrac{1}{\xi}\right)V^{(3)}.
\end{align}

\subsubsection{Quartic gauge self-coupling}

The quartic vertex (or four $\theta$ vertex) can be written as
\begin{eqnarray}
iV_{abcd}^{ABCD}\left(k_{1},k_{2},k_{3},k_{4}\right) & = & \frac{\partial^{4}\left(i\mathcal{L}_{\theta^{4}}\right)}{\partial\theta_{a}^{A}\left(k_{1}\right)\partial\theta_{b}^{B}\left(k_{2}\right)\partial\theta_{c}^{C}\left(k_{3}\right)\partial\theta_{d}^{D}\left(k_{3}\right)}\\
 & = & ig^{2}\left(\sum_{i=1}^{6}V{}_{\left(i\right)abcd}^{ABCD}+\left(1-\tfrac{1}{\xi}\right)\sum_{i=7}^{8}V{}_{\left(i\right)abcd}^{ABCD}\right)\label{eq:quartic theta vertex}
\end{eqnarray}
where we define 8 terms as 
\begin{align}
V_{(1)abcd}^{ABCD} & =-\frac{1}{4}f_{E}^{AB}f_{E}^{CD}\delta_{ac}\delta_{bd}\left(k_{1}\star_{a}k_{3}\right)\left(k_{2}\star_{b}k_{4}\right)+\mathrm{perms.}\\
V_{(2)abcd}^{ABCD} & =-\frac{1}{2}f_{E}^{AB}f_{E}^{CD}\delta_{bcd}\left(k_{1}\star_{a}\left(k_{3}+k_{4}\right)\right)\left(k_{2}\star_{b}k_{3}\right)+\mathrm{perms.}\\
V_{(3)abcd}^{ABCD} & =-\frac{1}{2}f_{E}^{AB}f_{E}^{CD}\delta_{acd}\left(k_{1}\star_{b}k_{2}\right)\left(k_{1}\star_{c}k_{3}\right)+\mathrm{perms.}\\
V_{(4)abcd}^{ABCD} & =+\frac{1}{2}f_{E}^{AB}f_{E}^{CD}\delta_{ab}\delta_{cd}\left(k_{1}\star_{b}k_{2}\right)\left(k_{1}\star_{c}k_{3}\right)+\mathrm{perms.}\\
V_{(5)abcd}^{ABCD} & =+\frac{1}{8}f_{E}^{AB}f_{E}^{CD}\delta_{abcd}\left(k_{1}+k_{2}\right)^{2}\left(k_{1}\star_{a}k_{3}\right)+\mathrm{perms.}\\
V_{(6)abcd}^{ABCD} & =+\frac{1}{6}f_{E}^{AB}f_{E}^{CD}\delta_{abcd}k_{1}^{2}\left(k_{1}\star_{a}k_{4}\right)+\mathrm{perms.}\\
V_{(7)abcd}^{ABCD} & =-\frac{1}{8}f_{E}^{AB}f_{E}^{CD}\delta_{ab}\delta_{cd}\left(k_{1}\star_{a}\left(k_{1}+k_{2}\right)\right)\left(k_{3}\star_{c}\left(k_{1}+k_{2}\right)\right)+\mathrm{perms.}\\
V_{(8)abcd}^{ABCD} & =-\frac{1}{6}f_{E}^{AB}f_{E}^{CD}\delta_{bcd}\left(k_{1}\star_{a}k_{1}\right)\left(k_{1}\star_{b}k_{4}\right)+\mathrm{perms.}
\end{align}
Here we are using the notation $f_{C}^{AB}=f^{CAB}=f^{ABC}$ for convenience.
The numbering here is organized according to powers of $A_{\mu}$
that contribute to these $\theta_{a}$ vertices in the following way:
\begin{align}
f_{E}^{AB}f_{E}^{CD}A^{\mu A}A_{\nu}^{\nu B}A_{\mu}^{C}A_{\nu}^{D} & \rightarrow V_{(1)}\\
f^{ABC}\partial^{\mu}A^{A\nu}A_{\mu}^{B}A_{\nu}^{C} & \rightarrow V_{(2)}+V_{(3)}+V_{(4)}\\
\partial_{\mu}A_{\nu}^{A}\partial^{\mu}A^{A\nu} & \rightarrow V_{(5)}+V_{(6)}\\
\left(1-\tfrac{1}{\xi}\right)\partial^{\mu}A_{\mu}^{A}\partial^{\nu}A_{\nu}^{A} & \rightarrow\left(1-\tfrac{1}{\xi}\right)\left(V_{(7)}+V_{(8)}\right).
\end{align}
Let's consider the evaluation of the permutations in each of these
terms.

Consider first $V_{\left(1\right)}$. Note that since $ABCD=BADC=CDAB=DCBA$,
we get a symmetry factor of $4$. This means we can write 
\begin{align}
V_{(1)abcd}^{ABCD}= & -f_{E}^{AB}f_{E}^{CD}\left(\delta_{ac}\delta_{bd}\left(k_{1}\star_{a}k_{3}\right)\left(k_{2}\star_{b}k_{4}\right)-\delta_{ad}\delta_{bc}\left(k_{1}\star_{a}k_{4}\right)\left(k_{2}\star_{b}k_{3}\right)\right)\nonumber \\
 & -f_{E}^{AC}f_{E}^{BD}\left(\delta_{ab}\delta_{cd}\left(k_{1}\star_{a}k_{2}\right)\left(k_{3}\star_{c}k_{4}\right)-\delta_{ad}\delta_{bc}\left(k_{1}\star_{a}k_{4}\right)\left(k_{2}\star_{b}k_{3}\right)\right)\nonumber \\
 & -f_{E}^{AD}f_{E}^{BC}\left(\delta_{ab}\delta_{cd}\left(k_{1}\star_{a}k_{2}\right)\left(k_{3}\star_{c}k_{4}\right)-\delta_{ac}\delta_{bd}\left(k_{1}\star_{a}k_{3}\right)\left(k_{2}\star_{b}k_{4}\right)\right).
\end{align}
If we assume a diagonal basis for $H^{a}$, this simplifies further
to 
\begin{align}
V_{(1)abcd}^{ABCD}= & -k_{1a}k_{2b}k_{3c}k_{4d}\left(f_{E}^{AB}f_{E}^{CD}\left(\delta_{ac}\delta_{bd}-\delta_{ad}\delta_{bc}\right)+f_{E}^{AC}f_{E}^{BD}\left(\delta_{ab}\delta_{cd}-\delta_{ad}\delta_{bc}\right)\right.\nonumber \\
 & \left.+f_{E}^{AD}f_{E}^{BC}\left(\delta_{ab}\delta_{cd}-\delta_{ac}\delta_{bd}\right)\right),
\end{align}
which takes on a form proportional to the quartic $A_{\mu}$ vertex
in the usual formalism. Similarly, we obtain other seven terms of
the quartic BTGT vertex by writing the rest of the permutations. The
full results can be found in Appendix \ref{sec:Feynman-Rules}.

\subsection{Generating function for BTGT}

The generating function for $A_{\mu}$ correlators in the usual formalism
is given by the path integral
\begin{equation}
Z\left[J\right]=\int\mathcal{D}A\mathcal{D}\bar{c}\mathcal{D}c\exp\left(iS\left[A,\bar{c},c\right]+i\int d^{4}xJ\cdot A\right),
\end{equation}
where 
\begin{equation}
S\left[A,\bar{c},c\right]=\int d^{4}x\left(-\frac{1}{4}F_{\mu\nu}^{A}F^{A\mu\nu}-\frac{1}{2\xi}\left(\partial\cdot A\right)^{2}-\partial^{\mu}\bar{c}^{A} D_{\mu}^{AB}c^{B}\right)\label{eq:Yang mills action}
\end{equation}
is the Yang-Mills action with gauge fixing and ghosts.

Now make $A_{\mu}^{A}\left(x\right)=A_{\mu}^{A}\left[\theta\left(x\right)\right]$
a composite operator as specified by Eq.\,(\ref{eq:nonabelianamu}).
This change affects both the action and the path measure. The generating
function is now 
\begin{equation}
Z\left[J\right]=\int\mathcal{D}\theta\mathcal{D}\bar{c}\mathcal{D}c\mathcal{D}\bar{d}\mathcal{D}d\,e^{iS\left[A[\theta],\bar{c},c\right]+iS_{{\rm gh2}}[\theta,\bar{d},d]+i\int d^{4}xJ\cdot A[\theta]},\label{eq:Z usual formalism}
\end{equation}
where $\bar{d}$, $d$ are the additional ghosts defined in Eq.\,(\ref{eq:Lghost2})
and the additional ghost action is $S_{{\rm gh2}}=\int d^{4}x\mathcal{L}_{{\rm gh2}}$. 

We will now construct a generating function for correlators of $A_{\mu}$ and $\theta_{a}$. We define $K_{a}^{A}$ as a source for
$\theta_{a}^{A}$ and define the new generating function as 
\begin{equation}
\bar{Z}\left[J,K\right]=\int\mathcal{D}\theta\mathcal{D}\bar{c}\mathcal{D}c\mathcal{D}\bar{d}\mathcal{D}de^{iS\left[A[\theta],\bar{c},c\right]+iS_{{\rm gh2}}[\theta,\bar{d},d]+i\int d^{4}x\left(J\cdot A[\theta]+K_{a}^{A}\theta_{a}^{A}\right)}.\label{eq:Zbar generating function}
\end{equation}
In this paper, Eq.\,(\ref{eq:Zbar generating function}) will be our
definition of the quantized theory and this will be used to calculate
both the $\theta_{a}$ and $A_{\mu}$ correlators. The difference
from the generating function of the $A_{\mu}$ formalism shown in
Eq.\,(\ref{eq:Z usual formalism}) is that $A_{\mu}$ is now a composite
operator in terms of $\theta_{a}$ fields and the path integral is
now over $\theta_{a}$ instead of $A_{\mu}$. We will find through
explicit computations below that $S_{{\rm gh2}}[\theta,\bar{d},d]$
(the action describing the ghosts coming from the transformation from
$A_{\mu}^{B}$ to $\theta_{a}^{A}$) does not contribute to the divergent
structure (in dimensional regularization) in the processes that we
compute in this paper. It would be interesting to elucidate this decoupling
in a future work.

For perturbative computations, we split apart the Yang-Mills action
Eq.\,(\ref{eq:Yang mills action}) in the following way
\begin{equation}
S\left[A[\theta],\bar{c},c\right]=S_{\mathrm{int}}\left[A[\theta],\bar{c},c\right]+\int d^{4}x\mathcal{L}_{\theta^{2}},
\end{equation}
where $\mathcal{L}_{\theta^{2}}$ is defined in Eq.\,(\ref{eq:L_theta^2}).
Then we can rewrite all powers of $\theta_{a}$ higher than quadratic
as functional derivatives with respect to $iK_{a}$. The generating
function Eq.\,(\ref{eq:Zbar generating function}) can then be written
as 
\begin{eqnarray}
\bar{Z}\left[J,K\right] & = & \int\mathcal{D}\theta\mathcal{D}\bar{c}\mathcal{D}c\mathcal{D}\bar{d}\mathcal{D}de^{iS_{int}\left[A[\theta],\bar{c},c\right]+iS_{{\rm gh2}}[\theta,\bar{d},d]+i\int d^{4}xJ\cdot A[\theta]}e^{i\int d^{4}x\left(\mathcal{L}_{\theta^{2}}+K_{a}^{A}\theta_{a}^{A}\right)}\\
 & = & \int\mathcal{D}\bar{c}\mathcal{D}c\mathcal{D}\bar{d}\mathcal{D}d\,e^{iS_{int}\left[A[\frac{\delta}{i\delta K}],\bar{c},c\right]+iS_{{\rm gh2}}[\frac{\delta}{i\delta K},\bar{d},d]+i\int d^{4}xJ\cdot A[\frac{\delta}{i\delta K}]}\nonumber \\
 &  & \times\int\mathcal{D}\theta e^{i\int d^{4}x\left(\mathcal{L}_{\theta^{2}}+K_{a}^{A}\theta_{a}^{A}\right)}\\
 & = & \mathcal{N}e^{i\int d^{4}xJ\cdot A[\frac{\delta}{i\delta K}]}\int\mathcal{D}\bar{c}\mathcal{D}c\mathcal{D}\bar{d}\mathcal{D}d\,e^{iS_{int}\left[A[\frac{\delta}{i\delta K}],\bar{c},c\right]+iS_{{\rm gh2}}[\frac{\delta}{i\delta K},\bar{d},d]}\nonumber \\
 &  & \times e^{i\int d^{4}xd^{4}yK_{a}^{A}(x)\Delta_{ab}^{AB}(x-y)K_{b}^{B}(y)}\label{eq:Zbar explicit}
\end{eqnarray}
where $\mathcal{N}$ is a normalization constant. Eq.\,(\ref{eq:Zbar explicit})
is what was used to derive the Feynman rules of non-Abelian BTGT,
which are presented in Appendix \ref{sec:Feynman-Rules}.

\section{Beta function computation}

In this section, we show that the beta function at one loop for non-Abelian
BTGT is
\begin{equation}
\beta\left(g\right)=-\frac{11}{6}C\left(A\right)\frac{g^{3}}{8\pi^{2}}\label{eq:beta_result}
\end{equation}
which is the same result as the usual $A_{\mu}$ formalism of Yang
Mills theory. This lends support to the quantum consistency of the
formalism and its faithful representation of the usual non-Abelian
gauge theory perturbative content. This result is achieved by computing
the renormalization constants of the counter-terms of the $\theta_{a}$
and ghost quadratic terms and the $\theta_{a}\bar{c}c$ ghost-gauge
vertex. The relevant terms in the Lagrangian are
\begin{eqnarray}
\mathcal{L} & \ni & -\frac{1}{2}Z_{\theta^{2}}\left(\partial_{\mu}\tilde{\partial}_{\nu}^{a}\theta_{a}^{A}-\partial_{\nu}\tilde{\partial}_{\mu}^{a}\theta_{a}^{A}\right)\partial^{\mu}\tilde{\partial}_{b}^{\nu}\theta_{b}^{A}-\frac{1}{2\xi}Z_{\frac{1}{\xi}\theta^{2}}\partial_{\nu}\tilde{\partial}_{\mu}^{a}\theta_{a}^{A}\partial^{\mu}\tilde{\partial}_{b}^{\nu}\theta_{b}^{A}\nonumber \\
 &  & -Z_{\bar{c}c}\partial_{\mu}\bar{c}\partial^{\mu}c+Z_{g\bar{c}c\theta}gf^{ABC}\partial_{\mu}\bar{c}^{A}c^{B}\tilde{\partial}_{a}^{\mu}\theta_{a}^{C}.
\end{eqnarray}

These renormalization constants are computed in $\overline{\mathrm{MS}}$
with $d=4-\epsilon$ dimensional regularization to be
\begin{gather}
Z_{\theta^{2}}=1+4C\left(A\right)\frac{g^{2}}{8\pi^{2}\epsilon}+O\left(g^{4}\right)\label{eq:Ztheta^2}\\
Z_{\bar{c}c}=1+\frac{1}{2}C\left(A\right)\frac{g^{2}}{8\pi^{2}\epsilon}+O\left(g^{4}\right)\label{eq:Zcc}\\
Z_{g\theta\bar{c}c}=1+\frac{2}{3}C\left(A\right)\frac{g^{2}}{8\pi^{2}\epsilon}+O\left(g^{4}\right),\label{eq:Zgcctheta}
\end{gather}
which implies Eq.\,(\ref{eq:beta_result}) since
\begin{equation}
Z_{g}=\frac{Z_{g\theta\bar{c}c}}{Z_{\theta^{2}}^{1/2}Z_{\bar{c}c}}=1-\frac{11}{6}C\left(A\right)\frac{g^{2}}{8\pi^{2}\epsilon}+O\left(g^{4}\right)\label{eq:Z_g}
\end{equation}

In the following subsections, we compute Eqs.\,(\ref{eq:Ztheta^2}),
(\ref{eq:Zcc}) and (\ref{eq:Zgcctheta}). We display a large amount
of details since this BTGT formalism is new and how the formalism
works is one of the main results of this paper. For convenience we
choose the Feynman gauge $\xi=1$ and we assume a diagonal basis for
$(H^{a})_{\mu\nu}$: $\left(H^{a}\right)_{\mu\nu}=g_{\mu a}g_{\nu a}g^{aa}\mbox{ (no sum over }a)$.
We will be using the minimal subtraction scheme and dimensional regularization
with $d=4-\epsilon$ to determine the renormalization constants. We
will also be using the shorthand
\begin{equation}
\int_{\ell}\equiv\int\frac{d^{d}\ell}{\left(2\pi\right)^{d}}
\end{equation}

In the computation below, many zeros appear for the following reasons.
In dimensional regularization, we utilize the identity
\begin{equation}
\int\frac{d^{n}\ell}{\left(2\pi\right)^{n}}\frac{1}{\ell^{n+k}}\propto\delta_{k0},\label{eq:dim reg identity}
\end{equation}
where $n>1,k$ are integers and where as is customary, we do not distinguish
raised or lowered indices on Kronecker delta functions whenever contextually
the Lorentzian metric information is irrelevant. Other diagrams are
zero due to the anti-symmetric nature of $f^{ABC}$. Yet other diagrams
are zero due to the identity
\begin{equation}
\delta_{ab}\left(1-\delta_{ab}\right)=\delta_{ab}-\delta_{ab}=0.\label{eq:delta_ab identity}
\end{equation}

\subsection{Computation of $Z_{\theta^{2}}$ and $Z_{\frac{1}{\xi}\theta^{2}}$
\label{sub: Theta self energy}}

\begin{figure}
\begin{centering}
\includegraphics[width=13cm]{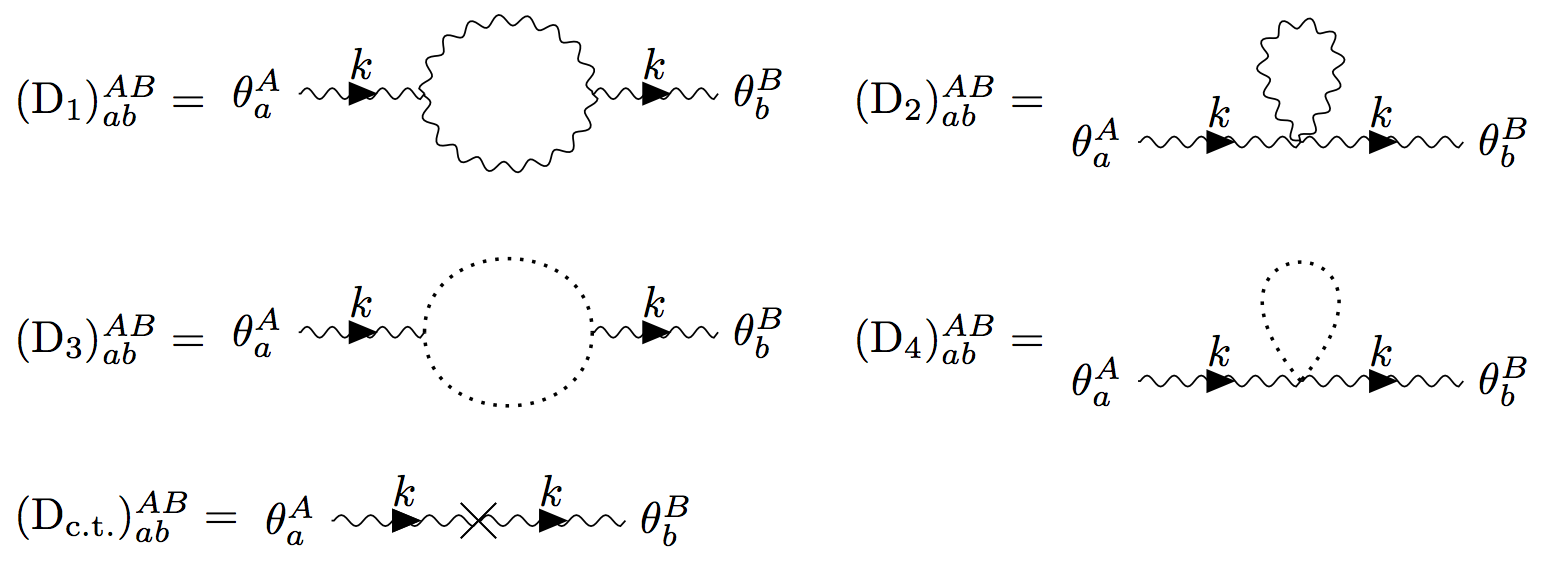}
\par\end{centering}

\caption{Self energy diagrams for $\theta_{a}$ \label{fig:Theta-self-energy-diagrams}}
\end{figure}

The relevant diagrams are defined in Fig.\,\ref{fig:Theta-self-energy-diagrams}.
It is understood that when we write symbols such as $\mathrm{D}_{1}$
without indices, the implicit indices are understood be of the form
$\mathrm{\left(\mathrm{D_{1}}\right)_{ab}^{AB}}\left(k\right)$. The
$\theta_{a}$ self energy can be written as 
\begin{align}
i\Pi_{ab}^{AB}\left(k\right) & =\sum_{i=1}^{4}\left(\mathrm{D_{i}}\right)_{ab}^{AB}\left(k\right)+\left(\mathrm{D_{c.t.}}\right)_{ab}^{AB}\left(k\right)
\end{align}

\subsubsection{$\theta$ self energy diagram 1}

Diagram 1 in Fig.\,\ref{fig:Theta-self-energy-diagrams} is given
by
\begin{eqnarray}
\left(\mathrm{D}_{1}\right)_{ab}^{AB} & = & \frac{1}{2}\int_{\ell}\sum_{cdef}\frac{\left(igV_{acd}^{ACD}\left(k,\ell\right)\right)\left(-i\delta^{CE}\delta_{ce}\right)\left(-i\delta^{DF}\delta_{ce}\right)\left(igV_{bef}^{BEF}\left(-k,-\ell\right)\right)}{\ell^{2}\ell_{c}^{2}\left(\ell+k\right)^{2}\left(\ell_{d}+k_{d}\right)^{2}}\\
 & = & \frac{g^{2}}{2}f^{ACD}f^{BCD}\sum_{i=1}^{2}\sum_{j=1}^{2}\int\frac{d^{4}\ell}{\left(2\pi\right)^{4}}\sum_{cd}\frac{V_{acd}^{(i)}\left(k,\ell\right)V_{bcd}^{(j)}\left(-k,-\ell\right)}{\ell^{2}\left(\ell+k\right)^{2}\ell_{c}^{2}\left(\ell_{d}+k_{d}\right)^{2}}\\
 & = & g^{2}C\left(A\right)\delta^{AB}\sum_{i=1}^{2}\sum_{j=1}^{2}\left(\mathrm{D}_{1}^{(i,j)}\right)_{ab}\label{eq:D1_sum}
\end{eqnarray}
where in the last line we define the sub-diagrams
\begin{equation}
\left(\mathrm{D}_{1}^{(i,j)}\right)_{ab}=\frac{1}{2}\int\frac{d^{4}\ell}{\left(2\pi\right)^{4}}\sum_{cd}\frac{V_{acd}^{(i)}\left(k,\ell\right)V_{bcd}^{(j)}\left(-k,-\ell\right)}{\ell^{2}\left(\ell+k\right)^{2}\ell_{c}^{2}\left(\ell_{d}+k_{d}\right)^{2}}
\end{equation}
The sums over $i$ and $j$ in Eq.\,(\ref{eq:D1_sum}) only go from
1 to 2 because $\left(1-\frac{1}{\xi}\right)V_{abc}^{(3)}=0$ in the
Feynman gauge. In the general $R_{\xi}$ gauge, the sums in Eq.\,(\ref{eq:D1_sum})
would go from 1 to 3. Due to the symmetry of the diagram, we also
know that
\begin{equation}
\left(\mathrm{D}_{1}^{(j,i)}\right)_{ab}^{AB}\left(k\right)=\left(\mathrm{D}_{1}^{(i,j)}\right)_{ba}^{BA}\left(-k\right)
\end{equation}
which means there are only three independent terms to compute in Eq.\,(\ref{eq:D1_sum}). 

We start with
\begin{align}
\left(\mathrm{D}_{1}^{(1,1)}\right)_{ab} & =\frac{1}{2}\int\frac{d^{4}\ell}{\left(2\pi\right)^{4}}\sum_{cd}\frac{V_{acd}^{\left(1\right)}\left(k,\ell\right)V_{bcd}^{\left(1\right)}\left(-k,-\ell\right)}{\ell^{2}\left(\ell+k\right)^{2}\ell_{c}^{2}\left(\ell_{d}+k_{d}\right)^{2}}\\
 & =\frac{1}{2}\int\frac{d^{4}\ell}{\left(2\pi\right)^{4}}\sum_{cd}\frac{k_{a}k_{b}\ell_{c}^{2}\left(\ell_{d}+k_{d}\right)^{2}N_{abcd}}{\ell^{2}\left(\ell+k\right)^{2}\ell_{c}^{2}\left(\ell_{d}+k_{d}\right)^{2}}\\
 & =\frac{1}{2}k_{a}k_{b}\int\frac{d^{4}\ell}{\left(2\pi\right)^{4}}\frac{\sum_{cd}N_{abcd}}{\ell^{2}\left(\ell+k\right)^{2}}\label{eq:D1(1,1)}
\end{align}
where the numerator is
\begin{eqnarray}
N_{abcd} & = & \left(-\delta_{cd}\left(2\ell_{a}+k_{a}\right)+\delta_{ad}\left(\ell_{c}+2k_{c}\right)+\delta_{ac}\left(\ell_{d}-k_{d}\right)\right)\nonumber \\
 &  & \times\left(-\delta_{cd}\left(2\ell_{b}+k_{b}\right)+\delta_{bd}\left(\ell_{c}+2k_{c}\right)+\delta_{bc}\left(\ell_{d}-k_{d}\right)\right).
\end{eqnarray}
Summing over $c$ and $d$ yields
\begin{eqnarray}
\sum_{cd}N_{abcd} & = & 10\ell_{a}\ell_{b}+5\ell_{a}k_{b}+5k_{a}\ell_{b}-2k_{a}k_{b}+\left(\left(\ell+2k\right)^{2}+\left(\ell-k\right)^{2}\right)\delta_{ab}
\end{eqnarray}
and applying this to Eq.\,(\ref{eq:D1(1,1)}) gives 
\begin{align}
\left(\mathrm{D}_{1}^{(1,1)}\right)_{ab} & =\frac{1}{2}k_{a}k_{b}\int_{\ell}\frac{10\ell_{a}\ell_{b}+5\ell_{a}k_{b}+5k_{a}\ell_{b}-2k_{a}k_{b}+\left(\left(\ell+2k\right)^{2}+\left(\ell-k\right)^{2}\right)\delta_{ab}}{\ell^{2}\left(\ell+k\right)^{2}}\\
 & =\frac{1}{2}\tilde{k}_{a}^{\mu}\tilde{k}_{b}^{\nu}\int_{\ell}\frac{10\ell_{\mu}\ell_{\nu}+5\ell_{\mu}k_{\nu}+5k_{\mu}\ell_{\nu}-2k_{\mu}k_{\nu}+\left(\left(\ell+2k\right)^{2}+\left(\ell-k\right)^{2}\right)g_{\mu\nu}}{\ell^{2}\left(\ell+k\right)^{2}}.\label{eq:D1 momentum integral}
\end{align}
The momentum integral of Eq.\,(\ref{eq:D1 momentum integral}) is
identical to the one that appears the usual non-Abelian $A_{\mu}$
formalism. We can evaluate it using the usual Feynman parameterization
technique to obtain
\begin{equation}
\left(\mathrm{D}_{1}^{(1,1)}\right)_{ab}=\frac{1}{2}\tilde{k}_{a}^{\mu}\tilde{k}_{b}^{\nu}\int_{0}^{1}dx\int\frac{d^{d}q}{\left(2\pi\right)^{d}}\frac{\left(\frac{9}{2}q^{2}+\left(5-2x+2x^{2}\right)k^{2}\right)g_{\mu\nu}-\left(2+10x-10x^{2}\right)k_{\mu}k_{\nu}}{\left[q^{2}+x\left(1-x\right)k^{2}\right]^{2}}
\end{equation}
We are only interested in the divergent part, which in dimensional
regularization with $d=4-\epsilon$ is 
\begin{equation}
\mathrm{div}\left(\left(\mathrm{D}_{1}^{(1,1)}\right)_{ab}\right)=\left(\frac{19}{12}k^{2}k_{a}^{2}\delta_{ab}-\frac{11}{6}k_{a}^{2}k_{b}^{2}\right)\frac{i}{8\pi^{2}\epsilon}\label{eq:divD1(1,1)}
\end{equation}
which has the same form numerically as the usual non-Abelian $A_{\mu}$
formalism.

We now compute
\begin{align}
\left(\mathrm{D}_{1}^{(2,1)}\right)_{ab} & =\frac{1}{2}\int\frac{d^{d}\ell}{\left(2\pi\right)^{d}}\sum_{cd}\frac{V_{acd}^{\left(2\right)}\left(k,\ell\right)V_{bcd}^{\left(1\right)}\left(-k,-\ell\right)}{\ell^{2}\left(\ell+k\right)^{2}\ell_{c}^{2}\left(\ell_{d}+k_{d}\right)^{2}}\\
 & =\frac{1}{4}\int\frac{d^{d}\ell}{\left(2\pi\right)^{d}}\frac{N_{ab}}{\ell^{2}\left(\ell+k\right)^{2}\ell_{a}^{2}\left(\ell_{a}+k_{a}\right)^{2}}\label{eq:D(2,1) integral}
\end{align}
where the numerator is
\begin{eqnarray}
N_{ab} & = & \left(\delta_{ab}-1\right)k_{b}\ell_{a}\left(\ell_{a}+k_{a}\right)\left(2\ell_{b}+k_{b}\right)\nonumber \\
 &  & \times\left(k^{2}k_{a}\left(2\ell_{a}+k_{a}\right)-\ell^{2}\ell_{a}\left(\ell_{a}+2k_{a}\right)+\left(\ell+k\right)^{2}\left(\ell_{a}^{2}-k_{a}^{2}\right)\right).
\end{eqnarray}
The divergent part of Eq.\,(\ref{eq:D(2,1) integral}) is
\begin{equation}
\mathrm{div}\left(\left(\mathrm{D}_{1}^{(2,1)}\right)_{ab}\right)=\frac{1}{4}\left(\delta_{ab}-1\right)k_{b}\left(4k^{2}k_{a}^{2}\delta_{ab}\frac{i}{8\pi^{2}\epsilon}\right)=0.\label{eq:divD1(2,1)}
\end{equation}
This is identically zero because of Eq.\,(\ref{eq:delta_ab identity}).
Due to the symmetry of the diagram we also know that
\begin{equation}
\mathrm{div}\left(\left(\mathrm{D}_{1}^{(1,2)}\right)_{ab}\right)=0.\label{eq:divD1(1,2)}
\end{equation}

Finally, we compute 
\begin{eqnarray}
\left(\mathrm{D}_{1}^{(2,2)}\right)_{ab} & = & \frac{1}{2}\int\frac{d^{d}\ell}{\left(2\pi\right)^{d}}\sum_{cd}\frac{V_{acd}^{\left(2\right)}\left(k,\ell\right)V_{bcd}^{\left(2\right)}\left(k,\ell\right)}{\ell^{2}\left(\ell+k\right)^{2}\ell_{c}^{2}\left(\ell_{d}+k_{d}\right)^{2}}\\
 & = & \frac{1}{8}\sum_{cd}\delta_{acd}\delta_{bcd}\int\frac{d^{d}\ell}{\left(2\pi\right)^{d}}\frac{n_{a}\left(k,\ell\right)n_{b}\left(k,\ell\right)}{\ell^{2}\left(\ell+k\right)^{2}\ell_{c}^{2}\left(\ell_{d}+k_{d}\right)^{2}}\\
 & = & \frac{1}{8}\delta_{ab}\int\frac{d^{d}\ell}{\left(2\pi\right)^{d}}\frac{n_{a}\left(k,\ell\right)^{2}}{\ell^{2}\left(\ell+k\right)^{2}\ell_{a}^{2}\left(\ell_{a}+k_{a}\right)^{2}}\label{eq:D1(2,2)integral}
\end{eqnarray}
where
\begin{equation}
n_{a}\left(k,\ell\right)=k^{2}k_{a}\left(2\ell_{a}+k_{a}\right)-\ell^{2}\ell_{a}\left(\ell_{a}+2k_{a}\right)+\left(\ell+k\right)^{2}\left(\ell_{a}^{2}-k_{a}^{2}\right)\,.
\end{equation}
The divergent part of Eq.\,(\ref{eq:D1(2,2)integral}) is 
\begin{equation}
\mathrm{div}\left(\left(\mathrm{D}_{1}^{(2,2)}\right)_{ab}\right)=\frac{i}{8\pi^{2}\epsilon}\left(\frac{5}{2}k^{2}k_{a}^{2}\delta_{ab}\right).\label{eq:divD1(2,2)}
\end{equation}

After summing the contributions from the sub-diagrams given by Eqs.\,(\ref{eq:divD1(1,1)}),
(\ref{eq:divD1(2,1)}), (\ref{eq:divD1(1,2)}), and (\ref{eq:divD1(2,2)}),
we find that divergent part of the first diagram is
\begin{equation}
\mathrm{div}\left(\left(\mathrm{D}_{1}\right)_{ab}^{AB}\right)=C\left(A\right)\frac{g^{2}}{8\pi^{2}\epsilon}\left(\frac{49}{12}i\delta^{AB}k^{2}k_{a}^{2}\delta_{ab}-\frac{11}{6}i\delta^{AB}k_{a}^{2}k_{b}^{2}\right).\label{eq:div(D1) theta self energy}
\end{equation}

\subsubsection{$\theta$ self energy diagram 2}

The second diagram is given by
\begin{align}
\left(\mathrm{D}_{2}\right)_{ab}^{AB} & =\frac{1}{2}\int\frac{d^{4}\ell}{\left(2\pi\right)^{4}}\sum_{cd}\left(\frac{-i\delta_{cd}\delta^{CD}}{\ell^{2}\ell_{c}^{2}}\right)iV_{abcd}^{ABCD}\left(k,-k,\ell,-\ell\right)\\
 & =\frac{g^{2}}{2}\int\frac{d^{4}\ell}{\left(2\pi\right)^{4}}\sum_{c}\sum_{i=1}^{6}\frac{V_{(i)abcc}^{ABCC}\left(k,-k,\ell,-\ell\right)}{\ell^{2}\ell_{c}^{2}};\label{eq:D2 self energy}
\end{align}
the seventh and eighth terms of Eq.\,(\ref{eq:D2 self energy}) don't
contribute because $\xi=1$. The following identity is useful in evaluating
the divergent part of Eq.\,(\ref{eq:D2 self energy}):
\begin{align}
\mathrm{div}\left(\int\frac{d^{4}\ell}{\left(2\pi\right)^{4}}\frac{\ell_{a}^{N_{a}}\ell_{b}^{N_{b}}}{\ell^{2}\ell_{a}^{2}}\right) & =\delta_{N_{a}0}\delta_{N_{b}0}\,\mathrm{div}\left(\int\frac{d^{4}\ell}{\left(2\pi\right)^{4}}\frac{1}{\ell^{2}\ell_{a}^{2}}\right)\\
 & =\delta_{N_{a}0}\delta_{N_{b}0}\frac{i\Gamma\left(\frac{\epsilon}{2}\right)\Gamma\left(-\frac{1}{2}\right)}{\left(4\pi\right)^{2}\Gamma\left(\frac{1}{2}\right)}\\
 & =\delta_{N_{a}0}\delta_{N_{b}0}\left(-\frac{i}{4\pi^{2}\epsilon}\right).\label{eq:int(1/l^2la^2)}
\end{align}
Since Eq.\,(\ref{eq:int(1/l^2la^2)}) is zero in dimensional regularization
unless $N_{a}=N_{b}=0$, we ignore any term in the numerator of Eq.\,(\ref{eq:D2 self energy})
that has any positive power of $\ell$ to find the divergence. We
need to ignore any term that has $k_{3}=+\ell$ or $k_{4}=-\ell$
since they proportional to $\ell$.

The divergent part of the first four terms of Eq.\,(\ref{eq:D2 self energy})
vanishes due to either Lorentz invariance or Eq.\,(\ref{eq:int(1/l^2la^2)}).
The only non zero divergent contributions come from the fifth term,
which is given by
\begin{eqnarray}
V_{(5)abcc}^{ABCC} & = & 0+\frac{1}{4}f_{E}^{AC}f_{E}^{BC}\delta_{abc}\left[\left(k_{1}+k_{3}\right)^{2}\left(k_{1}-k_{3}\right)_{\star a}\left(k_{2}-k_{4}\right)\right.\nonumber \\
 &  & \left.+\left(k_{1}+k_{4}\right)^{2}\left(k_{1}-k_{4}\right)_{\star a}\left(k_{2}-k_{3}\right)\right]\\
 & \rightarrow & \frac{1}{4}f_{E}^{AC}f_{E}^{BC}\delta_{abc}\left(\left(k_{1}\right)^{2}k_{1\star a}k_{2}+\left(k_{1}\right)^{2}\left(k_{1}\right)_{\star a}\left(k_{2}\right)\right)\\
 & \rightarrow & -\frac{1}{2}f_{E}^{AC}f_{E}^{BC}\delta_{ab}k^{2}k_{a}^{2},\label{eq:V(5)}
\end{eqnarray}
and the sixth term, given by
\begin{eqnarray}
V_{(6)abcc}^{ABCC} & = & \frac{1}{6}f_{E}^{AC}f_{E}^{BC}\delta_{abc}\left[k_{1}^{2}k_{1\star a}\left(k_{4}-k_{2}\right)+k_{3}^{2}k_{3\star a}\left(k_{2}-k_{4}\right)+k_{2}^{2}k_{2\star a}\left(k_{3}-k_{1}\right)+k_{4}^{2}k_{4\star a}\left(k_{1}-k_{3}\right)\right.\nonumber \\
 &  & \left.+k_{1}^{2}k_{1\star a}\left(k_{3}-k_{2}\right)+k_{4}^{2}k_{4\star a}\left(k_{2}-k_{3}\right)+k_{2}^{2}k_{2\star a}\left(k_{4}-k_{1}\right)+k_{3}^{2}k_{3\star a}\left(k_{1}-k_{4}\right)\right]\nonumber \\
 & \rightarrow & \frac{1}{6}f_{E}^{AC}f_{E}^{BC}\delta_{ab}\left(k_{1}^{2}k_{1\star a}\left(-k_{2}\right)+k_{2}^{2}k_{2\star a}\left(-k_{1}\right)+k_{1}^{2}k_{1\star a}\left(-k_{2}\right)+k_{2}^{2}k_{2\star a}\left(-k_{1}\right)\right)\\
 & = & +\frac{2}{3}f_{E}^{AC}f_{E}^{BC}\delta_{ab}k^{2}k_{a}^{2}.\label{eq:V(6)}
\end{eqnarray}

Applying the results from Eq.\,(\ref{eq:V(5)}) and Eq.\,(\ref{eq:V(6)})
to Eq.\,(\ref{eq:D2 self energy}) yields the following divergent
contribution 
\begin{equation}
\mathrm{div}\left(\left(\mathrm{D_{2}}\right)_{ab}^{AB}\right)=-\frac{1}{6}C\left(A\right)\frac{g^{2}}{8\pi^{2}\epsilon}\left(i\delta^{AB}k^{2}k_{a}^{2}\delta_{ab}\right).\label{eq:div(D2) theta self energy}
\end{equation}

\subsubsection{$\theta$ self energy diagram 3}

The ghost-loop diagram 3 of Fig.\,\ref{fig:Theta-self-energy-diagrams}
receives contributions from the ghosts of Eq.\,(\ref{eq:Lghost1}),
which we label as $\mathrm{D_{3}^{(gh1)}}$ and the ghosts of Eq.\,(\ref{eq:Lghost2}),
which we label as $\mathrm{D_{3}^{(gh2)}}$: 
\begin{equation}
\left(\mathrm{D_{3}}\right)_{ab}^{AB}=\left(\mathrm{D_{3}^{(gh1)}}\right)_{ab}^{AB}+\left(\mathrm{\mathrm{D_{3}^{(gh2)}}}\right)_{ab}^{AB}
\end{equation}
where
\begin{align}
\left(\mathrm{D_{3}^{(gh1)}}\right)_{ab}^{AB} & =\left(-1\right)\int_{p}igV_{a}^{A,CD}\left(k,p+k,p\right)\tfrac{1}{i}\Delta^{CF}\left(p+k\right)\tfrac{1}{i}\Delta^{DE}\left(p\right)igV_{b}^{B,EF}\left(-k,p,p+k\right)\\
 & =\left(-1\right)g^{2}\int\frac{d^{4}p}{\left(2\pi\right)^{4}}\frac{\left(f^{ACD}k_{\star a}\left(p+k\right)\right)\left(f^{BDC}\left(-k\right)_{\star b}p\right)}{p^{2}\left(p+k\right)^{2}}\\
 & =g^{2}f^{ACD}f^{BCD}\left(-\tilde{k}_{a}^{\mu}\tilde{k}_{b}^{\nu}\right)\int\frac{d^{4}p}{\left(2\pi\right)^{4}}\frac{\left(p+k\right)_{\mu}p_{\nu}}{p^{2}\left(p+k\right)^{2}}\label{eq:gh1integ}
\end{align}
and
\begin{align}
\left(\mathrm{D_{3}^{(gh2)}}\right)_{ab}^{AB} & =\left(-1\right)g^{2}\sum_{c,d}\int_{p}\frac{f^{ACD}\delta_{acd}\left(k-p\right)\star_{a}\left(p+k\right)f^{BDC}\delta_{bdc}\left(-k-p-k\right)\star_{a}p}{\left(p_{c}+k_{c}\right)^{2}p_{d}^{2}}\\
 & =-\frac{g^{2}}{4}C(A)\delta^{AB}\delta_{ab}\int_{p}\frac{\left(p_{a}-k_{a}\right)\left(p_{a}+k_{a}\right)\left(p_{a}+2k_{a}\right)p_{a}}{\left(p_{a}+k_{a}\right)^{2}p_{a}^{2}}.
\end{align}
Using the usual Feynman parameterization, the integral of Eq.\,(\ref{eq:gh1integ})
becomes

\begin{align}
\int\frac{d^{d}p}{\left(2\pi\right)^{d}}\frac{\left(p+k\right)_{\mu}p_{\nu}}{p^{2}\left(p+k\right)^{2}} & =\frac{i}{\left(4\pi\right)^{2}}\frac{2}{\epsilon}\int_{0}^{1}dx\left(-\frac{1}{2}g_{\mu\nu}x\left(1-x\right)k^{2}-x\left(1-x\right)k_{\mu}k_{\nu}\right)+\mathrm{finite}\\
 & =\frac{i}{8\pi^{2}\epsilon}\left(-\frac{1}{12}k^{2}g_{\mu\nu}-\frac{1}{6}k_{\mu}k_{\nu}\right)+\mathrm{finite}
\end{align}
and therefore 
\begin{equation}
\mathrm{div}\left(\left(\mathrm{D_{3}^{(gh1)}}\right)_{ab}^{AB}\right)=i\delta^{AB}C\left(A\right)\frac{g^{2}}{8\pi^{2}\epsilon}\left(\frac{1}{12}k^{2}k_{a}^{2}\delta_{ab}+\frac{1}{6}k_{a}^{2}k_{b}^{2}\right).\label{eq:D3(gh1)}
\end{equation}

The divergent part of $\mathrm{D_{3}^{(gh2)}}$ in dimensional regularization
is zero because of Eq.\,(\ref{eq:dim reg identity}) for $n=3$:
\begin{equation}
\mathrm{div}\left(\left(\mathrm{D_{3}^{(gh2)}}\right)_{ab}^{AB}\right)=0.\label{eq:D3(gh2)}
\end{equation}
As noted before, it is interesting that the ghosts arising from transforming
$A_{\mu}^{B}$ to $\theta_{c}^{A}$ do not contribute to the divergent
structure here. Combining these results, we conclude that 
\begin{equation}
\mathrm{div}\left(\left(\mathrm{D_{3}}\right)_{ab}^{AB}\right)=i\delta^{AB}C\left(A\right)\frac{g^{2}}{8\pi^{2}\epsilon}\left(\frac{1}{12}k^{2}k_{a}^{2}\delta_{ab}+\frac{1}{6}k_{a}^{2}k_{b}^{2}\right).\label{eq:div(D3) theta self energy}
\end{equation}
This ghost contribution will be important for restoring the transverse
structure of the gauge boson propagator.

\subsubsection{$\theta$ self energy diagram 4}

Similar to diagram 3, diagram 4 of Fig.\,\ref{fig:Theta-self-energy-diagrams}
describes ghost contributions to the propagator. These however do
not have any external momenta flowing through the ghost-lines. Just
as in diagram 3, this has a contribution coming from the usual gauge-fixing
ghost and the ghost associated with transforming the field coordinates
from $A_{\mu}^{B}$ to $\theta_{c}^{A}$: 
\begin{equation}
\left(\mathrm{D_{4}}\right)_{ab}^{AB}=\left(\mathrm{D_{4}^{(gh1)}}\right)_{ab}^{AB}+\left(\mathrm{D_{4}^{(gh2)}}\right)_{ab}^{AB}
\end{equation}

We find the first ghost contribution to be
\begin{align}
\left(\mathrm{D_{4}^{(gh1)}}\right)_{ab}^{AB} & =\left(-1\right)\int\frac{d^{4}p}{\left(2\pi\right)^{4}}ig^{2}V_{ab}^{AB,CD}\left(k,-k,p,p\right)\tfrac{1}{i}\Delta^{CD}\left(p\right)\\
 & =\left(-1\right)\frac{g^{2}}{2}\int\frac{d^{4}p}{\left(2\pi\right)^{4}}\frac{f^{ABE}f^{CCE}\delta_{ab}2k_{a}p_{a}}{p^{2}}\\
 & =0
\end{align}
and the second ghost contribution to be
\begin{align}
\left(\mathrm{D_{4}^{(gh2)}}\right)_{ab}^{AB} & =\left(-1\right)\sum_{c}\int\frac{d^{4}p}{\left(2\pi\right)^{4}}ig^{2}V_{ab,cd}^{AB,CD}\left(k,-k,p,p\right)\tfrac{1}{i}\Delta_{cd}^{CD}\left(p\right)\\
 & =\frac{-ig^{2}}{6}\delta_{ab}\sum_{c}\int\frac{d^{4}p}{\left(2\pi\right)^{4}}\frac{f^{ACE}f^{BCE}\delta_{abc}\left(\left(p+k\right)\star_{a}p+\left(p-k\right)\star_{a}p\right)}{p_{a}^{2}}\\
 & =\frac{-ig^{2}}{6}f^{ACE}f^{BCE}\delta_{ab}\int\frac{d^{4}p}{\left(2\pi\right)^{4}}\frac{2p_{a}^{2}}{p_{a}^{2}}.
\end{align}
Using the identity Eq.\,(\ref{eq:dim reg identity}), this also vanishes:
\begin{equation}
\mathrm{div}\left(\left(\mathrm{D_{4}^{(gh2)}}\right)_{ab}^{AB}\right)=0.
\end{equation}
Therefore, we conclude
\begin{equation}
\mathrm{div}\left(\left(\mathrm{D}_{4}\right)_{ab}^{AB}\right)=0\label{eq:div(D4) theta self energy}
\end{equation}
and thus there are no external momentum independent ghost contribution
to the divergent structure of the $\theta$ propagator in dimensional
regularization.

\subsubsection{$\theta$ self energy counter-term}

The counter-term diagram yields
\begin{eqnarray}
\left(\mathrm{D_{c.t.}}\right)_{ab}^{AB} & = & -i\delta^{AB}\left((Z_{\theta^{2}}-1)\delta^{AB}\left(k^{2}k_{a}^{2}\delta_{ab}-k_{a}^{2}k_{b}^{2}\right)+\frac{1}{\xi}(Z_{\frac{1}{\xi}\theta^{2}}-1)k_{a}^{2}k_{b}^{2}\right)\\
 & = & -i\delta^{AB}\left((Z_{\theta^{2}}-1)k^{2}k_{a}^{2}\delta_{ab}+(Z_{\frac{1}{\xi}\theta^{2}}-Z_{\theta^{2}})k_{a}^{2}k_{b}^{2}\right).
\end{eqnarray}
To have a finite self energy, we require the divergent parts of these
diagrams to cancel out. The sum of Eqs.\,(\ref{eq:div(D1) theta self energy}),
(\ref{eq:div(D2) theta self energy}), (\ref{eq:div(D3) theta self energy}),
and (\ref{eq:div(D4) theta self energy}) is 
\begin{equation}
\mathrm{div}\left(\sum_{i=1}^{4}\left(\mathrm{D}_{i}\right)_{ab}^{AB}\right)=i\delta^{AB}C\left(A\right)\frac{g^{2}}{8\pi^{2}\epsilon}\left(4k^{2}k_{a}^{2}\delta_{ab}-\frac{5}{3}k_{a}^{2}k_{b}^{2}\right)\label{eq:theta self energy}
\end{equation}
and therefore the renormalization constants are
\begin{align}
Z_{\theta^{2}} & =1+4C\left(A\right)\frac{g^{2}}{8\pi^{2}\epsilon}\label{eq:Ztheta^2 result}
\end{align}
and
\begin{equation}
Z_{\frac{1}{\xi}\theta^{2}}=1+\frac{7}{3}C\left(A\right)\frac{g^{2}}{8\pi^{2}\epsilon}.\label{eq:Z_xi^-1*theta^2 result}
\end{equation}

It is interesting that despite the nontransversality of the divergent
part of the $\theta$ propagator seen here, the divergent part of
the usual gauge field propagator when computed in the BTGT formalism
will maintain transversality, as we will demonstrate below.

\subsubsection{Comment on $Z_{\frac{1}{\xi}\theta^{2}}$}

Note that 
\begin{equation}
Z_{\xi}=\frac{Z_{\theta^{2}}}{Z_{\frac{1}{\xi}\theta^{2}}}=1+\frac{5}{3}C\left(A\right)\frac{g^{2}}{8\pi^{2}\epsilon}=Z_{A^{2}}=\frac{Z_{A^{2}}}{Z_{\frac{1}{\xi}A^{2}}}
\end{equation}
where $Z_{A^{2}}$ is gauge kinetic renormalization constant in the
usual gauge theory formalism. This is a nontrivial check of the theory.
It shows that $\xi_{B}=Z_{\xi}\xi_{R}$ has the same scaling behavior
in BTGT as in the usual formalism. It is interesting that while $Z_{\frac{1}{\xi}A^{2}}=1$
to all orders in $g$, $Z_{\frac{1}{\xi}\theta^{2}}-1\neq0$. This
does not indicate a violation of gauge invariance because the gauge
fixing parameter $\xi$ (parameterizing the coefficient of the gauge
fixing chosen to be of the same form as in ordinary gauge theories
with $A_{\mu}^{a}\rightarrow A_{\mu}^{a}[\theta]$) is still renormalized
by the same renormalization constant of $Z_{\xi}$ as in the ordinary
gauge theory formalism and $Z_{\theta^{2}}\neq Z_{A^{2}}$.

Another nontrivial check of the formalism would be to calculate $Z_{g\theta^{3}}$
and $Z_{\frac{1}{\xi}g\theta^{3}}$ and check that they satisfy 
\begin{equation}
\frac{Z_{g\theta^{3}}}{Z_{\frac{1}{\xi}g\theta^{3}}}=1+\frac{5}{3}C\left(A\right)\frac{g^{2}}{8\pi^{2}\epsilon}+O\left(g^{4}\right)=Z_{\xi},
\end{equation}
but we defer this to a future work.

\subsection{Computation of $Z_{\bar{c}c}$}

\begin{figure}
\begin{centering}
\includegraphics[width=13cm]{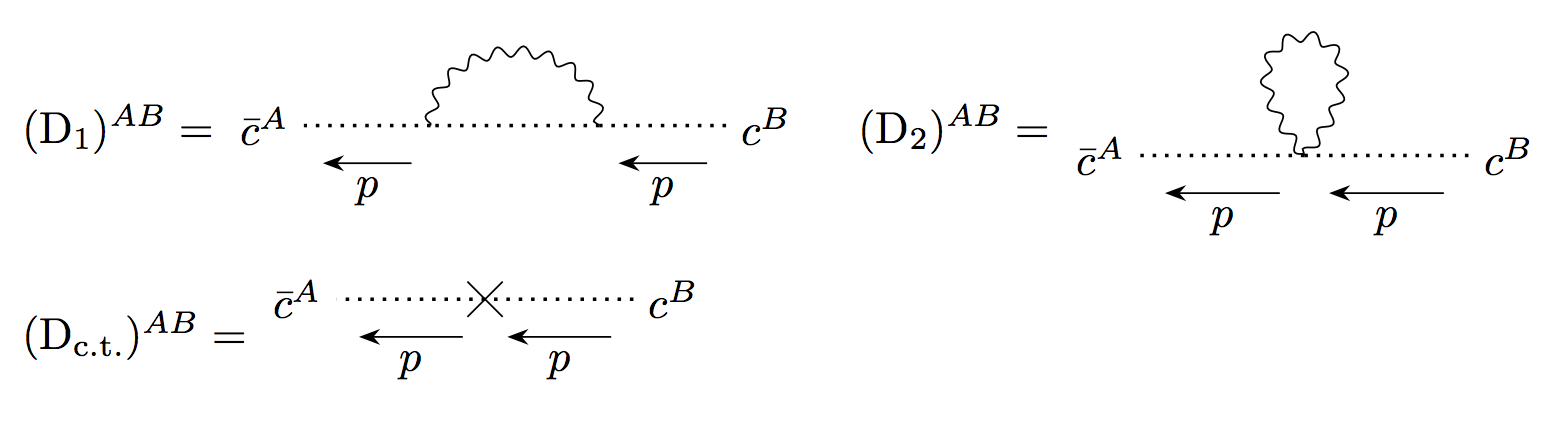}
\par\end{centering}

\caption{Ghost self energy diagrams \label{fig:Ghost-self-energy diagrams}}
\end{figure}

The renormalization constant $Z_{\bar{c}c}$ is determined by the
ghost self energy. The one loop diagrams that contribute to the ghost
self energy are given in Fig.\,\ref{fig:Ghost-self-energy diagrams}.

The first diagram in Fig.\,\ref{fig:Ghost-self-energy diagrams} is
\begin{align}
\left(\mathrm{D_{1}}\right)^{AB}= & g^{2}f^{CAD}f^{CDB}\sum_{c}\int_{\ell}\frac{\left(-\ell_{c}p_{c}\right)\ell_{c}\left(\ell_{c}+p_{c}\right)}{\ell_{c}^{2}\ell^{2}\left(\ell+p\right)^{2}}\\
= & g^{2}C\left(A\right)\delta^{AB}\int_{\ell}\frac{p^{2}+p\cdot\ell}{\ell^{2}\left(\ell+p\right)^{2}}\\
= & g^{2}C\left(A\right)\delta^{AB}p^{2}\int_{0}^{1}dx\left(1-x\right)\int_{q}\frac{1}{\left[q^{2}+x\left(1-x\right)p^{2}\right]^{2}}.
\end{align}
The divergent part of this is
\begin{align}
\mathrm{div}\left(\left(\mathrm{D_{1}}\right)^{AB}\right) & =-\frac{1}{2}C\left(A\right)\frac{g^{2}}{8\pi^{2}\epsilon}\left(-i\delta^{AB}p^{2}\right).
\end{align}

\noindent The second diagram in Fig.\,\ref{fig:Ghost-self-energy diagrams}
vanishes identically because of the anti-symmetric property of $f^{CDE}$:
\begin{align}
\left(\mathrm{D_{2}}\right)^{AB} & =g^{2}\sum_{c,d}\int_{\ell}V_{cd}^{CD,AB}\left(\ell,-\ell,p\right)\Delta_{cd}^{CD}\left(\ell\right)\\
 & =g^{2}\sum_{c,d}\delta_{cd}\left(\frac{1}{2}f_{E}^{CD}f_{E}^{AB}\left(\ell_{c}+\ell_{c}\right)p_{c}\right)\frac{\delta_{cd}\delta^{CD}}{\ell^{2}\ell_{c}^{2}}\\
 & =0.
\end{align}

\noindent The counter-term diagram is given by
\begin{equation}
\left(\mathrm{D_{\mathrm{c.t.}}}\right)^{AB}=-i\left(Z_{\bar{c}c}-1\right)\delta^{AB}p^{2}.
\end{equation}
In order to make the ghost self energy finite, we find that
\begin{equation}
Z_{\bar{c}c}=1+\frac{1}{2}C\left(A\right)\frac{g^{2}}{8\pi^{2}\epsilon}+O\left(g^{4}\right)\,.\label{eq:ZccComputed}
\end{equation}

Note that Eq.\,(\ref{eq:ZccComputed}) is the same result that is
obtained in the usual computation with $A_{\mu}^{a}$ fields. This
is most likely part of a general result discussed in more detail in
\ref{sub:Callan-Symanzik-Equation}.

\subsection{Computation of $Z_{g\theta\bar{c}c}$}

\begin{figure}
\begin{centering}
\includegraphics[width=13cm]{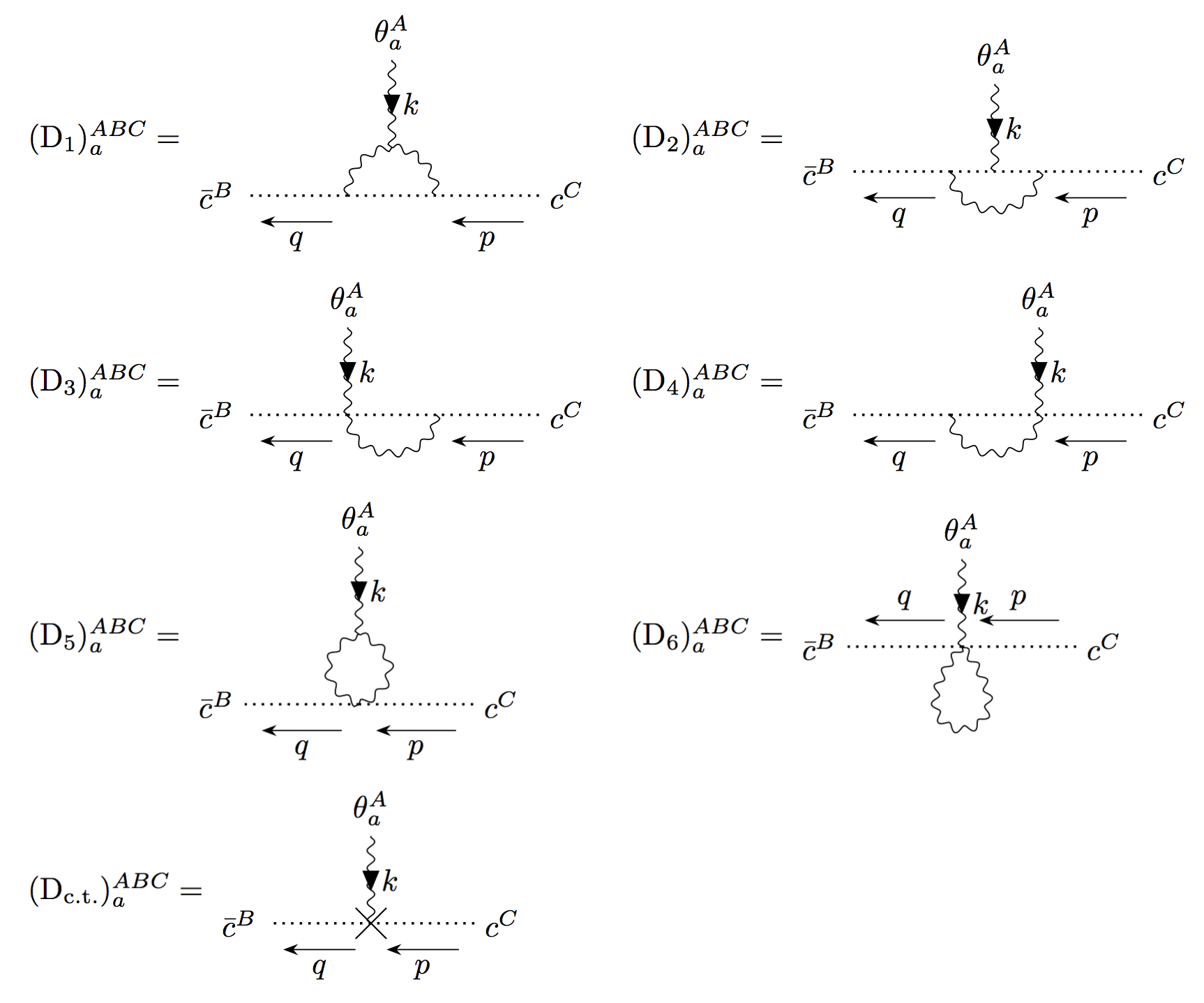}
\par\end{centering}

\caption{Ghost-$\theta$ vertex one loop diagrams \label{fig:Ghost-theta-vertex diagrams}}
\end{figure}

Let's now compute the $\theta_{a}$-ghost interaction in our continuing
efforts to derive Eq.\,(\ref{eq:beta_result}). The relevant diagrams
are defined in Fig.\,\ref{fig:Ghost-theta-vertex diagrams}.

One of the surprises in the computation below will be that the first
diagram $D_{1}$ of Fig.\,\ref{fig:Ghost-theta-vertex diagrams}
vanishes. This is in contrast with the case in which $\theta_{a}^{A}$
is replaced by $A_{\mu}^{A}$. Another interesting aspect of the computation
will be that diagrams $D_{3}$ and $D_{4}$ each violate the BTGT
symmetry in the divergence, but their sum has a cancellation that
thereby preserves the BTGT symmetry.

\subsubsection{Ghost-$\theta$ vertex diagrams 1 and 2}

Diagram 1 in Fig.\,\ref{fig:Ghost-theta-vertex diagrams} is given by
\begin{align}
\left(\mathrm{D_{1}}\right)_{a}^{ABC} & =g^{3}f^{EBD}f^{FDC}f^{AEF}\sum_{e,f}\int_{\ell}\frac{\left(-\ell_{e}q_{e}\right)\left(\ell_{f}+k_{f}\right)\left(\ell_{f}+q_{f}\right)\left(V_{aef}^{(1)}\left(k,\ell\right)+V_{aef}^{(2)}\left(k,\ell\right)\right)}{\ell^{2}\left(\ell+k\right)^{2}\left(\ell+q\right)^{2}\ell_{e}^{2}\left(\ell_{f}+k_{f}\right)^{2}}\\
 & =\left(\mathrm{D_{1}^{\left(1\right)}}\right)_{a}^{ABC}+\left(\mathrm{D_{1}^{\left(2\right)}}\right)_{a}^{ABC}
\end{align}
where we have denoted the $V_{aef}^{(n)}$ contributions as $D_{1}^{(n)}$
which we will evaluate separately. Through the identity 
\begin{equation}
f^{AEF}f^{EBD}f^{FDC}=-f_{A}^{FE}f_{B}^{ED}f_{C}^{DF}=-\frac{1}{2}f^{ABC}C\left(A\right),
\end{equation}
 the first contribution can be written as 
\begin{align}
\left(\mathrm{D_{1}^{\left(1\right)}}\right)_{a}^{ABC}= & -\frac{g^{3}}{2}f^{ABC}C\left(A\right)\sum_{e,f}\int_{\ell}\frac{\left(-\ell_{e}q_{e}\right)\left(\ell_{f}+k_{f}\right)\left(\ell_{f}+q_{f}\right)V_{aef}^{\left(1\right)}(k,\ell)}{\ell^{2}\left(\ell+k\right)^{2}\left(\ell+q\right)^{2}\ell_{e}^{2}\left(\ell_{f}+k_{f}\right)^{2}}\\
= & -\frac{g^{3}}{2}f^{ABC}C\left(A\right)k_{a}\nonumber \\
 & \times\int_{\ell}\frac{q_{a}\left(\ell+q\right)\cdot\left(k-\ell\right)-\left(\ell_{a}+q_{a}\right)q\cdot\left(\ell+2k\right)+\left(2\ell_{a}+k_{a}\right)q\cdot\left(\ell+q\right)}{\ell^{2}\left(\ell+k\right)^{2}\left(\ell+q\right)^{2}}\,.
\end{align}
A divergence only occurs when the numerator is at $\ell^{2}$ or higher
powers in $\ell$. There are no terms higher than $\ell^{2}$ and
therefore the maximum degree of divergence is zero. This means that
we can ignore the dependence on the external momenta in the denominator:
\begin{align}
\mathrm{div}\left(\left(\mathrm{D_{1}^{\left(1\right)}}\right)_{a}^{ABC}\right) & =-\frac{g^{3}}{2}f^{ABC}C\left(A\right)k_{a}\mathrm{div}\left(\int_{\ell}\frac{\left(\ell\cdot q\right)\ell_{a}-\ell^{2}q_{a}}{\ell^{2}\left(\ell+k\right)^{2}\left(\ell+q\right)^{2}}\right)\\
 & =-\frac{g^{3}}{2}f^{ABC}C\left(A\right)k_{a}\left(-\frac{3}{4}q_{a}\frac{i}{8\pi^{2}\epsilon}\right)\\
 & =+\frac{3}{8}C\left(A\right)\frac{g^{2}}{8\pi^{2}\epsilon}\left(igf^{ABC}k_{a}q_{a}\right)\,.\label{eq:D_1(1)}
\end{align}

The second contribution to this diagram is 
\begin{eqnarray}
\left(\mathrm{D_{1}^{\left(2\right)}}\right)_{a}^{ABC} & = & -\frac{g^{3}}{2}f^{ABC}C\left(A\right)\sum_{e,f}\int_{\ell}\frac{\left(-\ell_{e}q_{e}\right)\left(\ell_{f}+k_{f}\right)\left(\ell_{f}+q_{f}\right)V_{aef}^{\left(2\right)}(k,\ell)}{\ell^{2}\left(\ell+k\right)^{2}\left(\ell+q\right)^{2}\ell_{e}^{2}\left(\ell_{f}+k_{f}\right)^{2}}\\
 & = & \frac{1}{4}g^{3}f^{ABC}C\left(A\right)q_{a}\int_{\ell}\ell_{a}\left(\ell_{a}+k_{a}\right)\left(\ell_{a}+q_{a}\right)\nonumber \\
 &  & \times\frac{\left(k^{2}k_{a}\left(2\ell_{a}+k_{a}\right)-\ell^{2}\ell_{a}\left(\ell_{a}+2k_{a}\right)+\left(\ell+k\right)^{2}\left(\ell_{a}^{2}-k_{a}^{2}\right)\right)}{\ell^{2}\left(\ell+k\right)^{2}\left(\ell+q\right)^{2}\ell_{a}^{2}\left(\ell_{a}+k_{a}\right)^{2}}\,.
\end{eqnarray}
The divergent part evaluates to
\begin{align}
\mathrm{div}\left(\left(\mathrm{D_{1}^{\left(2\right)}}\right)_{a}^{ABC}\right) & =-\frac{3}{8}C\left(A\right)\frac{g^{2}}{8\pi^{2}\epsilon}\left(igf^{ABC}k_{a}q_{a}\right)\,.
\end{align}
Summing these contributions together gives
\begin{equation}
\mathrm{div}\left(\left(\mathrm{D_{1}}\right)_{a}^{ABC}\right)=\left(+\frac{3}{8}-\frac{3}{8}\right)C\left(A\right)\frac{g^{2}}{8\pi^{2}\epsilon}\left(igf^{ABC}k_{a}q_{a}\right)=0\,.\label{eq:D1 ghost gauge}
\end{equation}

The result of the diagram $D_{1}$ calculation with $\theta_{a}^{A}$
replaced with $A_{\mu}^{A}$ is equivalent to Eq.\,(\ref{eq:D_1(1)})
(see e.g.\,\cite{Grozin:2005yg}). The difference between this result
and Eq.\,(\ref{eq:D1 ghost gauge}) is a manifestation of how $\theta_{a}^{A}$
is different from $A_{\mu}^{A}$.

Diagram 2 is given by
\begin{align}
\left(\mathrm{D_{2}}\right)_{a}^{ABC} & =g^{3}\sum_{f}\int_{\ell}\frac{V_{f}^{F,BD}\left(-\ell+p,q\right)V_{a}^{A,DE}\left(k,\ell+k\right)V_{f}^{F,EC}\left(\ell-p,\ell\right)}{\left(\ell+k\right)^{2}\ell^{2}\left(\ell-p\right)^{2}\left(\ell_{f}-p_{f}\right)^{2}}\\
 & =g^{3}f^{FBD}f^{ADE}f^{FEC}\sum_{f}\int_{\ell}\frac{\left(-\ell_{f}+p_{f}\right)q_{f}k_{a}\left(\ell_{a}+k_{a}\right)\left(\ell_{f}-p_{f}\right)\ell_{f}}{\left(\ell+k\right)^{2}\ell^{2}\left(\ell-p\right)^{2}\left(\ell_{f}-p_{f}\right)^{2}}\\
 & =\frac{g^{3}}{2}C\left(A\right)f^{ABC}k_{a}\int_{\ell}\frac{\left(\ell_{a}+k_{a}\right)\sum_{f}q_{f}\ell_{f}}{\left(\ell+k\right)^{2}\ell^{2}\left(\ell-p\right)^{2}},
\end{align}
and the divergent part of this diagram is therefore
\begin{equation}
\mathrm{div}\left(\left(\mathrm{D_{2}}\right)_{a}^{ABC}\right)=+\frac{1}{8}C\left(A\right)\frac{g^{2}}{8\pi^{2}\epsilon}\left(if^{ABC}k_{a}q_{a}\right).\label{eq:D2 ghost gauge}
\end{equation}
The $1/8$ coefficient here is obtained when we replaces the $\theta_{a}^{A}$
with $A_{\mu}^{A}$ in the usual gauge theory.

\subsubsection{Ghost-$\theta$ vertex diagram 3 and 4}

Diagram 3 evaluates to
\begin{align}
\left(\mathrm{D_{3}}\right)_{a}^{ABC} & =g^{3}\int_{\ell}\sum_{d}\frac{V_{ad}^{AD,BE}\left(k,-\ell;q,\ell+p\right)V_{d}^{D,EC}\left(\ell;\ell+p,p\right)}{\ell^{2}\ell_{d}^{2}\left(\ell+p\right)^{2}}\\
 & =\frac{1}{2}g^{3}f_{F}^{AD}f_{F}^{BE}f^{DEC}\sum_{d}\int_{\ell}\frac{\delta_{ad}\left(k_{d}+\ell_{d}\right)q_{d}\ell_{d}\left(\ell_{d}+p_{d}\right)}{\ell^{2}\ell_{d}^{2}\left(\ell+p\right)^{2}}\\
 & =\frac{1}{4}g^{3}C\left(A\right)f^{ABC}q_{a}\int_{\ell}\frac{\left(\ell_{a}+k_{a}\right)\ell_{a}\left(\ell_{a}+p_{a}\right)}{\ell^{2}\left(\ell+p\right)^{2}\ell_{a}^{2}}
\end{align}
and after integrating, we find the divergent part is
\begin{align}
\mathrm{div}\left(\left(\mathrm{D_{3}}\right)_{a}^{ABC}\right) & =\frac{1}{4}C\left(A\right)\frac{g^{2}}{8\pi^{2}\epsilon}\left(igf^{ABC}\right)\left(\frac{1}{2}q_{a}k_{a}+\frac{1}{2}q_{a}^{2}\right).
\end{align}

Diagram 4 evaluates to
\begin{align}
\left(\mathrm{D_{4}}\right)_{a}^{ABC} & =g^{3}\int_{\ell}\sum_{d}\frac{V_{d}^{D,BE}\left(-\ell;q,\ell+p\right)V_{ad}^{AD,EC}\left(k,\ell;\ell+q,p\right)}{\ell^{2}\ell_{d}\left(\ell+q\right)^{2}}\\
 & =\frac{1}{2}g^{3}f^{DBE}f_{F}^{AD}f_{F}^{EC}\int_{\ell}\sum_{d}\frac{\left(-\ell_{d}q_{d}\right)\delta_{ad}\left(k_{a}-\ell_{a}\right)\left(\ell_{a}+q_{a}\right)}{\ell^{2}\left(\ell+q\right)^{2}\ell_{d}^{2}}\\
 & =-\frac{1}{4}g^{3}C\left(A\right)f^{ABC}q_{a}\int_{\ell}\frac{\ell_{a}\left(\ell_{a}-k_{a}\right)\left(\ell_{a}+q_{a}\right)}{\ell^{2}\left(\ell+q\right)^{2}\ell_{a}^{2}},
\end{align}
and the divergent part is 
\begin{align}
\mathrm{div}\left(\left(\mathrm{D_{4}}\right)_{a}^{ABC}\right) & =\frac{1}{4}C\left(A\right)\frac{g^{2}}{8\pi^{2}\epsilon}\left(igf^{ABC}\right)\left(q_{a}k_{a}-\frac{1}{2}q_{a}^{2}\right)\,.
\end{align}

Even though the divergent parts of $\mathrm{D_{3}}$ and $\mathrm{D_{4}}$
separately lead to new counter terms that would violate BTGT and gauge
invariance, their sum does not. The BTGT violating term proportional
to $q_{a}^{2}$ cancels and we are left with
\begin{equation}
\mathrm{div}\left(\left(\mathrm{D_{3}}\right)_{a}^{ABC}+\left(\mathrm{D_{4}}\right)_{a}^{ABC}\right)=\frac{3}{8}C\left(A\right)\frac{g^{2}}{8\pi^{2}\epsilon}\left(igf^{ABC}q_{a}k_{a}\right)\,.\label{eq:D3 + D4 ghost gauge}
\end{equation}
This contribution does not have an analog in the ordinary gauge theory
formalism in which there is no quartic coupling of the gauge sector
to the ghosts.

\subsubsection{Ghost-$\theta$ vertex diagram 5}

Diagram 5 in Fig.\,\ref{fig:Ghost-theta-vertex diagrams} is given
by 
\begin{align}
\mathrm{\left(D_{5}\right)}_{a}^{ABC} & =\frac{g^{3}}{2}\sum_{d,e}\int_{\ell}\frac{V_{ade}^{ADE}\left(k,\ell\right)V_{de}^{DE,BC}\left(-\ell,\ell+k;q,p\right)}{\ell^{2}\left(\ell+k\right)^{2}\ell_{d}^{2}\left(\ell_{e}+\ell_{e}\right)^{2}}\\
 & =\frac{g^{3}}{2}\sum_{d,e}\int_{\ell}\frac{f^{ADE}V_{ade}\left(k,\ell\right)\frac{1}{2}\delta_{de}f_{F}^{DE}f_{F}^{BC}q_{d}\left(-2\ell_{d}-k_{d}\right)}{\ell^{2}\left(\ell+k\right)^{2}\ell_{d}^{2}\left(\ell_{e}+\ell_{e}\right)^{2}}\\
 & =-\frac{g^{3}}{4}C\left(A\right)f^{ABC}\sum_{d}\int_{\ell}\frac{V_{add}\left(k,\ell\right)q_{d}\left(2\ell_{d}+k_{d}\right)}{\ell^{2}\left(\ell+k\right)^{2}\ell_{d}^{2}\left(\ell_{d}+k_{d}\right)^{2}}\\
 & =\left(\mathrm{D_{5}^{(1)}}\right)_{a}^{ABC}+\left(\mathrm{D_{5}^{(2)}}\right)_{a}^{ABC}\,.
\end{align}
 We find that
\begin{equation}
\mbox{div}\left(\left(\mathrm{D_{5}^{\left(1\right)}}\right)_{a}^{ABC}\right)=0,
\end{equation}
\begin{equation}
\mbox{div}\left(\mathrm{D_{5}^{\left(2\right)}}\right)=-\frac{3}{2}C\left(A\right)\frac{g^{2}}{8\pi^{2}\epsilon}\left(igf^{ABC}q_{a}k_{a}\right),
\end{equation}
and $\mathrm{D_{5}^{\left(3\right)}}=0$ in the Feynman gauge. The
total divergent component of this diagram is thus
\begin{equation}
\mathrm{div}\left(\mathrm{\left(D_{5}\right)}_{a}^{ABC}\right)=-\frac{3}{2}C\left(A\right)\frac{g^{2}}{8\pi^{2}\epsilon}\left(igf^{ABC}q_{a}k_{a}\right)\,.\label{eq:D5 ghost gauge}
\end{equation}
Diagram 6 in Fig.\,\ref{fig:Ghost-theta-vertex diagrams} is
\begin{align}
\left(\mathrm{D_{6}}\right)_{a}^{ABC} & =\frac{1}{2}g^{3}\sum_{d}\int_{\ell}\frac{V_{add}^{ADD,BC}\left(k,\ell,-\ell;p,q\right)}{\ell^{2}\ell_{d}^{2}}\\
 & =\frac{1}{12}g^{3}\sum_{d}\int_{\ell}\frac{\delta_{add}q_{a}f_{F}^{BC}\left(0+f_{G}^{FD}f_{G}^{DA}\left(k_{a}+\ell_{a}\right)+f_{G}^{FD}f_{G}^{AD}\left(\ell_{a}-k_{a}\right)\right)}{\ell^{2}\ell_{d}^{2}}\\
 & =\frac{1}{12}g^{3}f_{F}^{BC}\left(-C\left(A\right)\delta^{AF}\right)q_{a}\int_{\ell}\frac{2k_{a}}{\ell^{2}\ell_{a}^{2}}\,.
\end{align}
The divergent part turns out to be
\begin{equation}
\mathrm{div}\left(\left(\mathrm{D_{6}}\right)_{a}^{ABC}\right)=+\frac{1}{3}C\left(A\right)\frac{g^{2}}{8\pi^{2}\epsilon}\left(igf^{ABC}q_{a}k_{a}\right)\,.\label{eq:D6 ghost gauge}
\end{equation}

\subsubsection{Counter term and the conclusion of the explicit computation of the
beta function}

The counter term is
\begin{equation}
\left(\mathrm{D_{c.t.}}\right)_{a}^{ABC}=\left(Z_{g\theta\bar{c}c}-1\right)\left(igf^{ABC}q_{a}k_{a}\right)\,.
\end{equation}
After summing the contributions from Eqs.\,(\ref{eq:D1 ghost gauge}),
(\ref{eq:D2 ghost gauge}), (\ref{eq:D3 + D4 ghost gauge}), (\ref{eq:D5 ghost gauge}),
and (\ref{eq:D6 ghost gauge}), we immediately find the renormalization
constant
\begin{equation}
Z_{g\theta\bar{c}c}=1+\frac{2}{3}C\left(A\right)\frac{g^{2}}{8\pi^{2}\epsilon}+O\left(g^{4}\right).
\end{equation}

Hence, we have finally accomplished our computation of the $Z_{g}$
given by Eq.\,(\ref{eq:Z_g}) using the non-Abelian BTGT formalism.
Thus, as mentioned at the beginning of this section where we embarked
on an explicit computation of the beta function, it is gratifying
to see that the $\theta_{a}^{A}$ formalism can be used to reproduce
the perturbative results of the $A_{\mu}^{A}$ formalism. The true
physics advantage of using the non-Abelian BTGT formalism has yet
to be discovered, but its existence is expected since simple correlators
in $\theta_{a}^{A}$ will map to nonlinear and nonlocal $A_{\mu}^{B}$
correlators.

\subsection{Callan-Symanzik Equation and the beta function \label{sub:Callan-Symanzik-Equation}}

Here we give another perspective on the beta function computation
which we have explicitly carried out in the previous subsections.
We expect the correlator $\left\langle \Psi\overline{\Psi}\right\rangle $
to be independent of the gauge formalism chosen for any matter or
ghost field $\Psi$ because the change from the $A_{\mu}$ formalism
to $\theta_{a}$ formalism does not depend on $\Psi$. In other words,
assuming
\begin{equation}
\left\langle \Psi\overline{\Psi}\right\rangle ^{(A)}=\left\langle \Psi\overline{\Psi}\right\rangle ^{(\theta)},
\end{equation}
and using the Callan\textendash Symanzik equation 
\begin{equation}
\left[\frac{\partial}{\partial\ln\mu}+\beta\left(g\right)\frac{\partial}{\partial g}-\xi\frac{\partial\ln Z_{\xi}}{\partial\ln\mu}\frac{\partial}{\partial\xi}+\frac{\partial\ln Z_{\Psi\Psi}}{\partial\ln\mu}\right]\left\langle \Psi\overline{\Psi}\right\rangle =0,\label{eq:Callan-Symanzik}
\end{equation}
we infer that
\begin{equation}
\beta^{\left(A\right)}\left(g\right)=\beta^{\left(\theta\right)}\left(g\right),\label{eq:beta(A) is beta(theta)}
\end{equation}
\begin{equation}
Z_{\xi}^{(A)}=Z_{\xi}^{(\theta)},\label{eq:Z_xi is the same}
\end{equation}
and
\begin{equation}
Z_{\Psi\Psi}^{\left(A\right)}=Z_{\Psi\Psi}^{\left(\theta\right)}.\label{eq:Z_psipsi same}
\end{equation}
Even more generally, the anomalous dimension of any matter or ghost
field should be independent of the gauge formalism.

\section{Composite operator correlator }

One of the key differences of non-Abelian BTGT from Abelian BTGT is
the appearance of the nonlinearity in the map between the\textbf{
$\theta_{a}^{A}$} variable and the ordinary gauge field $A_{\mu}^{A}$
variable. Hence, any $A_{\mu}^{A}[\theta]$ correlator computation
in ordinary field theory turns into a composite operator correlation
computation beyond the leading order in the coupling constant expansion.
To demonstrate explicitly that we can recover the gauge dynamics of
$A_{\mu}^{A}$ at the quantum level using the non-Abelian BTGT formalism,
we give in this section an example of the requisite composite operator
renormalization. We will find that the transverse divergent structure
of the two-point function is recovered only after including the composite
operator renormalization, indicating the self-consistency of the formalism
and that ordinary gauge invariance is not spoiled by the nonlinear
field redefinition and the BTGT symmetry. We will also show in this
section that there is a sufficient number of counter term coefficients
to preserve finiteness of both $\theta_{a}^{A}$ and $A_{\mu}^{B}$
correlators without spoiling the gauge and BTGT symmetries, lending
further evidence that the $\theta_{a}^{A}$ theory is a consistent
rewriting of the $A_{\mu}^{A}$ theory.

More explicitly, define the two-point momentum space Green's function
by 
\begin{align}
G_{\mu\nu}^{AB}\left(k\right) & =\int d^{4}x\,e^{-ik\cdot x}\left\langle A_{\mu}^{A}\left(x\right)A_{\nu}^{B}\left(0\right)\right\rangle \\
 & =\int d^{4}x\,e^{-ik\cdot x}\left.\frac{\delta}{i\delta J^{\mu A}\left(x\right)}\frac{\delta}{i\delta J^{\nu B}\left(0\right)}\bar{Z}\left[J,K\right]\right|_{J=0,K=0}
\end{align}
where $\bar{Z}\left[J,K\right]$ is the generating function defined
in Eq.\,(\ref{eq:Zbar generating function}). The difference from
the usual generating function Eq.\,(\ref{eq:Z usual formalism}) is
that $A_{\mu}$ is now a composite operator in terms of $\theta_{a}$
fields and the path integral is now over $\theta_{a}$ instead of
$A_{\mu}$. Using dimensional regularization with $d=4-\epsilon$,
we will demonstrate below that the divergent part of the momentum
space Green's function for $A_{\mu}$ is transverse and exactly the
same as the typical formulation before introducing counter terms.
\begin{eqnarray}
\mathrm{div}\left(G_{\mu\nu}^{AB}\left(k\right)\right) & = & \frac{5}{3}C\left(A\right)\frac{g^{2}}{8\pi^{2}\epsilon}\frac{1}{i}\delta^{AB}\left(\frac{g_{\mu\nu}}{k^{2}}-\frac{k_{\mu}k_{\nu}}{k^{4}}\right)\nonumber \\
 &  & +\mathrm{div}\left(\left(\mathrm{D}_{c.t.1}\right)_{\mu\nu}^{AB}+\left(\mathrm{D}_{c.t.2}\right)_{\mu\nu}^{AB}+\left(\mathrm{D_{c.t.3}}\right)_{\mu\nu}^{AB}\right)
\end{eqnarray}
Furthermore, after introducing counter terms, we will find that both
$\left\langle \theta_{a}^{A}\theta_{b}^{B}\right\rangle $ and $\left\langle A_{\mu}^{A}A_{\nu}^{B}\right\rangle $
can be made finite without changing the symmetries of the theory.
The details of the $\left\langle A_{\mu}^{A}A_{\nu}^{B}\right\rangle $
computation are presented below.

This calculation simplifies significantly when using the Feynman gauge.
This is due to the gauge propagator becoming diagonal in the BTGT
indices, which greatly simplifies the sums.

\subsection{Tree level}

The tree level diagram for the two-point $A_\mu$ correlator in the Feynman gauge is
\begin{eqnarray}
\begin{gathered}
\includegraphics[width=4cm]{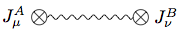}
\end{gathered}
& = &
\sum_{a,b}\left(-i\tilde{k}_{\mu}^{a}\right)\tfrac{1}{i}\Delta_{ab}^{AB}\left(k\right)\left(i\tilde{k}_{\nu}^{b}\right)\\ & = & -i\frac{\delta^{AB}}{k^{2}}\sum_{a}\frac{\tilde{k}_{\mu}^{a}\tilde{k}_{\nu}^{a}}{k_{a}^{2}}\\
 & = & -i\frac{\delta^{AB}}{k^{2}}\sum_{a}(H^{a})_{\mu\nu}\\
 & = & -i\frac{\delta^{AB}}{k^{2}}g_{\mu\nu}
\end{eqnarray}as expected. The structure is essentially identical to Abelian BTGT
at this level of approximation.

\subsection{Source operator terms}

\begin{figure}
\begin{centering}
\includegraphics[width=13cm]{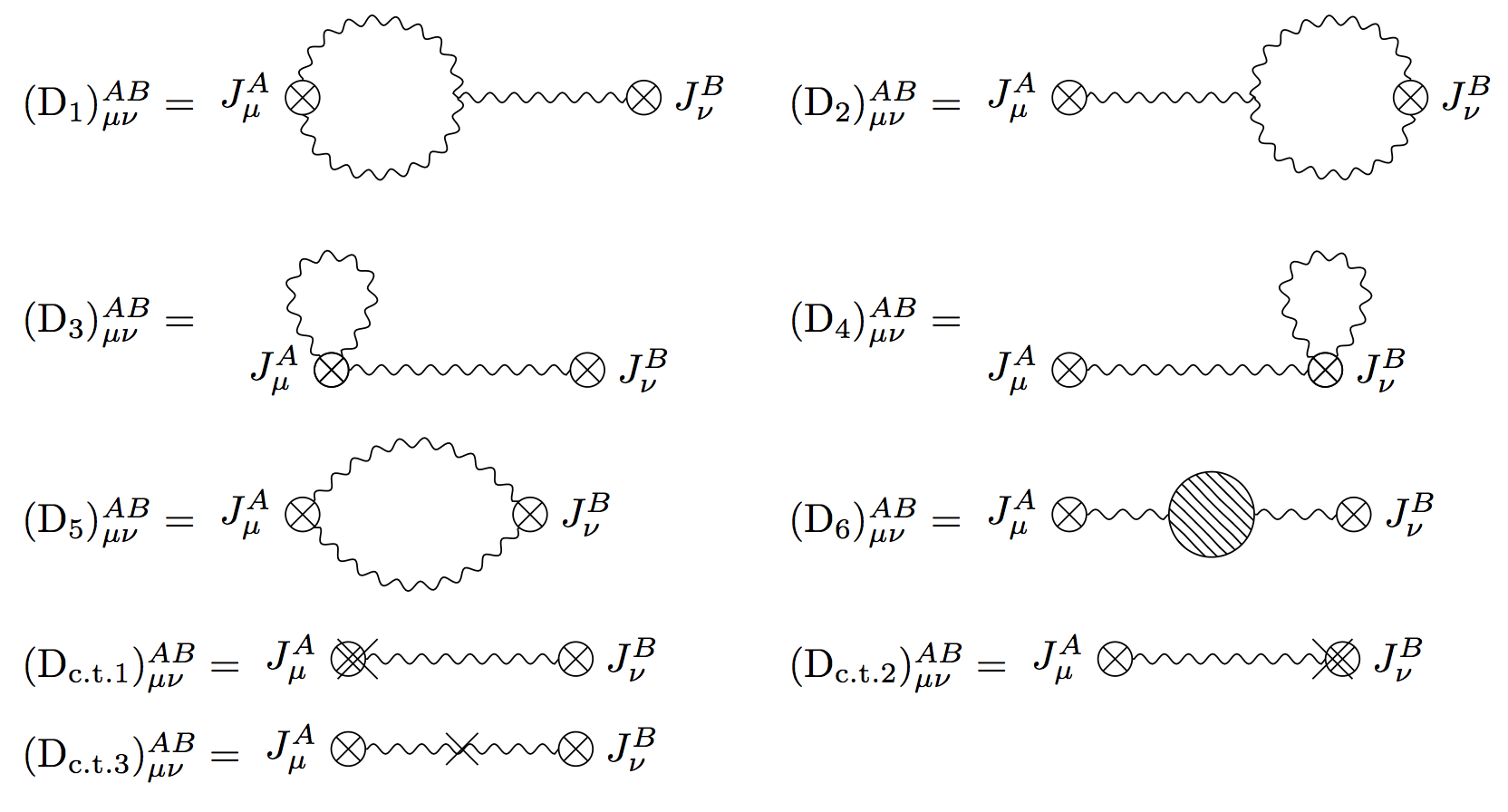}
\par\end{centering}

\caption{Diagram to compute the $A_{\mu}\left[\theta\right]$ two-point correlator.
The blob in $\mathrm{D}_{6}$ refers to all 1PI sub-diagrams and is
proportional to the $\theta_{a}$ self energy. A further breakdown
is shown in Fig.\,\ref{fig:D6 Self energy composite diagrams}.\label{Fig:Composite operator diagrams}}
\end{figure}

Next we consider the one-loop diagrams determining the composite operator
counter-terms. The diagrams involved in evaluating $\left\langle A_{\mu}^{A}A_{\nu}^{B}\right\rangle $
at one loop are shown in Fig.\,\ref{Fig:Composite operator diagrams}.

The first diagram in Fig.\,\ref{Fig:Composite operator diagrams}
is given by
\begin{equation}
\left(\mathrm{D}_{1}\right)_{\mu\nu}^{AB}=\sum_{i=1}^{2}\left(\mathrm{D_{1}^{(i)}}\right)_{\mu\nu}^{AB}
\end{equation}
where $i$ runs through the two possible terms of the $\theta^{3}$
vertex and
\begin{equation}
\left(\mathrm{D}_{1}^{(i)}\right)_{\mu\nu}^{AB}=\frac{g^{2}}{4}C\left(A\right)\delta^{AB}\sum_{a,b}\int\frac{d^{d}\ell}{\left(2\pi\right)^{d}}\frac{\left(2\tilde{\ell}_{\mu}^{a}+\tilde{k}_{\mu}^{a}\right)V_{baa}^{\left(i\right)}\left(k,\ell\right)}{\ell_{a}^{2}\ell^{2}\left(\ell_{a}+k_{a}\right)^{2}\left(\ell+k\right)^{2}}\frac{\tilde{k}_{\nu}^{b}}{k_{b}^{2}k^{2}}.\label{eq:source3loop}
\end{equation}
 The $\theta^{3}$ vertex in Eq.\,(\ref{eq:source3loop}) can be written
as 
\begin{eqnarray}
V_{baa}^{\left(1\right)}\left(k,\ell\right) & = & k_{b}\ell_{a}\left(\ell_{a}+k_{a}\right)\left(\delta_{ab}-1\right)\left(2\ell_{b}+k_{b}\right)\label{eq:V(1)aab}\\
V_{baa}^{\left(2\right)}\left(k,\ell\right) & = & \frac{1}{2}\delta_{ab}\left(k^{2}k_{a}\left(2\ell_{a}+k_{a}\right)-\ell^{2}\ell_{a}\left(\ell_{a}+2k_{a}\right)+\left(\ell+k\right)^{2}\left(\ell_{a}^{2}-k_{a}^{2}\right)\right)\label{eq:V(2)aab}
\end{eqnarray}
where there is no sum over $a$ or $b$. Using Eq.\,(\ref{eq:V(1)aab}),
we find 
\begin{align}
\left(\mbox{D}_{1}^{\left(1\right)}\right)_{\mu\nu}^{AB} & =\frac{g^{2}}{4}C\left(A\right)\delta^{AB}\sum_{a,b}\frac{k_{b}\tilde{k}_{\nu}^{b}}{k_{b}^{2}k^{2}}\left(\delta_{ab}-1\right)\int\frac{d^{d}\ell}{\left(2\pi\right)^{d}}\frac{\left(2\tilde{\ell}_{\mu}^{a}+\tilde{k}_{\mu}^{a}\right)\ell_{a}\left(\ell_{b}+k_{b}\right)\left(\ell_{a}-k_{a}\right)}{\ell_{a}^{2}\ell^{2}\left(\ell_{a}+k_{a}\right)^{2}\left(\ell+k\right)^{2}}\\
 & =\frac{g^{2}}{4}C\left(A\right)\delta^{AB}\sum_{a,b}\frac{k_{b}\tilde{k}_{b}^{\nu}}{k_{b}^{2}k^{2}}\left(\delta_{ab}-1\right)\left(2\delta_{\mu}^{a}\delta_{ab}\frac{i}{8\pi^{2}\epsilon}+\mathrm{finite}\right)\\
 & =0+\mathrm{finite}.
\end{align}
From Eq.\,(\ref{eq:V(2)aab}), we find 
\begin{eqnarray}
\left(\mathrm{D_{1}^{(2)}}\right)_{\mu\nu}^{AB} & = & \frac{g^{2}}{8}C\left(A\right)\delta^{AB}\sum_{a}\frac{\tilde{k}_{a}^{\nu}}{k_{a}^{2}k^{2}}\left(\left(12\tilde{k}_{a}^{\mu}\right)\frac{i}{8\pi^{2}\epsilon}+\mbox{finite}\right)\\
 & = & -\frac{3}{2}C\left(A\right)\frac{g^{2}}{8\pi^{2}\epsilon}\left(\frac{1}{i}\delta^{AB}\frac{g^{\mu\nu}}{k^{2}}\right)+\mbox{finite.}
\end{eqnarray}
Adding up the contributions gives
\begin{equation}
\mathrm{div}\left(\left(\mathrm{D_{1}}\right)_{\mu\nu}^{AB}\right)=-\frac{3}{2}C\left(A\right)\frac{g^{2}}{8\pi^{2}\epsilon}\left(\frac{1}{i}\delta^{AB}\frac{g^{\mu\nu}}{k^{2}}\right).\label{eq:div(D1) AA}
\end{equation}
The symmetry between diagrams 1 and 2 of Fig.\,\ref{Fig:Composite operator diagrams}
is given $\left\{ A,k\right\} \leftrightarrow\left\{ B,-k\right\} $,
and we can therefore conclude without computation 
\begin{equation}
\mathrm{div}\left(\left(\mathrm{D_{2}}\right)_{\mu\nu}^{AB}\right)=-\frac{3}{2}C\left(A\right)\frac{g^{2}}{8\pi^{2}\epsilon}\left(\frac{1}{i}\delta^{AB}\frac{g^{\mu\nu}}{k^{2}}\right).\label{eq:div(D2) AA}
\end{equation}

The third diagram in Fig.\,\ref{Fig:Composite operator diagrams}
is given by 
\begin{eqnarray}
\left(\mbox{D}_{3}\right)_{\mu\nu}^{AB} & = & \frac{1}{2}\sum_{b,b',c,d}\int_{\ell}ig^{2}V_{\mu,b'cd}^{A,B'CD}\left(k;-k,\ell,-\ell\right)\tfrac{1}{i}\Delta_{b'b}^{B'B}\left(k\right)\tfrac{1}{i}\Delta_{cd}^{CD}\left(\ell\right)i\tilde{k}_{\nu}^{b}\\
 & = & \frac{g^{2}}{12}\sum_{b,c}\frac{\tilde{k}_{\nu}^{b}}{k^{2}k_{b}^{2}}\int_{\ell}\delta_{bcc}\frac{0+f_{E}^{AC}f_{E}^{BC}\left(-\tilde{\ell}_{\mu}^{b}+\tilde{k}_{\mu}^{b}\right)+f_{E}^{AC}f_{E}^{BC}\left(\tilde{\ell}_{\mu}^{b}+\tilde{k}_{\mu}^{b}\right)}{\ell^{2}\ell_{c}^{2}}\\
 & = & +\frac{1}{3}C\left(A\right)\frac{g^{2}}{8\pi^{2}\epsilon}\left(\frac{1}{i}\delta^{AB}\frac{g_{\mu\nu}}{k^{2}}\right)+\mathrm{finite}.\label{eq:div(D3) AA}
\end{eqnarray}
Since the fourth diagram in Fig.\,\ref{Fig:Composite operator diagrams}
must be the same as $\mathrm{D}_{3}$ up to $\left\{ A,k\right\} \leftrightarrow\left\{ B,-k\right\} $,
we can immediately write 
\begin{equation}
\mathrm{div}\left(\left(\mathrm{D}_{4}\right)_{\mu\nu}^{AB}\right)=+\frac{1}{3}C\left(A\right)\frac{g^{2}}{8\pi^{2}\epsilon}\left(\frac{1}{i}\delta^{AB}\frac{g_{\mu\nu}}{k^{2}}\right).\label{eq:div(D4) AA}
\end{equation}

Diagram 5 in Fig.\,\ref{Fig:Composite operator diagrams} is
\begin{equation}
\left(\mathrm{D}_{5}\right)_{\mu\nu}^{AB}=\frac{1}{2}\sum_{a,b}\left(\frac{g}{2}\right)^{2}\int\frac{d^{d}\ell}{\left(2\pi\right)^{d}}\frac{f^{ACD}\left(-2\tilde{\ell}_{\mu}^{a}-\tilde{k}_{\mu}^{a}\right)f^{BCD}\left(2\tilde{\ell}_{\nu}^{b}+\tilde{k}_{\nu}^{b}\right)}{\ell^{2}\ell_{a}^{2}\left(\ell+k\right)^{2}\left(\ell_{b}+k_{b}\right)^{2}}.
\end{equation}
This momentum integral does not UV diverge for $d=4$: i.~e.~ 
\begin{equation}
\mathrm{div}\left(\left(\mathrm{D}_{5}\right)_{\mu\nu}^{AB}\right)=0.\label{eq:div(D5) AA}
\end{equation}

\subsection{$\theta$ self energy diagrams}

\begin{figure}
\begin{centering}
\includegraphics[width=15cm]{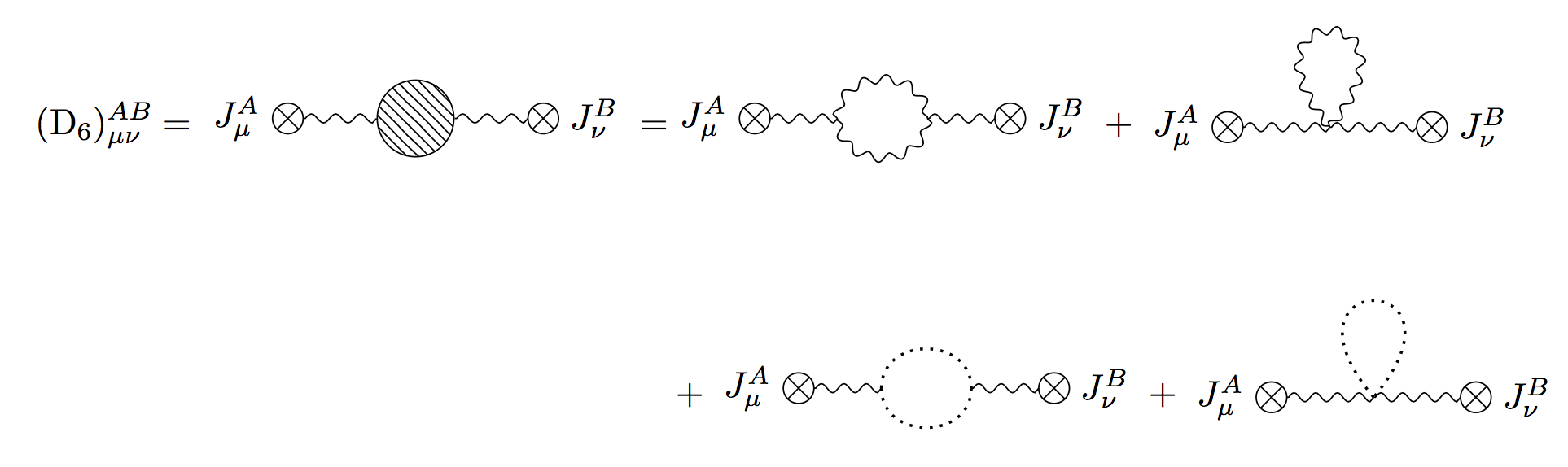}
\par\end{centering}

\centering{}\caption{Breakdown of $\left(\mathrm{D_{6}}\right)_{\mu\nu}^{AB}$ from Fig.\,\ref{Fig:Composite operator diagrams};
they are equivalent to the $\theta_{a}$ self energy diagrams of Fig.\,\ref{fig:Theta-self-energy-diagrams}.
\label{fig:D6 Self energy composite diagrams} }
\end{figure}
Diagram 6 in Fig.\,\ref{Fig:Composite operator diagrams} is the
sum of all 1PI sub-diagrams as shown in Fig.\,\ref{fig:D6 Self energy composite diagrams}.
Using the results of Section \ref{sub: Theta self energy}, we have
\begin{equation}
\mathrm{div}\left(\Pi_{ab}^{AB}\left(k\right)\right)=C\left(A\right)\frac{g^{2}}{8\pi^{2}\epsilon}\delta^{AB}\left(4k^{2}k_{a}^{2}\delta_{ab}-\frac{5}{3}k_{a}^{2}k_{b}^{2}\right)
\end{equation}
where $\Pi_{ab}^{AB}\left(k\right)$ is the $\theta_{a}$ self energy.
The divergent part of diagram 6 is given by
\begin{align}
\mathrm{div}\left(\left(\mathrm{D_{6}}\right)_{\mu\nu}^{AB}\right) & =\sum_{a,b,a',b'}\left(-i\tilde{k}_{\mu}^{a}\right)\tfrac{1}{i}\Delta_{aa'}^{AA'}\left(k\right)\mathrm{div}\left(i\Pi_{a'b'}^{A'B'}\left(k\right)\right)\tfrac{1}{i}\Delta_{b'b}^{B'B}\left(k\right)\left(i\tilde{k}_{\nu}^{b}\right)\\
 & =-\sum_{a,b}\frac{\tilde{k}_{\mu}^{a}\tilde{k}_{\nu}^{b}}{k^{4}k_{a}^{2}k_{b}^{2}}i\mathrm{div}\left(\Pi_{ab}^{AB}\left(k\right)\right)\\
 & =C\left(A\right)\frac{g^{2}}{8\pi^{2}\epsilon}\frac{1}{i}\delta^{AB}\left(4\frac{g_{\mu\nu}}{k^{2}}-\frac{5}{3}\frac{k_{\mu}k_{\nu}}{k^{4}}\right).\label{eq:div(D6) AA}
\end{align}
As expected, the divergences of Fig.\,\ref{fig:D6 Self energy composite diagrams}
are completely canceled out by the renormalization constants $Z_{\theta^{2}}$
and $Z_{\frac{1}{\xi}\theta^{2}}$ that arise from $\mathrm{D_{c.t.3}}$
in Fig.\,\ref{Fig:Composite operator diagrams}.

\subsection{Renormalization}

Adding up the contributions from the six diagrams of Fig.\,\ref{Fig:Composite operator diagrams},
given by Eqs.\,(\ref{eq:div(D1) AA}), (\ref{eq:div(D2) AA}), (\ref{eq:div(D3) AA}),
(\ref{eq:div(D4) AA}), (\ref{eq:div(D5) AA}), and (\ref{eq:div(D6) AA})
gives the divergent part of the two-point $A$ correlator before renormalization:
\begin{align}
\sum_{i=1}^{6}\mathrm{div}\left(\left(\mathrm{D_{i}}\right)_{\mu\nu}^{AB}\right) & =C\left(A\right)\frac{g^{2}}{8\pi^{2}\epsilon}\frac{1}{i}\delta^{AB}\left(\left(-3+\frac{2}{3}+4\right)\frac{g_{\mu\nu}}{k^{2}}-\frac{5}{3}\frac{k_{\mu}k_{\nu}}{k^{4}}\right)\\
 & =\frac{5}{3}C\left(A\right)\frac{g^{2}}{8\pi^{2}\epsilon}\frac{1}{i}\delta^{AB}\left(\frac{g_{\mu\nu}}{k^{2}}-\frac{k_{\mu}k_{\nu}}{k^{4}}\right).
\end{align}
It has the expected transverse property and the same numerical value
as in the usual $A_{\mu}$ formulation. While the $k_{\mu}k_{\nu}$
term receives a contribution from only diagram $\mathrm{D_{6}}$,
the $g_{\mu\nu}$ term receives contributions from six diagrams $\mathrm{D}_{1}$
through $\mathrm{D}_{6}$.

Now we need to renormalize both $\theta_{a}$ and the composite operator
$A_{\mu}[\theta]$ and show that both correlators are finite without
introducing any counter terms that spoil gauge invariance, BTGT invariance,
or Lorentz invariance. The composite operator counter terms in the
Lagrangian are of the form 
\begin{equation}
\mathcal{L}_{c.t.}\ni\left(Z_{J\theta}-1\right)J^{A\mu}\tilde{\partial}_{\mu}^{a}\theta_{a}^{A}+\left(Z_{gJ\theta^{2}}-1\right)\frac{g}{2}f^{ABC}J^{A\mu}\tilde{\partial}_{\mu}^{a}\theta_{a}^{B}\theta^{C}+\dots\label{eq:ZJsdefined}
\end{equation}
and to preserve BTGT invariance the counter terms have to obey certain
relations given by 
\begin{equation}
Z_{J}=\frac{Z_{J\theta}}{Z_{\theta^{2}}^{1/2}}=\frac{Z_{Jg\theta^{2}}}{Z_{g}Z_{\theta^{2}}}=\frac{Z_{Jg^{2}\theta^{3}}}{Z_{g}^{2}Z_{\theta^{2}}^{3/2}}=\dots\label{eq:ratios}
\end{equation}
where we have defined $Z_{J}$ to be the ratio of the bare source
$J_{0}$ to the renormalized source $J$: $J_{0}\equiv Z_{J}J$ .

The $Z_{J\theta}$ counter-term occurs in diagrams $\mathrm{D}_{c.t.1}$
and $\mathrm{D}_{c.t.2}$ of Fig.\,\ref{Fig:Composite operator diagrams},
which evaluate to
\begin{align}
\left(\mathrm{D}_{c.t.1}\right)_{\mu\nu}^{AB} & =\sum_{a,b}\left(-i\left(Z_{J\theta}-1\right)\tilde{k}_{\mu}^{a}\right)\tfrac{1}{i}\Delta_{ab}^{AB}\left(k\right)\left(i\tilde{k}_{\nu}^{b}\right)\\
 & =\left(Z_{J\theta}-1\right)\frac{1}{i}\frac{\delta^{AB}}{k^{2}}\sum_{a}\frac{\tilde{k}_{\mu}^{a}\tilde{k}_{\nu}^{a}}{k_{a}^{2}}\\
 & =\left(Z_{J\theta}-1\right)\left(-i\frac{\delta^{AB}}{k^{2}}g_{\mu\nu}\right)\,,
\end{align}
and
\begin{align}
\left(\mathrm{D}_{c.t.2}\right)_{\mu\nu}^{AB} & =\sum_{a,b}\left(-i\tilde{k}_{\mu}^{a}\right)\tfrac{1}{i}\Delta_{ab}^{AB}\left(k\right)\left(i\left(Z_{J\theta}-1\right)\tilde{k}_{\nu}^{b}\right)\\
 & =\left(Z_{J\theta}-1\right)\left(-i\frac{\delta^{AB}}{k^{2}}g_{\mu\nu}\right)\,.
\end{align}
Using the results of Section \ref{sub: Theta self energy}, we find
\begin{eqnarray}
\left(\mathrm{D_{c.t.3}}\right)_{\mu\nu}^{AB} & = & \sum_{a,b,a',b'}-i\tilde{k}_{\mu}^{a}\tfrac{1}{i}\Delta_{aa'}^{AA'}\left(k\right)\tfrac{1}{i}\delta^{AB}\left(\left(Z_{\theta^{2}}-1\right)k^{2}k_{a}^{2}\delta_{ab}+\left(Z_{\frac{1}{\xi}\theta^{2}}-Z_{\theta^{2}}\right)k_{a}^{2}k_{b}^{2}\right)\\
 &  & \times\tfrac{1}{i}\Delta_{b'b}^{B'B}\left(k\right)i\tilde{k}_{\nu}^{b}\nonumber \\
 & = & \sum_{a,b}\frac{\tilde{k}_{\mu}^{a}\tilde{k}_{\nu}^{b}}{k^{4}k_{a}^{2}k_{b}^{2}}i\delta^{AB}\left(4C\left(A\right)\frac{g^{2}}{8\pi^{2}\epsilon}k^{2}k_{a}^{2}\delta_{ab}-\frac{5}{3}C\left(A\right)\frac{g^{2}}{8\pi^{2}\epsilon}k_{a}^{2}k_{b}^{2}\right)+O\left(g^{4}\right)\\
 & = & C\left(A\right)\frac{g^{2}}{8\pi^{2}\epsilon}\frac{\delta^{AB}}{i}\left(-4\frac{g_{\mu\nu}}{k^{2}}+\frac{5}{3}\frac{k_{\mu}k_{\nu}}{k^{4}}\right)+O\left(g^{4}\right)\,.\label{eq:Dc.t.3. AA}
\end{eqnarray}

The divergence of all these diagrams cancel to make the two-point
$A_{\mu}$ correlator finite:
\begin{align}
\mathrm{div}\left(G_{\mu\nu}^{AB}\left(k\right)\right) & =\mathrm{div}\left(\sum_{i=1}^{6}\left(\mathrm{D}_{i}\right)_{\mu\nu}^{AB}+\left(\mathrm{D}_{c.t.1}\right)_{\mu\nu}^{AB}+\left(\mathrm{D}_{c.t.2}\right)_{\mu\nu}^{AB}+\left(\mathrm{D_{c.t.3}}\right)_{\mu\nu}^{AB}\right)\\
 & =\left(\frac{7}{3}C\left(A\right)\frac{g^{2}}{8\pi^{2}\epsilon}+2\mbox{div}\left(Z_{J\theta}\right)\right)\left(-i\frac{\delta^{AB}}{k^{2}}g_{\mu\nu}\right)\\
 & =0\,.
\end{align}
The renormalization constant $Z_{J\theta}$ is therefore
\begin{equation}
Z_{J\theta}=1+\frac{7}{6}C\left(A\right)\frac{g^{2}}{8\pi^{2}\epsilon}+O\left(g^{4}\right)\,.\label{eq:Z_Jtheta}
\end{equation}
Using this, Eq.\,(\ref{eq:Ztheta^2}), and Eq.\,(\ref{eq:ratios}),
we see that
\begin{equation}
Z_{J}^{-1}=\frac{Z_{\theta^{2}}^{1/2}}{Z_{J\theta}}=1+\frac{5}{6}C\left(A\right)\frac{g^{2}}{8\pi^{2}\epsilon}+O\left(g^{4}\right)=Z_{A^{2}}^{1/2}\,.
\end{equation}
The self-consistency of the renormalization, $Z_{\theta^{2}}^{1/2}/Z_{J\theta}$
equals $Z_{A^{2}}^{1/2}$, is as expected from the external source
coupling in the usual $A_{\mu}^{A}$ theory being of the form
\begin{equation}
\mathcal{L}_{J}\ni J^{\mu A}A_{\mu}^{A}
\end{equation}
 where $A_{\mu}^{A}$ are renormalized fields, while in the BTGT formulation
the source coupling is defined with a composite operator renormalization
constant $Z_{J\theta}$ as seen in Eq.\,(\ref{eq:ZJsdefined}).

\section{Counter term predictions and Slavnov-Taylor identities}

The Slavnov-Taylor identities have yet to be formally derived or shown
to exist for the BTGT formalism. This is an interesting area for future
study. The one loop calculations done thus far show that $g$ and
$\xi$ scale as expected, and $A$ scales as expected when written
as a composite operator of $\theta$. Assuming that the symmetries
in BTGT are preserved in a way similar to the explicitly computed
processes in this paper, we state in this section a set of concrete
generalizations for the one loop counter term factors for the $\theta^{n}$-vertex.

We expect the BTGT formulation of the Slavnov-Taylor identities to
show that the following holds 
\begin{align}
Z_{g^{n-2}\theta^{n}} & =Z_{g}^{n-2}Z_{\theta^{2}}^{n/2}\qquad\left(n\geq2\right)\label{eq:Ztheta^n SlavTaylor}\\
Z_{\xi^{-1}g^{n-2}\theta^{n}} & =Z_{\xi}^{-1}Z_{g}^{n-2}Z_{\theta^{2}}^{n/2}\qquad\left(n\geq2\right)\label{eq:xi Ztheta^n Slav Taylor}\\
Z_{g^{n}\theta^{n}\bar{c}c} & =Z_{\bar{c}c}Z_{g}^{n}Z_{\theta^{2}}^{n/2}\qquad\left(n\geq0\right).\label{eq:theta^n ghost Slav Taylor}
\end{align}
Based on calculated value in Eq.\,(\ref{eq:Ztheta^2 result}), the
predictions are
\begin{align}
Z_{g^{n-2}\theta^{n}} & =1+\frac{22+n}{6}C\left(A\right)\frac{g^{2}}{8\pi^{2}\epsilon}+O\left(g^{4}\right)\qquad\left(n\geq2\right)\label{eq:theta^n prediction}\\
Z_{\xi^{-1}g^{n-2}\theta^{n}} & =1+\frac{12+n}{6}C\left(A\right)\frac{g^{2}}{8\pi^{2}\epsilon}+O\left(g^{4}\right)\qquad\left(n\geq2\right)\label{eq:xi theta^n prediction}\\
Z_{g^{n}\theta^{n}\bar{c}c} & =1+\frac{3+n}{6}C\left(A\right)\frac{g^{2}}{8\pi^{2}\epsilon}+O\left(g^{4}\right)\qquad\left(n\geq0\right).\label{eq:ghost theta^n prediction}
\end{align}
We have explicitly computed the $n=2$ case of Eq.\,(\ref{eq:theta^n prediction})
and Eq.\,(\ref{eq:xi theta^n prediction}) and also the $n=0$ and
$n=1$ cases of Eq.\,(\ref{eq:ghost theta^n prediction}). An interesting
and nontrivial check of BTGT in the future is the $n=3$ case of Eq.\,(\ref{eq:theta^n prediction})
and Eq.\,(\ref{eq:xi theta^n prediction}), which is given by the
triple gauge $\theta^{3}$ vertex diagrams. Also of interest is the
$n=2$ case of Eq.\,(\ref{eq:ghost theta^n prediction}), which corresponds
to the $\theta^{2}\bar{c}c$ vertex.

The factors $Z_{g}$, $Z_{\xi}$ and $Z_{\bar{c}c}$ are unchanged
by the choice of using either the BTGT field $\theta_{a}$ or the
vector potential $A_{\mu}$ to describe the gauge boson sector. We
could have started by assuming that the following relations would
hold: 
\begin{align}
Z_{g}^{(\theta)}=Z_{g}^{(A)} & =1-\frac{11}{6}C\left(A\right)\frac{g^{2}}{8\pi^{2}\epsilon}+O\left(g^{4}\right)\\
Z_{\xi}^{(\theta)}=Z_{\xi}^{(A)} & =1+\frac{5}{3}C\left(A\right)\frac{g^{2}}{8\pi^{2}\epsilon}+O\left(g^{4}\right)\\
Z_{\bar{c}c}^{(\theta)}=Z_{\bar{c}c}^{(A)} & =1+\frac{1}{2}C\left(A\right)\frac{g^{2}}{8\pi^{2}\epsilon}+O\left(g^{4}\right),
\end{align}
where $Z^{(\theta)}$ is calculated in $\theta_{a}$ formalism and
$Z^{(A)}$ in the $A_{\mu}$ formalism. Therefore, $Z_{\theta^{2}}$
is the only a priori undetermined parameter in Eqs.\,(\ref{eq:Ztheta^n SlavTaylor}),
(\ref{eq:xi Ztheta^n Slav Taylor}), and (\ref{eq:ghost theta^n prediction})
. Since we have done four computations and there was only one a priori
undetermined parameter, we have done three independent nontrivial
checks of the gauge invariance of this theory at one loop level. This
result gives us confidence that gauge invariance in the BTGT formalism
is preserved in perturbation theory.

\section{\label{sec:Conclusions}Conclusions}

We have constructed a non-Abelian basis tensor gauge theory (BTGT)
which gives an alternate formulation of usual non-Abelian gauge theory
in terms of the vierbein analog for ordinary gauge bundles. For example,
the basis tensor that couples to matter transforming as $N$ of $SU(N)$
has the representation $\bar{N}$ and has the Lorentz transformation
properties of a rank 2 projection tensor. To match the usual gauge
theory formalism, the basis tensor must satisfy Eq.\,(\ref{eq:agrelation})
and the couplings must be symmetric under a non-gauge symmetry called
BTGT symmetry that is identical to the BTGT transformation of the
Abelian case. To have a simple match in the number of degrees of freedom
between the ordinary gauge theory formalism and the BTGT formalism,
we have decided to choose the scalar fields $\theta_{a}^{A}$ that
parameterize the basis tensor to be in the target space of the gauge
manifold just as in Abelian BTGT. As in the Abelian BTGT case, the
map between $\theta_{c}^{F}$ is a nonlocal functional of $A_{\mu}^{B}$.
More explicitly, $\theta_{c}$ is a type of path-ordered line integral
of $A_{\mu}$, and hence is related to Wilson lines. However, unlike
in the Abelian case, the map between $A_{\mu}^{B}$ and $\theta_{c}^{F}$
is nonlinear, where the nonlinearities form a power series of the
structure constants. This means that any $A_{\mu}^{B}$ correlator
computation is a composite operator correlator with respect to the
$\theta_{c}^{F}$ elementary field theory requiring composite operator
counter terms.

The Feynman rules for the 1-loop order and $O(g^{2})$ computations
were explicitly presented. We have tested non-Abelian BTGT to one-loop
and $O(g^{2})$ (where $g$ is the usual gauge coupling), using $\theta_{c}^{F}$
are the elementary field degrees of freedom, by computing the beta
function of the gauge coupling and finding it to be identical to the
usual formulation. We have also computed the gauge field 2-point function
to the same one-loop accuracy and found identical results as in the
usual gauge theory formulation. In particular, we found that the UV
divergent part of the correlator is transverse just as in the usual
gauge theory formulation. Furthermore, the composite operator counter
terms are sufficient to make both the $A_{\mu}^{B}$ correlator and
$\theta_{c}^{F}$ correlators finite. 

Through these explicit computations, we have also given several nontrivial
checks that the renormalization constants in the minimal subtraction
scheme are identical to those of the usual gauge theory formalism.
Although we defer a formal BRST construction for this theory to a
future work, the nontrivial checks indicate that there will be no
insurmountable obstacles to its formulation.

Although the nonlinearities in the map between $A_{\mu}^{B}$ and
$\theta_{a}^{C}$ might make this choice of formalism seem unnecessarily
complicated, it is a natural choice from several considerations. First,
it leads to a natural match in the number of functional degrees of
freedom of a gauge theory. Second, it is a continuous deformation
(as a function of group structure constants) of a simple linear map
in the case of Abelian theories. Third, its semblance with nonlinear
sigma-model parameterizations may allow several extensions of this
work using the techniques that have been developed for sigma models.
Fourth, the BTGT symmetry which stabilizes the Hamiltonian and the
gauge symmetry have elegant representations given by Eqs.\,(\ref{eq:gaugetransform})
and (\ref{eq:BTGTtransform}). Note also that from the perspective
of having a nontrivial transformation that may lead to new insights
into the usual gauge theory formulations, such nonlinear maps are
more promising. On the other hand, it is important to keep in mind,
just as in the usual sigma model parameterizations, this choice of
using $\theta_{a}^{C}$ is far from unique even though there is uniqueness
of the map between the vierbein-like field $\left[G_{(f)}(x)\right]_{\phantom{\gamma}\delta}^{\gamma}$
(which $\theta_{a}^{C}$ parameterizes) and the gauge field $A_{\mu}$
if we stipulate that the gauged matter kinetic term be locally gauge
equivalent to that without a gauge field.

Many extensions of this work on BTGT theory beyond explicit constructions
of BRST formalism are self-evident. To complete the tests of this
formalism's equivalence with the usual Standard Model formulation,
BTGT should also be tested in the contexts of spontaneous symmetry
breaking and curved spacetime. Since this is a formalism most naturally
suited for exploring Wilson lines, it would be interesting to reformulate
the Eikonal phase re-summing soft gluonic effects~\cite{Alday:2008yw,Korchemsky:1992xv,Korchemsky:1991zp,Frenkel:1984pz,Catani:1983bz}
in this formalism and investigate whether any new insights or simplifications
can arise. The enhanced local nature of BTGT for dealing with nonlocal
quantities such as Wilson lines also suggests exploring its applications
in lattice gauge theory~\cite{Kronfeld:2012uk,Brambilla:2014jmp}.
The gauge field representation $iU_{a}\tilde{\partial}_{\mu}^{a}U_{a}^{\dagger}$
also is reminiscent of the sigma model representation used in~\cite{Witten:1983tw}
to explore topological aspects of the theories with spontaneously
broken global symmetries. This suggests there may be a way to more
conveniently explore the topological aspects of gauge theories using
the BTGT formalism. The precise connection between the generalized
global symmetries of~\cite{Gaiotto:2014kfa} and the symmetries of
BTGT remains to be clarified. For physics beyond the standard model,
it would be interesting to see if the gauge fields can be interpreted
as Nambu-Goldstone bosons of a spontaneously broken theory since $A_{\mu}=iU_{a}\tilde{\partial}_{\mu}^{a}U_{a}^{\dagger}$
are suggestive of a sigma model.
\begin{acknowledgments}
DJHC was supported in part by the DOE through grant DE-SC0017647.
DJHC would like to thank Lisa Everett for comments on the manuscript.
All of our Feynman diagrams in this paper were made using the help
of TikZ-Feynman~\cite{Ellis:2016jkw}. 
\end{acknowledgments}

\appendix

\section{\label{sec:Notation} Relevant Notation}

This section lists the various notations and conventions used throughout
this paper. The metric signature chosen was 
\begin{equation}
g_{\mu\nu}=\mathrm{diag}(-,+,+,+)\,\,.
\end{equation}
If $\psi_{(a)}^{\mu}$ for $a\in\{0,1,2,3\}$ are 4 orthonormal Lorentz
4-vectors, we can write an explicit representation of the projection
tensors $(H^{a})_{\phantom{\mu}\nu}^{\mu}$ as 
\begin{equation}
(H^{a})_{\phantom{\mu}\nu}^{\mu}=\psi_{(a)}^{\mu}\psi_{(a)\nu}g^{aa}\,.
\end{equation}
The $H^{a}$ matrices are commutative.

Using these projection tensors $(H^{a})_{\phantom{\mu}\nu}^{\mu}$,
we define the following notation related to them. We define the tilde
notation as 
\begin{equation}
\tilde{A}_{a}^{\mu}\equiv(H^{a})_{\phantom{\mu}\nu}^{\mu}A^{\nu}
\end{equation}
to denote the contraction between $H^{a}$ and any 4-vector $A^{\mu}$.
Note that $\tilde{A}^{a\mu}=\tilde{A}_{a}^{\mu}$ because there is
no covariant/contravariant distinction for the BTGT index unlike a
Lorentz index $\mu$. Also, we define the star product as
\begin{equation}
A\star_{a}B\equiv\left(H^{a}\right)_{\mu\nu}A^{\mu}B^{\nu}
\end{equation}
for any two 4-vectors $A^{\mu}$ and $B^{\mu}$. Using the tilde notation
defined above, we have the following identities 
\begin{equation}
A\star_{a}B=\tilde{A}_{a}^{\mu}B_{\mu}=A_{\mu}\tilde{B}_{a}^{\mu}=g_{\mu\nu}\tilde{A}_{a}^{\mu}\tilde{B}_{a}^{\nu}
\end{equation}
We define the product of two Kronecker deltas as
\begin{equation}
\delta_{abc}\equiv\delta_{ab}\delta_{bc}=\delta_{ac}\delta_{bc}=\delta_{ab}\delta_{ac}\qquad\mbox{no sum over \ensuremath{a,b,c}}\,.
\end{equation}
 moving to Euclideanized space via Wick rotation , we can unambiguously
define for any four vector $p^{\mu}$
\begin{equation}
p_{a}\equiv\psi_{(a)}^{\mu}p_{\mu}
\end{equation}
that satisfies
\begin{equation}
p\star_{a}p=p_{a}^{2}\quad\mbox{and}\quad\sum_{a=0}^{3}p_{a}^{2}=p^{2}\,.
\end{equation}

The group structure constant $f^{ABC}$ is defined by the Lie bracket
\begin{equation}
\left[T^{A},T^{B}\right]=if^{ABC}T^{C}
\end{equation}
where $T^{A}$ are basis elements of the Lie algebra such that $e^{iT^{A}\Gamma^{A}}$
are group elements for some function $\Gamma^{A}\left(x\right)$.
We take the basis of generators such that $f^{ABC}$ is completely
anti-symmetric. Given this anti-symmetry, we can define without ambiguity
the following
\begin{equation}
f_{C}^{AB}=f_{AB}^{C}=f^{ABC}\,\,.
\end{equation}
Note that $f_{C}^{AB}=f_{B}^{CA}=f_{A}^{BC}$.

Note that Ref.\,\cite{Chung:2016lhv} uses the notation of having
the basis tensor index $c$ of $\theta^{c}$ (with $c\in\{1,2,3,4\}$)
instead of $\theta_{c}^{B}$ (with $c\in\{0,1,2,3\}$) as in Eq.\,(\ref{eq:nonabelianvierbein}).
Also, the sign convention for $\theta$ has been flipped between Eq.\,(23)
of Ref.\,\cite{Chung:2016lhv} and Eq.\,(\ref{eq:nonabelianvierbein}).

In the Feynman diagrams, all momenta that flow into a vertex are assigned
a positive value.

\section{\label{sec:Construction-of-Nonabelian}The relationship between non-Abelian
basis tensor and ordinary gauge fields $A_{\mu}$ }

Here we follow the equivalence-principle-like procedure of~\cite{Chung:2016lhv}
to construct the relationship of non-Abelian basis tensor and the
ordinary non-Abelian gauge field $A_{\mu}(x)$.

Start with a gauge frame such that the Lagrangian at spacetime point
$x_{1}$ looks like there is no gauge field (i.e.~trivial Chern-Simons
number vacuum): 
\begin{equation}
\mathcal{L}_{\phi}(x_{1})=\partial_{\mu}\tilde{\phi}^{a}\partial^{\mu}\tilde{\phi}^{*a}(x_{1}).
\end{equation}
We demand in this special gauge frame that the vierbein-like tensor
field has the following value at point $x_{1}$: 
\begin{equation}
\tilde{G}_{\alpha\beta}(x_{1})=S_{\alpha\beta}(x_{1}).\label{eq:specialframe-1}
\end{equation}
Upon making a gauge transformation to move to the general frame, we
have 
\begin{equation}
\phi(x)=e^{i\theta^{C}(x)T^{C}}\tilde{\phi}(x),\label{eq:phaseshift-1}
\end{equation}
The gauge field in the new frame is 
\begin{equation}
\tilde{D}_{\mu}\tilde{\phi}=\tilde{g}^{-1}D_{\mu}\tilde{g}\tilde{g}^{-1}\phi
\end{equation}
where 
\begin{equation}
\tilde{g}=e^{i\theta^{C}(x)T^{C}}.
\end{equation}
Hence, we find 
\begin{equation}
\partial_{\mu}\tilde{\phi}=\tilde{g}^{-1}(\partial_{\mu}-iA_{\mu})\phi
\end{equation}
where the right hand side can be also be written in terms of $\tilde{\phi}$
as 
\begin{equation}
\partial_{\mu}\tilde{\phi}=\left[\partial_{\mu}+\tilde{g}^{-1}\partial_{\mu}\tilde{g}-i\tilde{g}^{-1}A_{\mu}\tilde{g}\right]\tilde{\phi}.
\end{equation}
This implies 
\begin{equation}
0=\left[\tilde{g}^{-1}\partial_{\mu}\tilde{g}-i\tilde{g}^{-1}A_{\mu}\tilde{g}\right]\tilde{\phi}
\end{equation}
or equivalently 
\begin{equation}
A_{\mu}(x_{1})=-i\left[\partial_{\mu}\tilde{g}(x_{1})\right]\tilde{g}^{-1}(x_{1})\label{eq:puregagueatonepoint-1}
\end{equation}
which is pure gauge only at a single point $x_{1}$ and not for all
spacetime (just as in the Abelian construction).

We can use Eq.\,(\ref{eq:puregagueatonepoint-1}) to find the map
between $G_{\alpha\beta}$ and $A_{\mu}$. Since $G_{\alpha\beta}$
is \textbf{defined} to obey the transformation rule of Eq.\,(\ref{eq:Gtransform}):
\begin{equation}
G_{(f)\,\beta}^{\alpha}(x_{1})\phi(x_{1})=\tilde{G}_{(f)\,\beta}^{\alpha}(x_{1})\tilde{\phi}(x_{1})
\end{equation}
where 
\begin{equation}
\phi(x_{1})=\tilde{g}(x_{1})\tilde{\phi}(x_{1}).
\end{equation}
This means 
\begin{equation}
G_{(f)\,\beta}^{\alpha}(x_{1})=\tilde{G}_{(f)\,\beta}^{\alpha}(x_{1})\tilde{g}^{-1}(x_{1}).\label{eq:rank2-1}
\end{equation}
Similarly as in~\cite{Chung:2016lhv}, choose $\partial_{\alpha}\tilde{G}_{(f)\,\beta}^{\alpha}(x_{1})=0$.
To solve for the right hand side of Eq.\,(\ref{eq:puregagueatonepoint-1}),
we take the derivative 
\begin{equation}
\left[G_{(f)\beta\mu}\right]^{m}\left[\partial_{\alpha}\tilde{g}\right]^{ml}+\left[\partial_{\alpha}G_{(f)\beta\mu}\right]^{m}\left[\tilde{g}\right]^{ml}=0\label{eq:derivative-1}
\end{equation}
Let 
\begin{equation}
\delta^{ks}=\sum_{f}^{{\rm dim}R}\xi_{(f)}^{k}\xi_{(f)}^{*s}
\end{equation}
where the $\xi_{(f)}$ are constant vectors in the group representation
space. This allows us to rewrite Eq.\,(\ref{eq:derivative-1}) as
\begin{equation}
\xi_{(f)}^{*s}\left[G_{\beta\mu}\right]^{sm}\left[\partial_{\alpha}\tilde{g}\right]^{ml}+\xi_{(f)}^{*s}\left[\partial_{\alpha}G_{\beta\mu}\right]^{sm}\left[\tilde{g}\right]^{ml}=0
\end{equation}
where 
\begin{equation}
\xi_{(f)}^{*s}\left[G_{\beta\mu}\right]^{sm}\equiv\left[G_{(f)\beta\mu}\right]^{m}.\label{eq:relatebasiswithmatrix}
\end{equation}
Multiplying both sides by $\xi_{(f)}^{q}$ and summing, we find 
\begin{equation}
\sum_{f}\xi_{(f)}^{q}\xi_{(f)}^{*s}\left[G_{\beta\mu}\right]^{sm}\left[\partial_{\alpha}\tilde{g}\right]^{ml}=-\sum_{f}\xi_{(f)}^{q}\xi_{(f)}^{*s}\left[\partial_{\alpha}G_{\beta\mu}\right]^{sm}\left[\tilde{g}\right]^{ml}
\end{equation}
to arrive at 
\begin{equation}
\left[G_{\beta\mu}\right]^{qm}\left[\partial_{\alpha}\tilde{g}\right]^{ml}=-\left[\partial_{\alpha}G_{\beta\mu}\right]^{qm}\left[\tilde{g}\right]^{ml}\label{eq:justsbeforeproperty}
\end{equation}
\textbf{Require} that the inverse of $\left[G_{\beta\mu}\right]^{qm}$
exists such that 
\begin{equation}
\left[G^{-1\lambda\beta}\right]^{bq}\left[G_{\beta\mu}\right]^{qm}=\delta_{\,\,\,\,\,\,\mu}^{\lambda}\delta^{bm}\label{eq:property}
\end{equation}
Eq.\,(\ref{eq:justsbeforeproperty}) then becomes 
\begin{equation}
\delta_{\,\,\,\,\,\,\mu}^{\lambda}\left[\partial_{\alpha}\tilde{g}\right]^{bl}\left[\tilde{g}^{-1}\right]^{ls}=-\left[G^{-1\lambda\beta}\right]^{bq}\left[\partial_{\alpha}G_{\beta\mu}\right]^{qs}
\end{equation}
After setting $\lambda=\alpha$, we sum over $\alpha$ to obtain 
\begin{equation}
A_{\mu}=i\left[G^{-1\alpha\beta}\right]\left[\partial_{\alpha}G_{\beta\mu}\right]\label{eq:gaugefieldintermsofG}
\end{equation}
where Eq.\,(\ref{eq:relatebasiswithmatrix}) gives the explicit relationship
to the basis tensor as 
\begin{equation}
\left[G_{\beta\mu}\right]^{qm}=\sum_{f}^{{\rm dim}R}\xi_{(f)}^{q}\left[G_{(f)\beta\mu}\right]^{m}.
\end{equation}
Eq.\,(\ref{eq:gaugefieldintermsofG}) can also be expressed in terms
of derivative of the basis tensor $G_{(f)\beta\mu}$ as 
\begin{equation}
A_{\mu}=i\sum_{f}^{{\rm dim}R}\left[G^{-1\alpha\beta}\right]^{bq}\xi_{(f)}^{q}\left[\partial_{\alpha}G_{(f)\beta\mu}\right]^{s}
\end{equation}
where one notes $\left[G^{-1\alpha\beta}\right]^{bq}\xi_{(f)}^{q}$
is an object that satisfies the identity 
\begin{equation}
\sum_{f}^{{\rm dim}R}\left[G^{-1\alpha\beta}\right]^{bq}\xi_{(f)}^{q}\left[G_{(f)\beta\mu}\right]^{s}=\delta^{bs}\delta_{\,\,\,\,\,\,\mu}^{\alpha}.
\end{equation}

One can check that the non-Abelian basis tensor of Eq.\,(\ref{eq:nonabelianvierbein})
satisfies Eq.\,(\ref{eq:property}). Using the identity 
\begin{equation}
\frac{d}{dx}\exp\left[O(x)\right]=\int_{0}^{1}dy\exp\left[(1-y)O(x)\right]\frac{dO(x)}{dx}\exp\left[yO(x)\right]
\end{equation}
for a matrix $O$, we can evaluate Eq.\,(\ref{eq:gaugefieldintermsofG})
as 
\begin{equation}
A_{\mu}^{Q}(x)=\sum_{c}\left(\left(\left[\theta_{c}^{J}f^{J}\right]^{-1}\right)^{QR}\left(e^{\theta_{c}^{K}f^{K}}-1\right)^{RB}\tilde{\partial}_{\mu}^{c}\theta_{c}^{B}\right)\label{eq:nonabelianamu-1}
\end{equation}
where $f^{J}$ is a structure constant matrix having the components
$(f^{J})^{AB}=f^{JAB}$.

\section{\label{sec:Gauge-and-BTGT}Gauge and BTGT transforms}

In this appendix, we derive an explicit expression for the finite
and linearized gauge and BTGT transforms of the $\theta_{a}^{A}$
field. The key simplification occurs from the fact the $\theta_{a}^{A}$
parameterizes the group manifold. As a result $U_{a}\equiv e^{i\theta_{a}}$
has a relatively simple transformation law governed by a first order
differential equation. The result is
\begin{equation}
U_{a}\rightarrow e^{i\Gamma}U_{a}e^{iZ_{a}}\label{eq:Ua gauge BTGT transform}
\end{equation}
The BTGT symmetry can then be seen as a result of the constant of
integration. The BTGT symmetry in Eq.\,(\ref{eq:Ua gauge BTGT transform})
can also be viewed as the symmetry inherent in the covariant derivative
as defined by 
\begin{equation}
D_{\mu}\left(\cdot\right)=\sum_{a}U_{a}\tilde{\partial}_{\mu}^{a}(U_{a}^{\dagger}\,\cdot)\,\,\,.
\end{equation}

Let's start with the vector potential given by 
\begin{equation}
\left(A_{\mu}\right)_{ij}=i\sum_{a}\left(e^{i\theta_{a}}\tilde{\partial}_{\mu}^{a}e^{-i\theta_{a}}\right)_{ij}=i\sum_{a}\left(U_{a}\tilde{\partial}_{\mu}^{a}U_{a}^{\dagger}\right)_{ij}
\end{equation}
where $\theta_{a}\equiv\theta_{a}^{A}T^{A}$ and $U_{a}\equiv e^{i\theta_{a}}$.
From now on the group indexes $i,j$ will be dropped and implied by
matrix multiplication. In this appendix, repeated lower-case Latin
indices will not be implicitly summed. Under an infinitesimal gauge
transformation parameterized by $\Gamma^{A}$, we have 
\begin{equation}
\delta\tilde{A}_{\mu}^{a}=(H^{a})_{\phantom{\nu}\mu}^{\nu}\delta A_{\nu}=\left[\tilde{D}_{\mu}^{a},\Gamma\right]=\tilde{\partial}_{\mu}^{a}\Gamma-i\left[\tilde{A}_{\mu}^{a},\Gamma\right]
\end{equation}
where $\Gamma\equiv T^{A}\Gamma^{A}$ and $D_{\mu}=\partial_{\mu}-iA_{\mu}$.
In terms of $U_{a}$ this is
\begin{eqnarray}
\delta\left(iU_{a}\tilde{\partial}_{\mu}^{a}U_{a}^{\dagger}\right) & = & \tilde{\partial}_{\mu}^{a}\Gamma+\left[U_{a}\tilde{\partial}_{\mu}^{a}U_{a}^{\dagger},\Gamma\right]\\
 & = & U_{a}\tilde{\partial}_{\mu}^{a}\left(U_{a}^{\dagger}\Gamma U_{a}\right)U_{a}^{\dagger}\,\,\,.\label{eq:deltaUeqRHS}
\end{eqnarray}
 To first order in variations, unitarity implies $\delta U_{a}^{\dagger}=-U_{a}^{\dagger}\delta U_{a}U_{a}^{\dagger}$
(which is equivalent to keeping all $\theta_{a}^{A}$ real). This
can be used to reexpress the left hand side of Eq.\,(\ref{eq:deltaUeqRHS})
as 
\begin{align}
\delta\left(U_{a}\tilde{\partial}_{\mu}^{a}U_{a}^{\dagger}\right) & =\delta U_{a}\tilde{\partial}_{\mu}^{a}U_{a}^{\dagger}+U_{a}\tilde{\partial}_{\mu}^{a}\left(\delta U_{a}^{\dagger}\right)\\
 & =-U_{a}\tilde{\partial}_{\mu}^{a}\left(U_{a}^{\dagger}\delta U_{a}\right)U_{a}^{\dagger}\,\,.\label{eq:deltaUeqLHS}
\end{align}

Combining Eqs.\,(\ref{eq:deltaUeqRHS}) and (\ref{eq:deltaUeqLHS}),
we arrive at the following first order differential equation: 
\begin{equation}
\tilde{\partial}_{\mu}^{a}\left(-iU_{a}^{\dagger}\delta U_{a}\right)=\tilde{\partial}_{\mu}^{a}\left(U_{a}^{\dagger}\Gamma U_{a}\right)\,\,.\label{eq:GaugeBTGTDiffEqu}
\end{equation}
The general solution to Eq.\,(\ref{eq:GaugeBTGTDiffEqu}) is 
\begin{equation}
-iU_{a}^{\dagger}\delta U_{a}=U_{a}^{\dagger}\Gamma U_{a}+Z_{a}\quad\mbox{(no sum over a}),\label{eq:DiffEqSolution}
\end{equation}
where $Z_{a}$ is an infinitesimal zero mode that satisfies 
\begin{equation}
\tilde{\partial}_{\mu}^{a}Z_{a}=0.\quad\mbox{(no sum over a})\label{eq:Zero-mode-condition}
\end{equation}
Inhomogeneously transforming $\theta_{a}$ by this zero mode is the
BTGT symmetry of Eq.\,(\ref{eq:BTGTtransform}).

Since $-iU_{a}^{\dagger}\delta U_{a}$ is an element of the Lie algebra
spanned by $T^{A}$ and $U_{a}$ is unitary, we choose the boundary
conditions of Eq.\,(\ref{eq:GaugeBTGTDiffEqu}) such that $Z_{a}\equiv Z_{a}^{A}T^{A}$
for some real components $Z_{a}^{A}$ that each satisfy the zero mode
equation. Thus we have the result 
\begin{equation}
-i\delta U_{a}U_{a}^{\dagger}=\Gamma+U_{a}Z_{a}U_{a}^{\dagger}\,\,.
\end{equation}

To solve for the components $\delta\theta_{a}^{A}$ 
\begin{align}
-i\delta U_{a}U_{a}^{\dagger} & =-i\delta\left(e^{i\theta_{a}}\right)e^{-i\theta_{a}}\\
 & =\int_{0}^{1}dt\,e^{it\theta_{a}}\delta\theta_{a}e^{i\left(1-t\right)\theta_{a}}e^{-i\theta_{a}}\\
 & =\int_{0}^{1}dt\,e^{it\theta_{a}}\delta\theta_{a}e^{-it\theta_{a}}\\
 & =\int_{0}^{1}dt\sum_{n=0}^{\infty}\frac{\left(-it\right)^{n}}{n!}\left[\left[\dots\left[\left[\delta\theta_{a},\theta_{a}\right],\theta_{a}\right]\dots\right],\theta_{a}\right]\\
 & =\int_{0}^{1}dt\sum_{n=0}^{\infty}\frac{\left(-it\right)^{n}}{n!}\left[\left[\dots\left[\left[T^{B},T^{C_{1}}\right],T^{C_{2}}\right]\dots\right],T^{C_{n}}\right]\theta_{a}^{C_{1}}\cdots\theta_{a}^{C_{n}}\delta\theta_{a}^{B}\label{eq:prevstep}
\end{align}
where we made use of 
\begin{equation}
e^{-C}Be^{C}=1+\left[B,C\right]+\frac{1}{2}\left[\left[B,C\right],C\right]+\frac{1}{6}\left[\left[\left[B,C\right],C\right],C\right]+\dots\,\,.
\end{equation}
Note that
\begin{align}
\left[T^{B},T^{C_{1}}\right] & =iT^{D}f_{C_{1}}^{DB}=iT^{A}\left(f^{C_{1}}\right)^{AB}\\
\left[\left[T^{B},T^{C_{1}}\right],T^{C_{2}}\right] & =i\left[T^{D},T^{C_{2}}\right]f^{DBC_{1}}=i^{2}T^{A}f^{ADC_{2}}f^{DBC_{1}}=i^{2}T^{A}\left(f^{C_{2}}f^{C_{1}}\right)^{AB}.
\end{align}
Using iteration it is straight forward to show that
\begin{equation}
\left[\left[\dots\left[T^{B},T^{C_{1}}\right]\dots\right],T^{C_{n}}\right]=i^{n}T^{A}\left(f^{C_{n}}\cdots f^{C_{1}}\right)^{AB}
\end{equation}
such that the Eq.\,(\ref{eq:prevstep}) becomes 
\begin{equation}
-i\delta U_{a}U_{a}^{\dagger}=T^{A}\int_{0}^{1}dt\,\sum_{n=0}^{\infty}\frac{t^{n}}{n!}\left(f^{C_{n}}\cdots f^{C_{1}}\right)^{AB}\theta_{a}^{C_{1}}\cdots\theta_{a}^{C_{n}}\delta\theta_{a}^{B}=T^{A}\left(\frac{e^{f\cdot\theta_{a}}-1}{f\cdot\theta_{a}}\right)^{AB}\delta\theta_{a}^{B}\,\,.\label{eq:-idUUdag}
\end{equation}
Another useful identity in solving for $\delta\theta_{a}^{A}$ is
\begin{align}
U_{a}Z_{a}U_{a}^{\dagger} & =e^{i\theta_{a}}Z_{a}e^{-i\theta_{a}}\\
 & =\sum_{n=0}^{\infty}\frac{1}{n!}\left[\left[\dots\left[\left[Z_{a},\theta_{a}\right],\theta_{a}\right]\dots\right],\theta_{a}\right]\\
 & =T^{A}\sum_{n=0}^{\infty}\frac{1}{n!}\left(f^{C_{n}}\cdots f^{C_{1}}\right)^{AB}\theta_{a}^{C_{1}}\cdots\theta_{a}^{C_{n}}Z_{a}^{B}\\
 & =T^{A}\left(e^{f\cdot\theta_{a}}\right)^{AB}Z_{a}^{B}\,\,.\label{eq:UZUdag}
\end{align}

We can eliminate $T^{A}$ from both Eq.\,(\ref{eq:-idUUdag}) and
Eq.\,(\ref{eq:UZUdag}) to obtain 
\begin{equation}
\left(\frac{e^{f\cdot\theta_{a}}-1}{f\cdot\theta_{a}}\right)^{AB}\delta\theta_{a}^{B}=\Gamma^{A}+\left(e^{f\cdot\theta_{a}}\right)^{AB}Z_{a}^{B}\,\,.
\end{equation}
From here, we can immediately solve for $\delta\theta_{a}^{A}$ as
\begin{equation}
\delta\theta_{a}^{A}=\left(\frac{f\cdot\theta_{a}}{e^{f\cdot\theta_{a}}-1}\right)^{AB}\Gamma^{B}+\left(\frac{f\cdot\theta_{a}}{1-e^{-f\cdot\theta_{a}}}\right)^{AB}Z_{a}^{B}\,\,.\label{eq:thetaBTGTgaugetransformation}
\end{equation}
Again, both $\Gamma^{A}$ and $Z_{a}^{A}$ are infinitesimal parameters
in Eq.\,(\ref{eq:thetaBTGTgaugetransformation}). 

Next, we will express the finite gauge and BTGT transformations as
a left and right multiplication of a group element representation.
Start by writing the condition for $\delta U_{a}$ as 
\begin{equation}
\delta U_{a}=i\left(\epsilon\Gamma U_{a}+U_{a}\epsilon Z_{a}\right)=\epsilon\left(i\Gamma U_{a}+iU_{a}Z_{a}\right)
\end{equation}
where we added $\epsilon$ to $\Gamma$ and $Z$ to emphasize that
the transformation is infinitesimal. We can then rewrite the infinitesimal
transformation using the exponential map as 
\begin{align}
U_{a}\rightarrow U_{a}' & =U_{a}+i\epsilon\Gamma U_{a}+iU_{a}\epsilon Z_{a}\\
 & =\left(1+i\epsilon\Gamma\right)U_{a}\left(1+i\epsilon Z_{a}\right)+O\left(\epsilon^{2}\right)\\
 & =e^{i\epsilon\Gamma}U_{a}e^{i\epsilon Z_{a}}+O\left(\epsilon^{2}\right)\,\,.
\end{align}
Next, if we apply the infinitesimal transformation twice, we see 
\begin{equation}
U_{a}''=e^{i\epsilon\Gamma}U_{a}'e^{i\epsilon Z_{a}}=e^{i\epsilon\Gamma}e^{i\epsilon\Gamma}U_{a}e^{i\epsilon Z_{a}}e^{i\epsilon Z_{a}}=e^{2i\epsilon\Gamma}U_{a}e^{2i\epsilon Z_{a}}\,\,.
\end{equation}
Thus, we can then iterate this for $N=\frac{1}{\epsilon}$ times to
obtain the finite gauge transformation 
\begin{equation}
U_{a}\rightarrow e^{i\Gamma}U_{a}e^{iZ_{a}}
\end{equation}
which gives an elegant finite gauge and BTGT transformation expression.
This can also be expressed as 
\begin{equation}
e^{i\theta_{a}}\rightarrow e^{i\Gamma}e^{iU_{a}Z_{a}U_{a}^{\dagger}}e^{i\theta_{a}}\,\,.
\end{equation}

\section{Feynman Rules\label{sec:Feynman-Rules}}

\begin{figure}
\begin{centering}
\includegraphics[width=15cm]{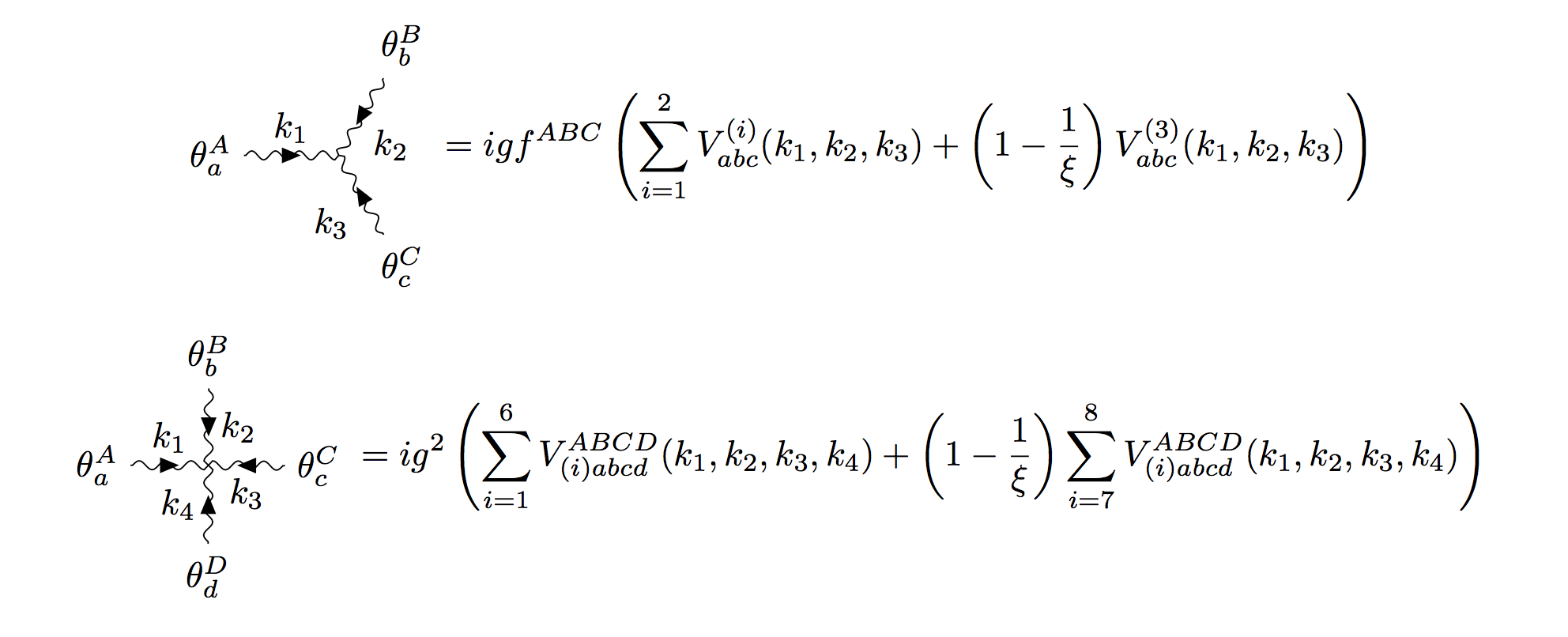}
\par\end{centering}

\caption{Gauge interaction vertices up to quartic order in $\theta$. \label{fig:Rules-theta-interaction-vertices}}
\end{figure}

The Feynman rules for non-Abelian BTGT are given in the following
figures. Fig.\,\ref{fig:Rules-Propagators} shows the propagators
for the gauge field $\theta_{a}^{A}$ and ghost fields $c^{A}$and
$d_{a}^{A}$. Fig.\,\ref{fig:Rules-theta-interaction-vertices} shows
the first three $\theta^{n}$ vertices that exist for all integer
$n\ge3$. There are an infinite number of such vertices, but they
are suppressed by higher powers of the gauge coupling $g$. The explicit
form of the $\theta^{5}$ vertex is not given in this paper because
it was lengthy to show and was not necessary for the computations
shown in this paper. It can be derived by expanding the Yang-Mills
actions written in terms of $A\left[\theta\right]$ and keeping the
$\theta^{5}$ terms.

\begin{figure}
\begin{centering}
\includegraphics[width=15cm]{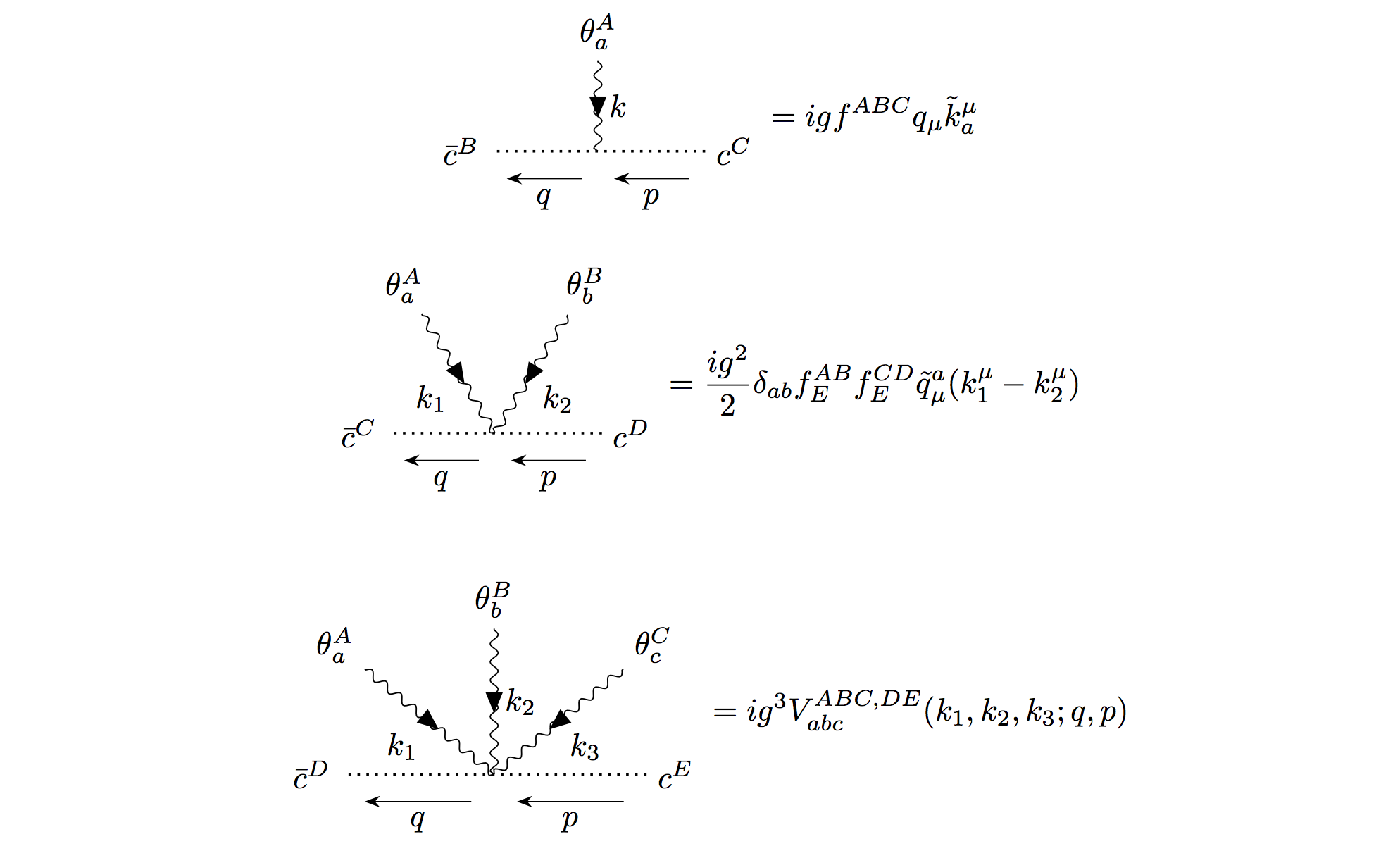}
\par\end{centering}

\caption{Ghost gauge vertices up to third order in $\theta$ \label{fig:Rules-ghost-gauge-vertices}}
\end{figure}

Fig.\,\ref{fig:Rules-ghost-gauge-vertices} shows the first three
ghost gauge interaction terms. Qualitatively, they are of the form
$V_{\theta^{n}\bar{c}c}\sim g^{n}\theta^{n}\bar{c}c$ for all $n\ge1$.
Like in the case of $V_{\theta^{n}}$, there are an infinite number
of such vertices but are suppressed by higher power of $g$.

\begin{figure}
\begin{centering}
\includegraphics[width=15cm]{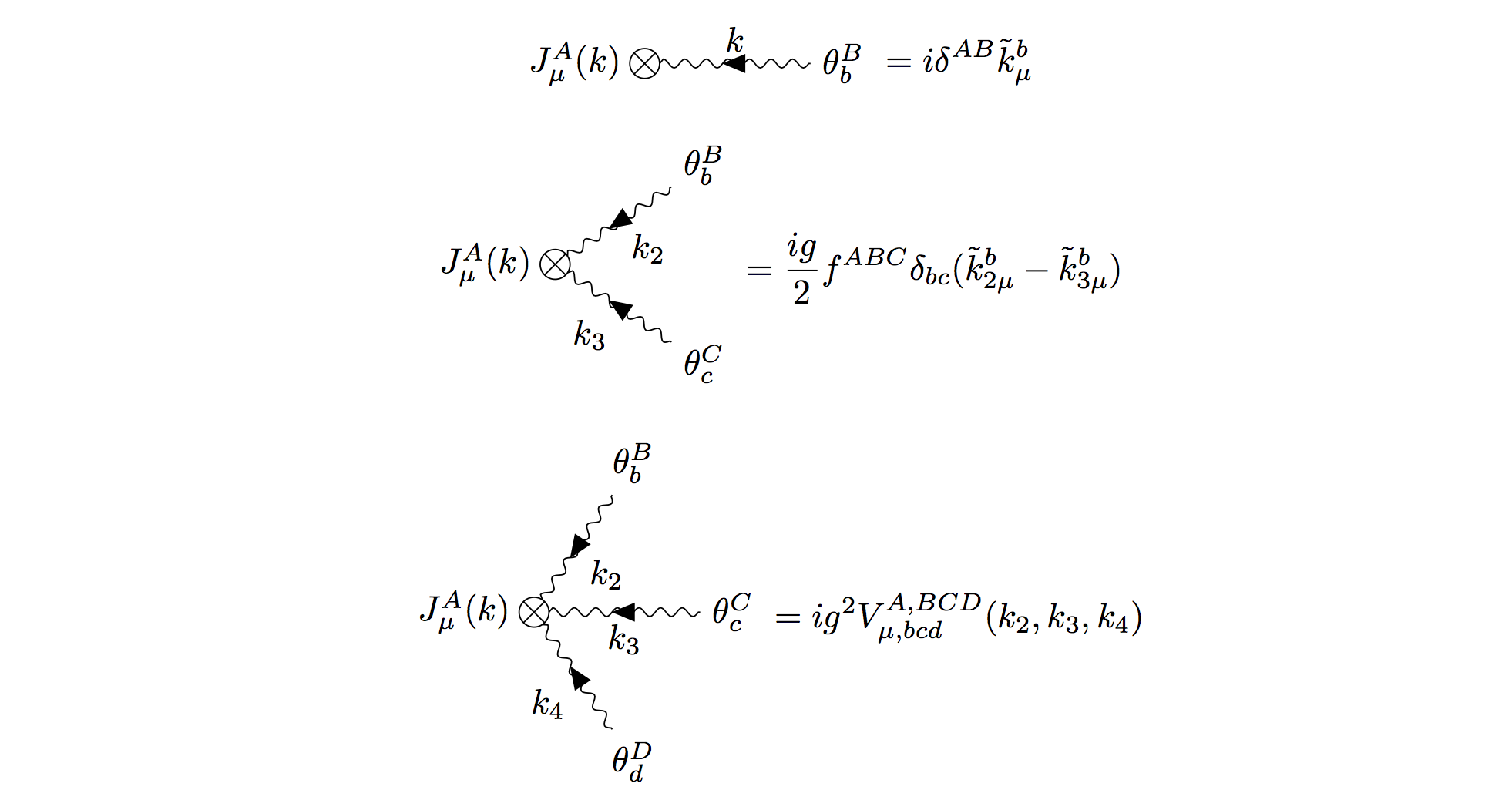}
\par\end{centering}

\caption{Composite operator vertices up to third order in $\theta$\label{fig:Rules-Composite-operator-vertices}}
\end{figure}

The composite operator $A_{\mu}^{A}\left[\theta\right]$ defined in
Eq.\,(\ref{eq:A=00005Btheta=00005D}) can be computed using the vertices
of Fig.\,\ref{fig:Rules-Composite-operator-vertices}.

\begin{figure}
\begin{centering}
\includegraphics[width=15cm]{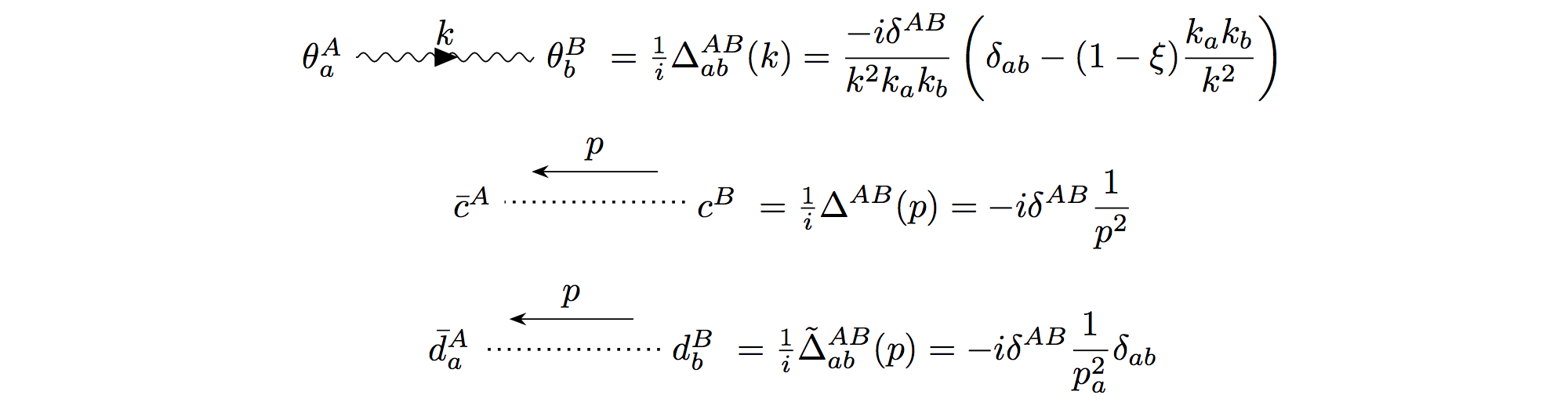}
\par\end{centering}

\caption{Propagators \label{fig:Rules-Propagators}}
\end{figure}

\begin{figure}
\begin{centering}
\includegraphics[width=15cm]{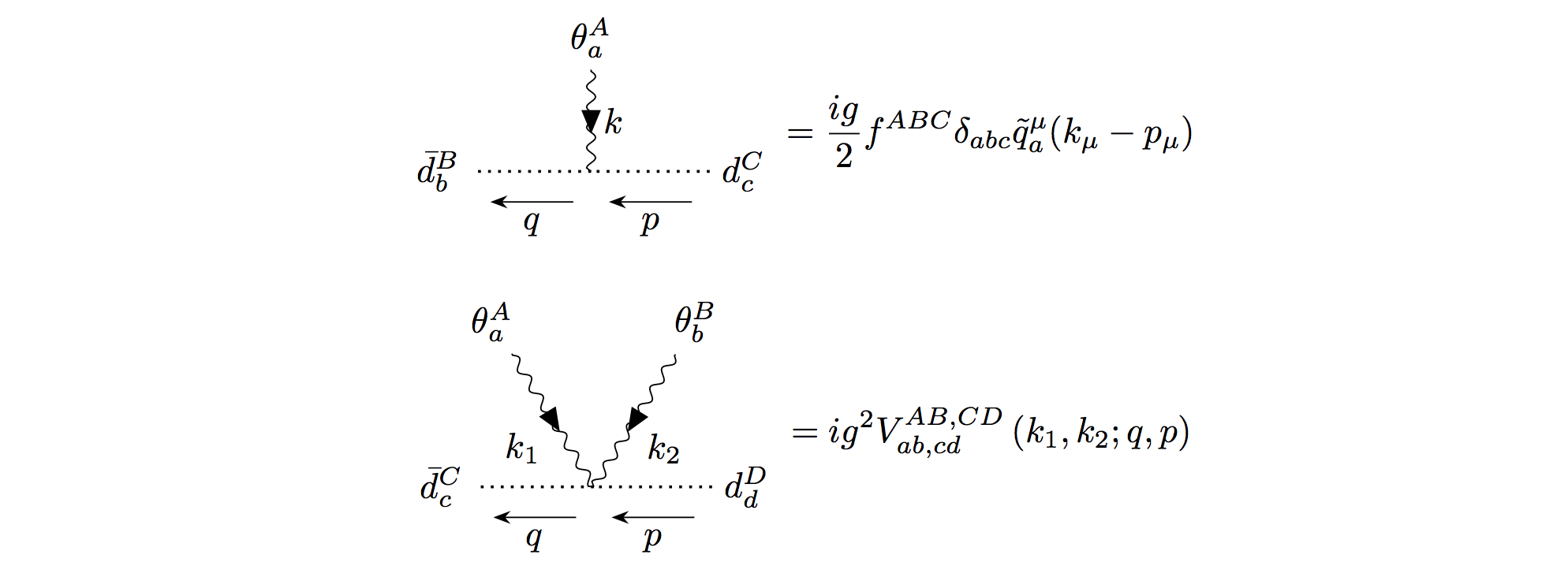}
\par\end{centering}

\caption{Additional ghost gauge vertices up to second order in $\theta$ \label{fig:Rules-additional-ghost-gauge} }
\end{figure}

\begin{figure}
\begin{centering}
\includegraphics[width=15cm]{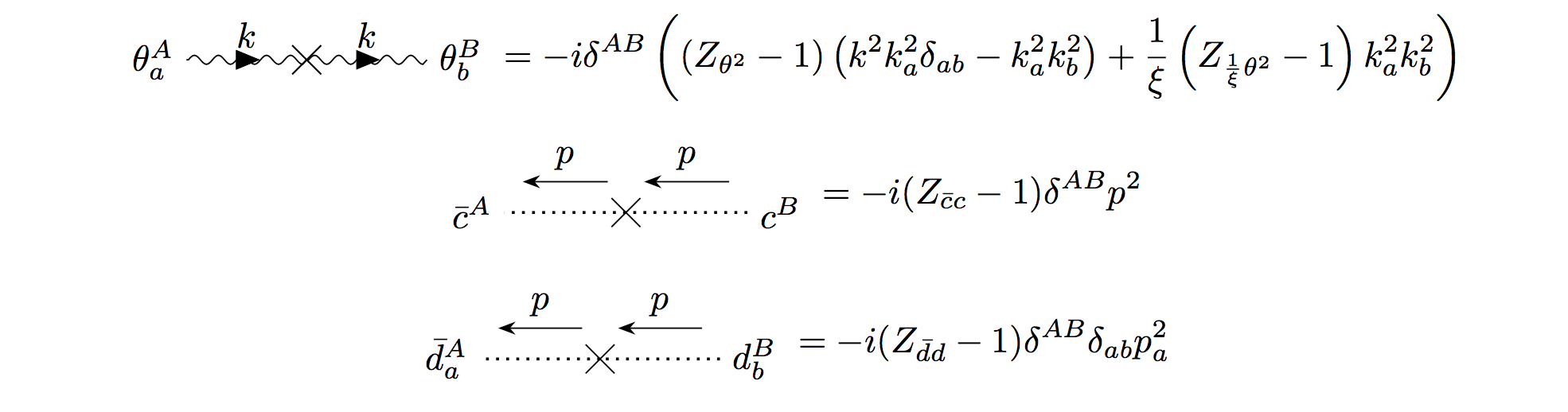}
\par\end{centering}

\caption{Quadratic counter-terms}
\end{figure}

\begin{figure}
\begin{centering}
\includegraphics[width=15cm]{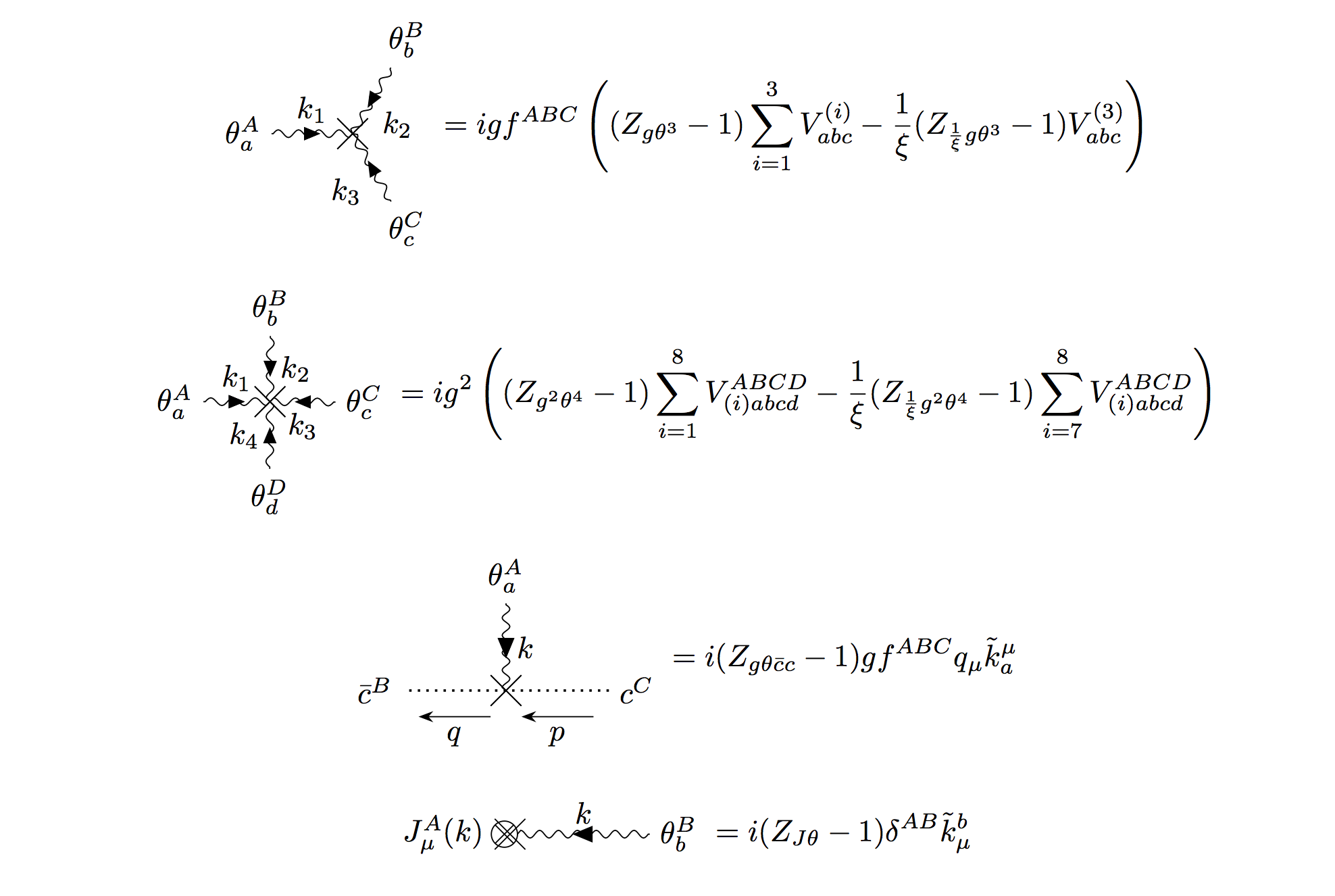}
\par\end{centering}

\caption{Interaction vertex counter-terms }
\end{figure}

\subsection{Explicit Vertex Expressions}

This section contains vertex expressions that were defined in the Feynman rules figures. The $\theta^{3}\bar{c}c$ vertex
$V_{abc}^{ABC,DE}\left(k_{1},k_{2},k_{3};q\right)$ defined in Figure
\ref{fig:Rules-ghost-gauge-vertices} is 
\begin{equation}
iV_{abc}^{ABC,DE}=\frac{i}{6}\delta_{abc}\tilde{q}_{\mu}^{a}f_{F}^{DE}\left(f_{G}^{FA}f_{G}^{BC}\left(k_{3}^{\mu}-k_{2}^{\mu}\right)+f_{G}^{FB}f_{G}^{CA}\left(k_{1}^{\mu}-k_{3}^{\mu}\right)+f_{G}^{FC}f_{G}^{AB}\left(k_{1}^{\mu}-k_{2}^{\mu}\right)\right),
\end{equation}
where the momenta are constrained to satisfy $q=k_{1}+k_{2}+k_{3}+p$.
The $\theta^{2}\bar{d}d$ vertex $V_{ab,cd}^{AB,CD}\left(k_{1},k_{2};q,p\right)$
defined in Fig.\,\ref{fig:Rules-additional-ghost-gauge} with $q=k_{1}+k_{2}+p$
is 
\begin{align}
iV_{ab,cd}^{AB,CD}= & \frac{i}{6}\delta_{abcd}\tilde{q}_{\mu}^{a}\left(f_{E}^{AB}f_{E}^{CD}(k_{1}^{\mu}-k_{2}^{\mu})+f_{E}^{AC}f_{E}^{BD}(p^{\mu}-k_{2}^{\mu})+f_{E}^{AD}f_{E}^{BC}(p^{\mu}-k_{1}^{\mu})\right).
\end{align}
The $J\theta^{3}$ vertex $V_{\mu,bcd}^{A,BCD}\left(k_{2},k_{3},k_{4}\right)$
defined in Fig.\,\ref{fig:Rules-Composite-operator-vertices} is
\begin{equation}
iV_{\mu,bcd}^{A,BCD}=\frac{i}{6}\delta_{bcd}\left(f_{E}^{AB}f_{E}^{CD}(\tilde{k}_{4\mu}^{b}-\tilde{k}_{3\mu}^{b})+f_{E}^{AC}f_{E}^{BD}(\tilde{k}_{4\mu}^{b}-\tilde{k}_{2\mu}^{b})+f_{E}^{AD}f_{E}^{BC}(\tilde{k}_{3\mu}^{b}-\tilde{k}_{2\mu}^{b})\right),
\end{equation}
where the composite operator momentum is $k=-k_{2}-k_{3}-k_{4}$.

\subsection{Quartic vertex terms}

Here, we are using the notation $\delta_{bcd}=\delta_{bc}\delta_{cd}$
such as to avoid confusion regarding summation. The quartic BTGT gauge
vertex in Fig.\,\ref{fig:Rules-theta-interaction-vertices} is given
by
\begin{equation}
iV_{abcd}^{ABCD}=ig^{2}\left(\sum_{i=1}^{6}V_{(i)abcd}^{ABCD}\left(k_{1},k_{2},k_{3},k_{4}\right)+\left(1-\tfrac{1}{\xi}\right)\sum_{i=7}^{8}V_{(i)abcd}^{ABCD}\left(k_{1},k_{2},k_{3},k_{4}\right)\right)
\end{equation}
where the momenta $k_{i}$ must sum to zero. In a diagonal basis for
$H^{a}$, the eight terms are given by
\begin{align}
V_{(1)abcd}^{ABCD}= & -k_{1a}k_{2b}k_{3c}k_{4d}\left(f_{E}^{AB}f_{E}^{CD}\left(\delta_{ac}\delta_{bd}-\delta_{ad}\delta_{bc}\right)+f_{E}^{AC}f_{E}^{BD}\left(\delta_{ab}\delta_{cd}-\delta_{ad}\delta_{bc}\right)\right.\nonumber \\
 & \left.+f_{E}^{AD}f_{E}^{BC}\left(\delta_{ab}\delta_{cd}-\delta_{ac}\delta_{bd}\right)\right)
\end{align}
\begin{align}
V_{(2)abcd}^{ABCD}= & -\frac{1}{2}\left\{ f_{E}^{AB}f_{E}^{CD}\left[\delta_{bcd}k_{1a}\left(k_{3a}+k_{4a}\right)k_{2b}\left(k_{3b}-k_{4b}\right)+\delta_{acd}k_{2b}\left(k_{3b}+k_{4b}\right)k_{1a}\left(k_{4a}-k_{3a}\right)\right.\right.\nonumber \\
 & \left.+\delta_{abd}k_{3c}\left(k_{1c}+k_{2c}\right)k_{4a}\left(k_{1a}-k_{2a}\right)+\delta_{abc}k_{4d}\left(k_{1d}+k_{2d}\right)k_{3a}\left(k_{2a}-k_{1a}\right)\right]\nonumber \\
 & +f_{E}^{AC}f_{E}^{BD}\left[\delta_{bcd}k_{1a}\left(k_{2a}+k_{4a}\right)k_{3b}\left(k_{2b}-k_{4b}\right)+\delta_{abd}k_{3c}\left(k_{2c}+k_{4c}\right)k_{1a}\left(k_{4a}-k_{2a}\right)\right.\nonumber \\
 & \left.+\delta_{acd}k_{2b}\left(k_{1b}+k_{3b}\right)k_{4d}\left(k_{1d}-k_{3d}\right)+\delta_{abc}k_{4\star d}\left(k_{1d}+k_{3d}\right)k_{2b}\left(k_{3b}-k_{1b}\right)\right]\nonumber \\
 & +f_{E}^{AD}f_{E}^{BC}\left[\delta_{bcd}k_{1a}\left(k_{2a}+k_{3a}\right)k_{4d}\left(k_{2d}-k_{3d}\right)+\delta_{abc}k_{4c}\left(k_{2c}+k_{3c}\right)k_{1a}\left(k_{3a}-k_{2a}\right)\right.\nonumber \\
 & \left.+\delta_{acd}^{\phantom{A}}\left.k_{2b}\left(k_{1b}+k_{4b}\right)k_{3c}\left(k_{1c}-k_{4c}\right)+\delta_{abd}k_{3c}\left(k_{1c}+k_{4c}\right)k_{2b}\left(k_{4b}-k_{1b}\right)\right]\right\} 
\end{align}
\begin{align}
V_{(3)abcd}^{ABCD}= & -\frac{1}{2}\left\{ f_{E}^{AB}f_{E}^{CD}\left[\delta_{acd}k_{1b}k_{2b}k_{1a}\left(k_{3a}-k_{4a}\right)+\delta_{bcd}k_{1a}k_{2a}k_{2b}\left(k_{4b}-k_{3b}\right)\right.\right.\nonumber \\
 & +\left.\delta_{abc}k_{3d}k_{4d}k_{3a}\left(k_{1a}-k_{2a}\right)+\delta_{abd}k_{3c}k_{4c}k_{4a}\left(k_{2a}-k_{1a}\right)\right]\nonumber \\
 & +f_{E}^{AC}f_{E}^{BD}\left[\delta_{abd}k_{1c}k_{3c}k_{1a}\left(k_{2a}-k_{4a}\right)+\delta_{bcd}k_{1a}k_{3a}k_{3b}\left(k_{4b}-k_{2b}\right)\right.\nonumber \\
 & \left.+\delta_{abc}k_{2d}k_{4d}k_{2a}\left(k_{1a}-k_{3a}\right)+\delta_{acd}k_{2b}k_{4b}k_{4a}\left(k_{3a}-k_{1a}\right)\right]\nonumber \\
 & +f_{E}^{AD}f_{E}^{BC}\left[\delta_{abc}k_{1d}k_{4d}k_{1a}\left(k_{2a}-k_{3a}\right)+\delta_{bcd}k_{1a}k_{4a}k_{4b}\left(k_{3b}-k_{2b}\right)\right.\nonumber \\
 & \left.+\delta_{abd}^{\phantom{A}}\left.k_{2c}k_{3c}k_{2a}\left(k_{1a}-k_{4a}\right)+\delta_{acd}k_{2b}k_{3b}k_{3a}\left(k_{3a}-k_{1a}\right)\right]\right\} 
\end{align}
\begin{align}
V_{(4)abcd}^{ABCD}= & \frac{1}{2}\left\{ f_{E}^{AB}f_{E}^{CD}\delta_{ab}\delta_{cd}\left[k_{1a}k_{2a}(k_{1c}-k_{2c})(k_{3c}-k_{4c})\right.\right.\nonumber \\
 & \left.+k_{3c}k_{4c}(k_{1a}-k_{2a})(k_{3a}-k_{4a})\right]\nonumber \\
 & +f_{E}^{AC}f_{E}^{BD}\delta_{ac}\delta_{bd}\left[k_{1a}k_{3a}\left(k_{1b}-k_{3b}\right)\left(k_{2b}-k_{4b}\right)\right.\nonumber \\
 & \left.+k_{2b}k_{4b}\left(k_{1a}-k_{3a}\right)\left(k_{2a}-k_{4a}\right)\right]\nonumber \\
 & +f_{E}^{AD}f_{E}^{CD}\delta_{ad}\delta_{bc}\left[k_{1a}k_{4a}(k_{1b}-k_{4b})(k_{2b}-k_{3b})\right.\nonumber \\
 & \left.+k_{2b}^{\phantom{A}}\left.k_{3b}(k_{1a}-k_{4a})(k_{2a}-k_{3a})\right]\right\} 
\end{align}
\begin{align}
V_{(5)abcd}^{ABCD}= & \frac{1}{4}\delta_{abcd}\left\{ f_{E}^{AB}f_{E}^{CD}\left(k_{1}+k_{2}\right)^{2}\left(k_{1a}-k_{2a}\right)\left(k_{3a}-k_{4a}\right)\right.\nonumber \\
 & +f_{E}^{AC}f_{E}^{BD}\left(k_{1}+k_{3}\right)^{2}\left(k_{1a}-k_{3a}\right)\left(k_{2a}-k_{4a}\right)\nonumber \\
 & \left.+f_{E}^{AD}f_{E}^{BC}\left(k_{1}+k_{4}\right)^{2}\left(k_{1a}-k_{4a}\right)\left(k_{2a}-k_{3a}\right)\right\} 
\end{align}
\begin{align}
V_{(6)abcd}^{ABCD}= & \frac{1}{6}\delta_{abcd}\left\{ f_{E}^{AB}f_{E}^{CD}\left[\left(k_{1}^{2}k_{1a}-k_{2}^{2}k_{2a}\right)\left(k_{4a}-k_{3a}\right)+\left(k_{3}^{2}k_{3a}-k_{4}^{2}k_{4a}\right)\left(k_{2a}-k_{1a}\right)\right]\right.\nonumber \\
 & +f_{E}^{AC}f_{E}^{BD}\left[\left(k_{1}^{2}k_{1a}-k_{3}^{2}k_{3a}\right)\left(k_{4a}-k_{2a}\right)+\left(k_{2}^{2}k_{2a}-k_{4}^{2}k_{4a}\right)\left(k_{3a}-k_{1a}\right)\right]\nonumber \\
 & \left.+f_{E}^{AD}f_{E}^{BC}\left[\left(k_{1}^{2}k_{1a}-k_{4}^{2}k_{4a}\right)\left(k_{3a}-k_{2a}\right)+\left(k_{2}^{2}k_{2a}-k_{3}^{2}k_{3a}\right)\left(k_{4a}-k_{1a}\right)\right]\right\} 
\end{align}
\begin{align}
V_{(7)abcd}^{ABCD}= & -\frac{1}{4}\left\{ f_{E}^{AB}f_{E}^{CD}\delta_{ab}\delta_{cd}\left(k_{1a}-k_{2a}\right)\left(k_{1a}+k_{2a}\right)\left(k_{3c}-k_{4c}\right)\left(k_{1c}+k_{2c}\right)\right.\nonumber \\
 & +f_{E}^{AC}f_{E}^{BD}\delta_{ac}\delta_{bd}\left(k_{1a}-k_{3a}\right)\left(k_{1a}+k_{3a}\right)\left(k_{2b}-k_{4b}\right)\left(k_{1b}+k_{3b}\right)\nonumber \\
 & +\left.f_{E}^{AD}f_{E}^{BC}\delta_{ad}\delta_{bc}\left(k_{1a}-k_{4a}\right)\left(k_{1a}+k_{4a}\right)\left(k_{2b}-k_{3b}\right)\left(k_{1b}+k_{4b}\right)\right\} 
\end{align}
\begin{alignat}{1}
V_{(8)abcd}^{ABCD}= & -\frac{1}{6}\left\{ f_{E}^{AB}f_{E}^{CD}\left[\delta_{bcd}k_{1a}^{2}k_{1b}\left(k_{4b}-k_{3b}\right)+\delta_{acd}k_{2b}^{2}k_{2a}\left(k_{3a}-k_{4a}\right)\right.\right.\nonumber \\
 & \left.+\delta_{abd}k_{3c}^{2}k_{3a}\left(k_{2a}-k_{1a}\right)+\delta_{abc}k_{4d}^{2}k_{4a}\left(k_{1a}-k_{2a}\right)\right]\nonumber \\
 & +f_{E}^{AC}f_{E}^{BD}\left[\delta_{bcd}k_{1a}^{2}k_{1b}\left(k_{4b}-k_{2b}\right)+\delta_{abd}k_{3c}^{2}k_{3a}\left(k_{2a}-k_{4a}\right)\right.\nonumber \\
 & +\left.\delta_{acd}k_{2b}^{2}k_{2a}\left(k_{3a}-k_{1a}\right)+\delta_{abc}k_{4d}^{2}k_{4a}\left(k_{1a}-k_{3a}\right)\right]\nonumber \\
 & +f_{E}^{AD}f_{E}^{BC}\left[\delta_{bcd}k_{1a}^{2}k_{1b}\left(k_{3b}-k_{2b}\right)+\delta_{abc}k_{4d}^{2}k_{4a}\left(k_{2a}-k_{3a}\right)\right.\nonumber \\
 & \left.+\delta_{acd}^{\phantom{A}}\left.k_{2b}^{2}k_{2a}\left(k_{4a}-k_{1a}\right)+\delta_{abd}k_{3c}^{2}k_{3a}\left(k_{1a}-k_{4a}\right)\right]\right\} \,\,.
\end{alignat}

\bibliographystyle{JHEP2}
\bibliography{nonabelian-btgtpaper}

\end{document}